\documentclass{aa} 

\usepackage{threeparttable}
\usepackage{booktabs}  
\usepackage{siunitx}
\usepackage{rotating}
\usepackage{float}
\usepackage{orcidlink}
\usepackage{placeins}

\newcommand{\HH}{H$_2$}
\newcommand{\micron}{$\mu$m}  
\newcommand{\tmicron}{\text{\micron}}  
\renewcommand\makeLineNumber{}

\begin{document}

\title{JWST Observations of Photo-dissociation Regions. \\
II. Aliphatic/Aromatic Carbonaceous Dust, Ices, and Gas Phase Spectral Line Inventory}

\author{K. Misselt\inst{1}        \and 
        A.~N. Witt\inst{2}         \and 
        K.~D. Gordon\inst{3, 4, \orcidlink{0000-0001-5340-6774}}    \and 
        D. Van De Putte\inst{5,3} \and
        B. Trahin\inst{3, 6}      \and
        A. Abergel\inst{6}        \and 
        A. Noriega-Crespo\inst{3} \and 
        P. Guillard\inst{7}       \and 
        M. Zannese\inst{6}        \and
        P. Dell'ova\inst{6}       \and     
        M. Baes\inst{4, \orcidlink{0000-0002-3930-2757}}           \and
        P. Klaassen\inst{8}       \and
        N. Ysard\inst{9,6}}

\institute{ 
Steward Observatory, University of Arizona, Tucson, AZ 85721-0065, USA \label{inst1} \and
Ritter Astrophysical Research Center, University of Toledo, Toledo, OH 43606, USA \label{inst2} \and
Space Telescope Science Institute, 3700 San Martin Drive, Baltimore, MD, 21218, USA \label{inst3} \and
Sterrenkundig Observatorium, Universiteit Gent, Krijgslaan 281 S9, B-9000 Gent, Belgium \label{inst4} \and
Department of Physics \& Astronomy, The University of Western Ontario, London ON N6A 3K7 Canada\label{inst5} \and
Institut d'Astrophysique Spatiale, Université Paris-Saclay, CNRS, 91405 Orsay, France \label{inst6} \and
Sorbonne Universit\'{e}, CNRS, Institut d'Astrophysique de Paris, 98\,bis bd Arago, 75014 Paris, France\label{inst7} \and
United Kingdom Astronomy Technology Centre, Edinburgh, GB, UK\label{inst8} \and
Institut de Recherche en Astrophysique et Plan\'etologie, Universit\'e Toulouse III - Paul Sabatier, CNRS, CNES, 9 Av. du colonel Roche, 31028 Toulouse, France\label{inst9}
}

\abstract
{}
{This paper provides an overview of the spectroscopic data obtained by the JWST 
Guaranteed Time Observations (GTO) program 1192, "The Physics and Chemistry of PDR Fronts",
including an inventory of the spatially resolved dust, gas, 
and molecular content in the Horsehead nebula and the NW filament of NGC~7023. 
We demonstrate the unique capability of this high spatial resolution data set 
to elucidate the evolution of gas and dust at the interface between stars 
and their natal clouds at the scale at which the physics and chemistry occur.}
{The Disassociation Regions (PDRs) in the Horsehead nebula and the North West (NW) 
filament of NGC~7023 were mapped at a spectral resolving power between 1000-3000 and 
a spatial resolution of 2-7$\times$10$^{-4}$~pc between 0.97-28~\micron. The data were
obtained as long, narrow strips with the long axis aligned with the line of sight
between the exciting start and perpendicular to the PDR front.}
{Spectra extracted from the template regions yield a large number of atomic, ionized,
and molecular lines. Full line lists and extracted spectra for all 10 regions are 
provided through CDS. Absorption from H$_2$O, CO$_2$, and CO ices are identified
in 3 regions in NGC~7023. In this overview, we have focused on the spectral region 
between 3 and 5~\micron\ which is dominated by emission from aromatic and aliphatic
carbon bonds to illustrate the power of the data set. We confirm the entrainment of
aromatic carbonaceous species in the photo-evaporative flow from the PDR surface
into the H\,{\sc{ii}} region in the Horsehead. No aliphatic emission is present in
the outflow, indicating the complete removal of aliphatic bonds when exposed to strong
UV fields. There is a clear detection of deuterium substitution in the carbon bonds.
Aliphatic D-substitution is more efficient relative to aromatic D-substitution,
ranging from N$_{D}$/N$_{H}\sim$ 0.1-0.3  for aliphatics compared to  $\sim$0.03
for the aromatics.}
{}

\keywords{Infrared: ISM: NGC~7023, Horsehead, dust, extinction, molecules, lines and bands, photon-dominated region (PDR), H\,II regions, Techniques: photometric, spectroscopy, Methods: observational, data analysis}

\titlerunning{NGC7023 and Horsehead Spectra}

\maketitle

\section{Introduction} \label{sec:intro}
Photo-dissociation regions (PDRs), alternatively referred to as photon-dominated regions, are regions of 
interstellar gas and dust that are permeated by far-ultraviolet (FUV)  continuum radiation (6 – 13.6 eV) 
from hot stars. This radiation is sufficiently energetic to dissociate common interstellar molecules such 
as \HH\ and CO and to ionize many abundant atomic species with the notable exception of hydrogen and helium. 
The physical conditions and the chemistry of PDRs are largely determined by the strength of the incident FUV radiation field. In cases where the FUV radiation originates with a single star or small group of stars near the surface of a dense molecular cloud, a PDR presents a multilayered structure with different zones characterized by varying degrees of dissociation and ionization as the intensity and hardness of 
the incident radiation diminishes with the depth of penetration into the cloud 
\citep{1985ApJ...291..722T,1999RvMP...71..173H,2022ARA&A..60..247W,2023ASSP...59..129T}.

Two PDRs, the Horsehead nebula, irradiated by the (O9.5V + B0.2V) binary star Sigma Orionis ($T_{\text{eff}}\simeq32000$~K), and the
North West (NW) PDR in the reflection nebula NGC 7023, irradiated by the B2Ve binary star HD 200775 ($T_{\text{eff}}\simeq18000$~K), are cases where
the stellar radiation impinges on the PDR surfaces at angles close to 90 degrees with respect to our line
of sight, thus offering the possibility of studying the multilayered structure of these PDRs. Importantly, 
the two PDRs differ significantly in the intensity and hardness of their respective radiation fields. 
\citet{2015A&A...577A..16P} estimate a FUV flux of $\simeq2600 G_0$ at the surface of the NGC~7023 NW PDR, where $G_0$
is the Habing unit of FUV flux, with $G_0 = 1$ corresponding to $1.59\times10^{-3}$~erg~s$^{-1}$~cm$^{-2}$, the average
interstellar UV field in the ISM in the solar vicinity \citep{1968BAN....19..421H}.
The FUV field at the surface of the Horsehead PDR is estimated to be $\sim100 G_0$
\citep{2005A&A...437..177H}.  While less intense than the FUV field at the NW PDR of NGC~7023, the FUV
field at the surface of the Horsehead nebula is substantially harder, being dominated by the O9.5V component 
of Sigma Ori ($T_{\text{eff}}\simeq32000$~K) compared to the B2Ve spectrum of HD~200775 
($T_{\text{eff}}\simeq18000$~K) illuminating the NGC~7023 PDR.
Given their near-ideal geometry and relative proximity -- $\sim$355~pc \citep{2020yCat.1350....0G} 
and $\sim$400~pc \citep{1982AJ.....87.1213A}, respectively -- the NGC~7023 and Horsehead PDRs have been 
extensively studied. 

\citet{1988ApJ...334..803C} presented the first model of the NGC~7023 PDR,
based on Kuiper Airborne Observatory observations of the far-infrared fine-structure lines of O\,{\sc{i}} (63~\micron) 
and C\,{\sc{ii}} (158~\micron). \citet{2000ESASP.456...95F}, using ISO SWS and LWS observations, reported strong 
[Si\,{\sc{ii}}] (34.8~\micron) emission in the nebular cavity in front of the NW PDR, peaking near the position of 
HD~200775, indicative of partial photo-evaporation of silicate grains. \citet{2006ApJ...636..303W} used the high spatial
resolution afforded by the ACS and NICMOS cameras onboard the Hubble Space Telescope (HST) to map the distribution of extended red emission (ERE, $\sim$0.54-0.9~\micron) and the 1-0~S(1) H$_2$ line in the NW PDR of NGC 7023. Based on the widths of emission filaments in both features, their
spatial correlation, and the known excitation energy of the H$_2$ line, they concluded that FUV photons initiate the ERE
process in the $\sim$10.5-13.6~eV energy range.
\citet{2013A&A...552A..15M} modeled the evolution of poly-cyclic aromatic hydrocarbons (PAHs) in the 
NW PDR of NGC 7023, examining their hydrogenation and charge states. They found that PAHs with fewer than 
$N_{\text{C}} \approx 50$ carbon atoms are unlikely to survive in this PDR, and small carbon clusters are suggested to be the likely end 
stage of the PAH photo-dissociation process. \citet{2014A&A...569A.109K} conducted a detailed analysis 
of the CO-gas and dust emission in NGC 7023, using instruments onboard the Herschel Space Observatory.
Their observations revealed a steep density gradient at the front of the PDR, rising to densities of 
$10^5-10^6$~H~cm$^{-3}$ in the NW PDR. \citet{2016A&A...590A..26C} mapped the PAH sizes across the NW PDR
using SOFIA and archival data. They found PAH sizes with $N_{\text{C}}\approx 70$ in the cavity in front of 
the PDR and PAHs with $N_{\text{C}} \approx 50$ in the PDR surface regions. 
    
Similar studies have been carried out on the Horsehead nebula PDR. \citet{2005A&A...437..177H} carried out 
a detailed study of the density structure of the Horsehead PDR in nearly the same region that is to be 
discussed in the present paper, albeit with a resolution limited by ground-based observations. They 
detected a steep density gradient at the cloud’s edge with a scale length of $\sim$0.02 pc, leading to 
densities of $n_{\text{H}} \sim 10^{4}-10^{5}$~cm$^{-3}$ in the front portions of the PDR. \citet{2007A&A...464L..41P} 
reported on deuterium fractionation found through observations of DCO+ with the IRAM 30-m telescope. 
\citet{2020A&A...639A.144S} investigated dust evolution across the Horsehead PDR within the context of the 
THEMIS dust model \citep{2024A&A...684A..34Y}, employing observations with the Spitzer and Herschel space observatories. 
They found a factor of 6-10 reduction in the abundance and a 2–2.25 increase in the minimum size of 
carbonaceous nano-grains relative to conditions in the diffuse ISM in the front portion of the PDR. 
A similar study by \citet{elyajouri2024} but based on JWST observations from this program
revealed a less steep size distribution of carbonaceous nano-grains compared to the diffuse ISM, consistent 
with photo-processing of nano-grains in a PDR with moderate UV illumination. A presentation of first results 
from a multi-band near- and mid-IR imaging study of the Horsehead PDR with the JWST NIRCam and MIRI 
instruments carried out during our current JWST program has been published by \citet{2024A&A...687A...4A}.

In the present paper, we provide a spectral feature inventory obtained from NIRSpec 
and MIRI Integral Field Unit (IFU) \citep{2008SPIE.7010E..11C,2015PASP..127..646W} 
spectra of the NGC 7023 NW PDR and the Horsehead PDR during our JWST Guaranteed Time
Observing (GTO) program, PID-1192,
which was constructed with time contributions from three GTO stakeholders,  NIRCam, MIRI-US, and MIRI-EC.
We review the observing program,  data reduction, and data processing in Section 2, followed by a description 
of the line extraction and identification in Section 3. In Section 4, we provide a broad overview of several 
interesting results derived from the current data set, illustrating the power of combined JWST data sets to 
illuminate physical processes in PDRs. These results include 
line strength ratios of H\,{\sc{i}}, the plethora of ro-vibrational and pure rotational lines of \HH, the detection of 
other molecular species including CO and CH$^+$, as well as H$_2$O, CO, and CO$_2$ ices, dust outflow in the 
Horsehead, the relative stability of aromatic and aliphatic hydrocarbon bonds when exposed to increasing 
UV fields and the detection of deuterated aromatic and aliphatic hydrocarbons, indicative of 
high degrees of deuterium fractionation.

We note that, for the remainder of the paper, we refer to the broad IR emission features, commonly referred
to as PAHs, with the  more general term 'carbonaceous features'.
While the IR emission features are firmly associated with carbonaceous material, the specific identification with 
PAHs is less clear. Some dust models explicitly associate the family of IR emission features with true PAH molecules 
\citep[e.g.][]{Zubko04,Draine07model,2023ApJ...948...55H} while others \citep[e.g.][]{2013A&A...558A..62J,2024A&A...684A..34Y}
do not include PAH molecules but rather ascribe the emission to aromatic and alaphatic units within a
disordered carbon structure (a-C(:H)). In addition to unambiguously aliphatic features which are not produced by 
strictly planar PAH molecules (e.g. 3.4~\micron\ feature), there are, especially in the JWST era, multiple 
unidentified substructures in the various emission features. For those reasons we prefer not to restrict the 
discussion to a particular class of materials.

\section{Data \label{sec:data}}
\subsection{Observations \label{ssec:obs}}
The observations are part of the JWST Guaranteed Time Observing (GTO) program 1192.
Here, we describe the details of the spectroscopic data acquisition and reduction; for 
details on the imaging data, see \citet{2024A&A...687A...4A}.
Due to the limited field of view of the MIRI 
and NIRSpec IFUs, the IFU maps were restricted to a smaller subset of the region of the PDR covered by the 
imaging data. To resolve the transition from ionized to neutral to molecular regions, the IFU mosaics
were constructed as narrow rectangles with the long axis oriented along the line of sight from the exciting 
star, roughly perpendicular to the main PDR front, and was centered on the peak 
of the emission seen in previous datasets \citep[e.g.,][]{2003A&A...410..577A, 2015A&A...577A..16P, 2020A&A...639A.144S}. 
With this construction, spectra can be extracted as a function of depth into the molecular cloud to resolve
the evolution of the PDR emission with depth.  To achieve this coverage with the IFU, we designed the mosaic using 6 pointings in the row direction of the IFU detector with a 10\% overlap between the individual
pointings.  When combined with stringent position angle constraints (\ang{135} or \ang{315} for NGC~7023 
and \ang{68} or \ang{248} for the Horsehead), the resulting mosaic is a strip approximately 
$(3\farcs3-8\arcsec) \times (21\arcsec-26\arcsec)$\footnote{Ranges account for the variation in the fields of view for the
IFU channels.} perpendicular to the PDR front with the line of sight to the exciting star aligning
roughly with a detector axis.  The final orientation of the IFU mosaics with respect to the larger
Horsehead and NGC~7023 complexes, as well as imaging footprints, is shown in Fig.~\ref{fig:img_regions}.

We estimated exposure times with the ETC\footnote{\url{https://jwst.etc.stsci.edu/}}
using spectra available in the literature
\citep{2000ApJ...544..859G,2004ApJS..154..309W,2007A&A...471..205C}, merged and scaled to the
brightness of the peak of Spitzer IRAC imaging in our IFU target region.
This resulted in a depth of $\sim$122 seconds per pixel in all MIRI channels for NGC7023 (11 groups in FASTR1 at
4 dither positions) and $\sim$144 seconds per pixel for the Horsehead (26 groups in FASTR1, 2 dither positions). 
The 2 dither positions of the Horsehead background were a compromise to optimize the observations given the 
allocated time (and overheads) in the GTO program. Given the complexity of the Horsehead surroundings, 
the depth of the background observations has shown (see section 2.2) to be a limiting factor on accurately 
measuring the continuum. For NIRSpec, a 3-pointing cycling dither strategy was employed.  For NGC7023, 4-5 
groups in NRSIRS2RAPID were obtained for a depth of $\sim$250 seconds per pixel. For the fainter Horsehead, we 
obtained 4 groups in NRSIRS2 mode (5 frame coadds per down-linked group) for a total depth of $\sim$875 seconds per 
pixel in each grating.

To account for background emission, we specified backgrounds for both the NIRSpec and MIRI IFUs. 
Regions that were relatively free of emission in Spitzer and WISE images were selected for each object and 
dedicated exposures with an identical configuration to a single on-source mosaic pointing were obtained.  
For NIRSpec, full leak calibrations were obtained for both the on-source mosaics and the dedicated background 
pointings. While this doubles the exposure time for NIRSpec, it was considered crucial to remove potential 
MSA leak-through on bright extended source pointings, especially in the early mission, where the importance
of MSA leaks was not fully characterized. 

\subsection{Reduction \label{ssec:red}}
While our IFU data were automatically processed within the pipeline environment, for reasons
explained below, we downloaded `raw' files (uncal.fits) from MAST and reprocessed them
locally.  Our NGC~7023 MIRI and NIRSpec data were processed with pipeline version 1.14.0 and
CRDS context 1242 and 1241, respectively.  The Horsehead NIRSpec data were also processed with
1.14.0/1242 while the MIRI data were processed with 1.17.0/1321.  Specific deviations from 
the pipeline processing are noted below.

\subsubsection{MIRI \label{sssec:miri}}
The MIRI data (source and background) were processed through level-1b (rate). 
Before running the level-2 (cal) pipeline, we implemented additional pixel
flagging using "Background Bad Pixel
Update"\footnote{\url{https://github.com/STScI-MIRI/MRS-ExampleNB/blob/main/D2P_Notebooks/MRS_Flag_Badpix.ipynb}}. 
Briefly, additional pixels in the science
data were flagged as DO\_NOT\_USE using both the background and science images to identify additional bad pixels not 
present in the flight bad pixel reference files. This procedure has the effect of reducing large positive 
and negative artifacts in the final mosaics. We elected to use image-to-image background subtraction - 
e.g., background subtraction is applied at level 2 as this resulted in better instrumental artifact mitigation 
than the master background subtraction at level 3. In the case of the Horsehead, our dedicated background pointing 
contained significant H\,{\sc{ii}} and fine-structure line contamination as it fell on a faint H\,{\sc{ii}} region near
the Horsehead, resulting in over-subtraction of those lines. 
To remove the line contamination 
from the background before subtraction, we mapped the wavelength coordinate system 
onto the rate images to associate a wavelength with each pixel, identified
the central wavelength of the contaminating lines, and interpolated across the lines in the wavelength 
dimension of the rate image. A careful examination of the background data showed no significant source 
continuum contribution outside of the lines. In the case of NGC~7023, no background source contamination   
was identified.  After these custom steps, the data were processed through levels 2 and 3 normally, with 
residual fringe correction and outlier detection enabled. The mosaic was created in the "ifualign" cube
coordinate system.  With our mosaic configuration, this results in the long axis of the mosaic
being aligned with one of the primary (x,y) image axes. Cubes were generated with {\it output\_type=band} to 
generate a single cube for each MIRI channel.

\subsubsection{NIRSpec \label{sssec:nirspec}}
The NIRSpec data (source, leak cal, and background) were processed through level-1b (rate).
The default pipeline does not properly handle grouped data with a small number of groups (4 or less)
in the jump detection step.  Briefly, co-adding frames on board has the effect of `smearing' a 
cosmic ray impact across 2 co-added groups and the algorithm implemented (2-pt difference) for 
detecting jumps will often identify the wrong group when 4 or fewer groups are available. Additionally, 
the default algorithm does not drop the group immediately following the detected group.  Combined,
this process resulted in including samples impacted by cosmic rays in the rate calculation for a large
number of pixels. 
We modified the detection algorithm to treat the 4-group case in the same fashion as a 3-group case, which 
resulted in more robust detection of the correct cosmic ray group and significantly reduced the number 
of spurious artifacts in the resulting mosaics.  As these data were processed before the NSClean 1/f 
noise removal algorithm \citep{2024PASP..136a5001R} was included in the default pipeline, we ran NSClean offline on the 
rate files. Additionally, we utilized an adaptation of MRS\_Flag\_Badpix algorithm for MIRI 
(see \ref{sssec:miri}) that includes the leak cal observations in addition to the backgrounds to identify
bad pixels not included in the default reference file.  After these custom processing steps, the rate 
files were reinserted into the pipeline for level 2 processing. For NIRSpec processing, we found that master
background subtraction during level 3 produced the best results, so background subtraction was performed
during level 3. As for MIRI, we found substantial line contamination (with no apparent continuum source
contamination) in our dedicated background for the Horsehead. For master background subtraction in level 3 processing, a 
2-dimensional background is created by replicating the extracted 1-dimensional background produced during 
level 2 processing of the dedicated background images (s1d.fits products). So to mask the lines, we extracted
the binary tables from the s1d product and interpolated across the contaminating lines,
replacing the 'BACKGROUND' column with the interpolant.  Level 3 processing proceeded with background 
subtraction using these modified backgrounds, and the `ifualign' flag was set to construct the 
final mosaic with the long axis aligned to an image coordinate axis. 

\subsection{WCS and Imaging Cross Calibration\label{ssec:wcsup}}
A comparison of the IFU mosaics to the corresponding NIRCam and MIRI images revealed that small
offsets in IFU WCS were present. As the imaging WCS were refined using the GAIA-DR3 catalog, we took
the imaging WCS as the `correct' reference and adjusted the IFU WCS to match in the following fashion. 
First, synthetic wide band images were generated from the IFU cubes by integrating the cube in wavelength space over
all imaging bandpasses that overlapped with the IFU wavelength coverage - e.g., the NIRSpec IFU
cubes were integrated over all overlapping NIRCam filter transmission curves to generate synthetic
NIRCam images and the MIRI IFU cubes integrated over the overlapping MIRI transmission curves. 
The imaging data for each filter were then re-projected onto the spatial WCS of the corresponding
synthetic IFU image.
Since our IFU mosaics were generated with the long axis aligned to an image (x,y) axis, we simply
cross-correlated every line of the synthetic image from IFU with the corresponding reprojected image
data to determine, line by line, the optimal pixel shift in the IFU WCS to align with the imaging WCS. 
We took the shift to apply to a given instrument IFU - we derived corrections for each grating and
channel separately (3 for NIRSpec, 4 for MIRI) - by averaging all the shifts derived from each 
synthetic image line for each instrument.  With this procedure, for NGC 7023, we derived a shift of
0.7 and -6 pixels for MIRI and NIRSpec, respectively, where the shift is defined as positive 
towards the exciting star.  For the Horsehead, the MIRI data were shifted
by 1.5 pixels and no shift was applied to the NIRSpec data. 

In addition to verifying the WCS of the IFU cubes, the spatially coincident imaging provides a 
cross-check on the absolute flux calibration of the IFUs.  Utilizing the same synthetic IFU images
and spatially matched, reprojected imaging data as described above for the WCS update, we compare
the measured flux on a per-pixel basis for each matching imaging filter and synthetic IFU image pair. 
We only considered image/synthetic image pairs for which the IFU wavelength coverage in individual 
gratings (NIRSpec - G140M, G235M, and G395M) and channels (MIRI - 1, 2, 3, and 4) overlapped fully 
with a corresponding imaging filter bandpass.  For NIRSpec, this results in all observed NIRCam 
filters, with 
the exception of F070W and F090W, having synthetic images computed.  Note that filters falling near the 
edge of a NIRSpec grating response (e.g. F187N and G140M, F300M and G235M) were analyzed but not 
included in the final result. For MIRI, neither the F770W nor F1800W filter fell completely within 
a MIRI IFU channel and so were not included. 

For each resultant image/IFU synthetic image pair, the matched pixel ensemble was fit with a linear
function
\begin{equation}
F_{i,j}^{\text{band}} = A_{\text{band}} + B_{\text{band}} F_{i,j}^{\text{synth}},
\end{equation}
where $F_{i,j}^{\text{band}}$ is the imaging pixel, $F_{i,j}^{\text{synth}}$ the matching IFU synthetic pixel.
If the imaging and IFU calibrations were identical, we would expect $A_{\text{band}}=0$ and 
$B_{\text{band}}=1$. We take $A_{\text{band}}$ to represent differences in background estimation/correction between 
the imaging and IFU data and don't consider it further. The slope, $B_{band}$, represents the scaling between 
the imaging and IFU data flux calibrations.  For both NIRSpec/NIRCam and the MIRI imager/IFU, we take the imaging 
calibration as the reference and derive a `correction' to the IFU calibration.  

For NIRSpec, comparisons in all 3 gratings across both objects yield similar results for the scale factor, 
roughly 0.88. The G140M grating has a slightly smaller factor of 0.76 but is of lower signal-to-noise 
as well as overlapping with two narrow-band filters (which tend to have a larger 
dispersion than medium and wide-band filters in the fit) and a single medium-band filter. 
Averaging all medium band filters across all gratings and both objects, we derive a single factor
of 0.87, consistent with that found by \citet{2024A&A...685A..74P} for the NIRSpec 
high-resolution gratings. The 0.87 multiplicative scaling was applied to all extracted 
NIRSpec spectra.
For MIRI, while the dispersion is large, we find no systematic 
scaling between the imaging and IFU across channels or between objects and do not scale the 
MIRI IFU cubes.  

\subsection{Region extraction \label{ssec:extraction}}
\begin{figure*}[h!]
   \centering
   \includegraphics[width=\textwidth]{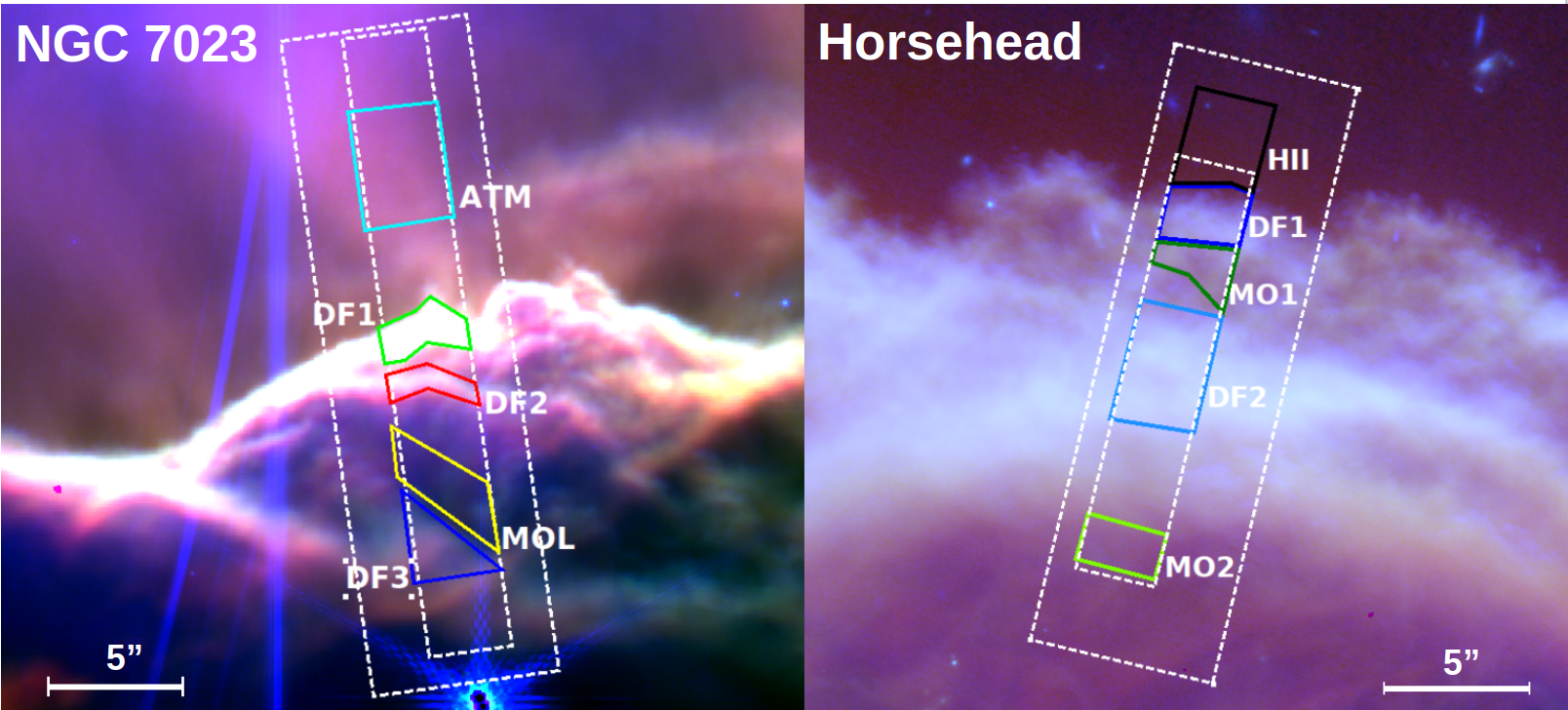}
   \caption{Extractions regions on images (blue = NRC-F210M, green = NRC-F335M, red = MIR-F770W) 
   of NGC~7023 (left) and the Horsehead (right). The 
   exciting star is off the top of the image in both cases. The color used to designate the regions here will be 
   used in spectral plots for those regions, the remainder of the text. The white dashed rectangles define the 
   minimum (NIRSpec) and maximum (MIRI channel 4) IFU mosaic coverage.
   \label{fig:img_regions}}
\end{figure*}

\begin{figure*}[h!]
   \centering
   \includegraphics[width=\textwidth]{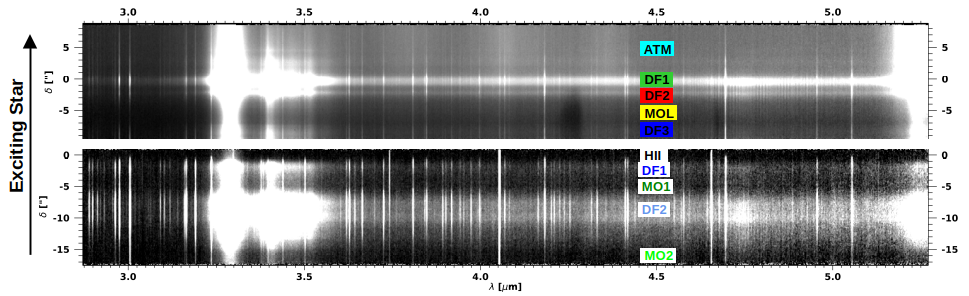}
   \caption{Pseudo long-slit (on G395M; see text) visualization of region definitions for spectral extraction for both 
   NGC~7023 (top) and the Horsehead (bottom). The exciting star is at the top, and the wavelength increases to the right. 
   The origin of the y-axis (depth into the PRD) is defined at 
   ($\alpha$,$\delta$) = (\ang{5;40;53.11},\ang{-2;28;06.37}) and
   (\ang{21;01;31.82},\ang{68;10;23.61}) for the Horsehead and NGC~7023, respectively.
   \label{fig:ls_regions}}
\end{figure*}

While we have spectra for each pixel in the IFUs ("spaxel"), for purposes of this paper, we 
extract average spectra over 5 larger regions in each object (see Fig. \ref{fig:img_regions}). Detailed 
"feature maps" (spatial maps over specific spectral features) will be presented in an upcoming paper 
(Van De Putte et. al. in prep.) and are touched on briefly in Sect. \ref{ssec:hi} and Sect. \ref{sssec:alistability}.
The region locations were selected to map physical regions at increasing `depth'
into the molecular cloud past the primary PDR front. We note that we have used the term `depth'
to refer to depth as projected onto the plane of the sky; owing to the non-trivial geometry in both objects,
this may not truly reflect a physically rigorous definition of 
depth into the molecular cloud material traversed by an exciting photon. We defined `physical regions' based 
on the continuum and line flux levels inferred from the imaging \citep{2024A&A...687A...4A} as well as 
directly from the IFUs by constructing pseudo long-slit images.  As the IFU mosaics were constructed
such that a detector axis was aligned with the line connecting the exciting star to the PDR front, 
we can collapse the IFU cube on the spatial axis perpendicular to the front.  This results in a
two-dimensional spectrum with the spatial dimension lying along the line of sight to the exciting 
star - see  Fig. \ref{fig:ls_regions} for an example using the NIRSpec G395M grating. We do not use 
these long-slit images directly for analysis as, by construction, they will potentially mix emission
from distinct physical regions as the physical transitions in the molecular cloud do not align perfectly
with the mosaic axis; however, they do provide a high SN visualization of line and 
continuum strength as a function of depth into the PDR. Combined with imaging in all the
band-passes, we can define distinct regions of high and low line and continuum flux for extraction, 
with the imaging allowing us to largely avoid mixing distinct physical regions by utilizing 
irregular region shapes (see Fig. \ref{fig:img_regions}). Even though we tag regions as
dissociation front (DF) and molecular (MO) based mostly on position and continuum levels, we note
that the geometry is sufficiently complex that the names are largely notional. The defined regions 
integrate emissions from regions with mixed physical properties.

In the Horsehead, the IFU mosaics extend into the H\,{\sc{ii}} region, albeit minimally for the NIRSpec
 mosaic (Fig. \ref{fig:img_regions}). In the long slit visualization  (Fig. \ref{fig:ls_regions}),
we see a relatively sharp edge to the H$_2$ lines and the atomic lines 
extending into the H\,{\sc{ii}} region and a well-defined continuum just inside, especially around the 
3.3~\micron\ carbonaceous features
complex. Note also the clear extension of only the 3.3~\micron\ carbonaceous emission 
into the H\,{\sc{ii}} region (see Sect. \ref{sssec:outflow}). From these images, we define the H\,{\sc{ii}} and DF1 regions. 
Inside of DF1, there is a trough in both the continuum (imaging and long slit) and line emission (long slit), 
which we define as MO1 - the region boundary is constructed to avoid complex substructure in the imaging
data. Further into the PDR (projected distance), there is another region of elevated continuum and line 
emission - this is likely material that appears deeper into the PDR in projection but is directly 
illuminated by the exciting radiation.  While this region resolves into two distinct structures in the
long slit visualization, for SN reasons, we combine them into a single region designated DF2.  Finally, 
there is a distinct region of low continuum and line emission near the end of our IFU mosaic, which we define 
as MO2. 

For NGC~7023, we follow a similar procedure.  Here, there is no true H\,{\sc{ii}} region; outside of the PDR front, 
we define the ATM (atomic) region. This region is not a pure atomic region but rather
samples the inner cavity of the HD~200775 bubble as well as PDR emission from the inside 
face of the far side of the cavity.  
The brightest emission in 
NGC~7023 is extracted as DF1 and is constructed to follow the strongest emission in the broad-band    
images. Immediately inside of DF1, there is a drop in the emission followed by another strong front feature. 
The latter is extracted as DF2, while the drop in emission immediately behind DF1 is small enough that we 
do not extract a high SN region; the detailed behavior of the PDR across the transition will be explored in 
a future paper.  Starting in DF2, hints of ice absorption at $\sim$4.3~\micron\ (CO$_2$) and $\sim$4.7~\micron\
(CO) are apparent in the long slit visualization, corresponding to a region of low emission in the broad-band 
imaging (see Sect. \ref{ssec:ices}). In both the imaging and long slit visualizations, 
a sharp increase in the continuum and line emission is seen at the edge of the IFU coverage. These regions
are extracted as MOL and DF3, respectively.
Note that subtle bands at 
$\sim$4.4, 3.4, and 4.7~\micron\ with various strengths are visible on the long slit
visualization; these bands are discussed in more detail in Sect. \ref{sssec:alistability} and
\ref{sssec:dsubstitution}.

\begin{figure*}[!h]
   \centering
   \includegraphics[width=\columnwidth]{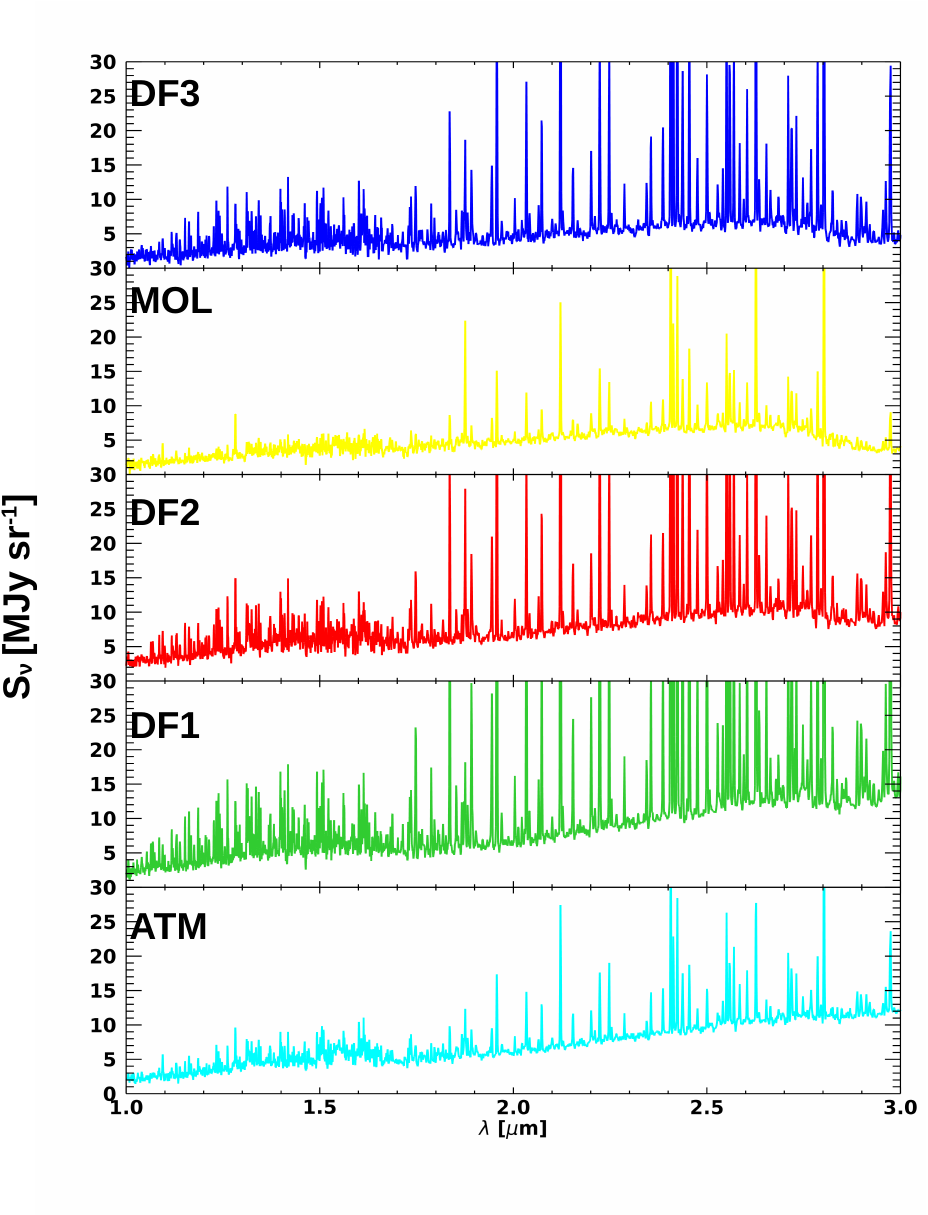}
   \caption{Spectral extractions from the five regions defined for NGC~7023 over the 
   wavelength range $1.0 < \lambda < 3.0$~\micron. Regions are color coded as in Fig. \ref{fig:img_regions} with 
   DF3, MOL, DF2, DF1, and ATM displayed top to bottom. Similar figures for all regions in both objects covering the 
   full spectral range are given in Appendix \ref{ap:spectralplots}.
   \label{fig:ngc7023_1p0_3p0_spec}}
\end{figure*}

For each of these 10 defined regions, we extract spectra averaged spatially over all pixels that
fell within the region boundaries.  For the remainder of the paper, unless explicitly stated
otherwise (e.g. carbonaceous features - Sect. \ref{ssec:carb_ss}), features and line identifications
will be extracted from these region-averaged spectral templates rather than per "spaxel". 
Tables of all extracted spectra (flux and continuum) are available at CDS\footnote{\url{https://cds.unistra.fr/}}, and a 
sample of the table format is provided in Table \ref{tbl:spectra_example}.  A sample set of spectra 
is shown in Fig. \ref{fig:ngc7023_1p0_3p0_spec} for 1.0-3.3~\micron\ in NGC~7023, and plots
of all extracted spectra are shown in Figs. \ref{fig:ngc7023_2p8_5p5_spec}-\ref{fig:horsehead_9p5_28p0_spec}.

\section{Line Extraction \label{sec:lines}}

\subsection{Finding the Lines\label{ssec:line_finding}}
\begin{figure*}
   \centering
   \includegraphics[width=\textwidth]{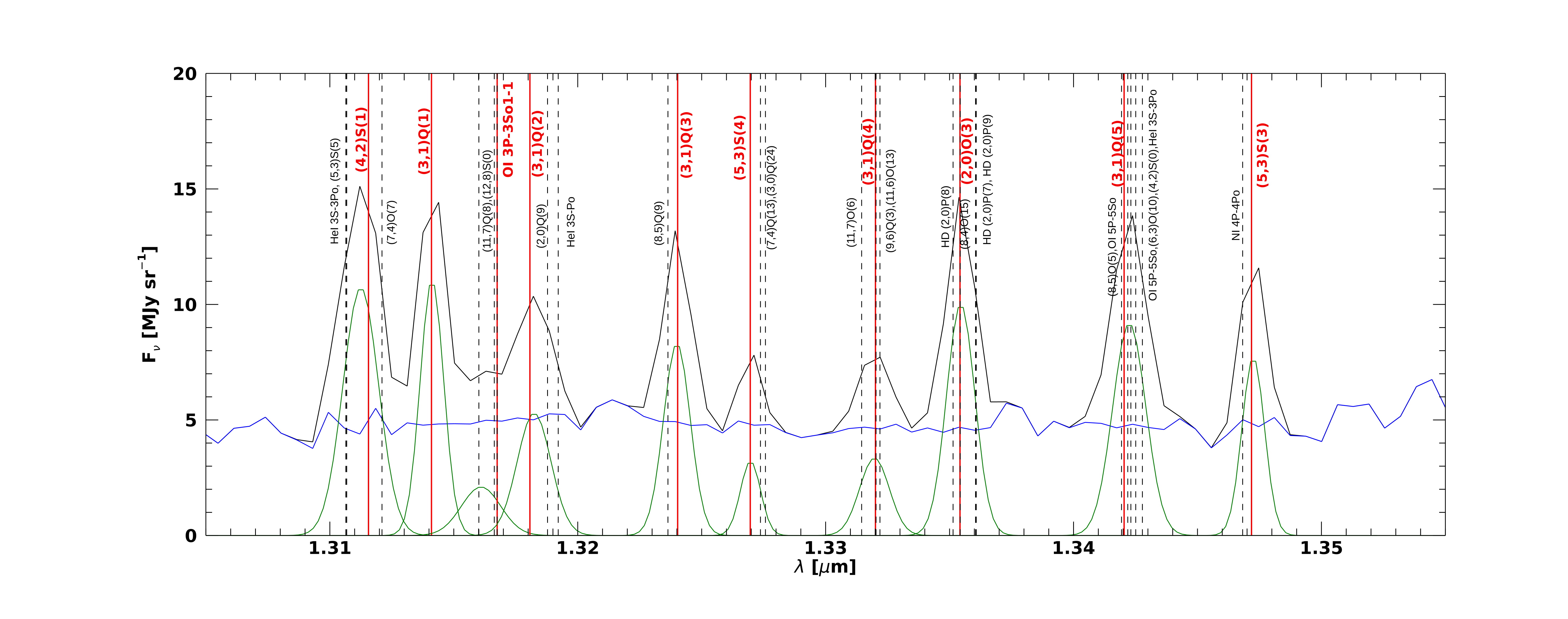}
   \caption{Plot of a narrow wavelength slice of the NIRSpec G140M spectrum in DF1 qualitatively 
   illustrating line identification.  The black line is the 
   extracted spectrum, green the Gaussian fits to identified lines, and blue the continuum after subtracting the green fits
   from the extracted spectrum.  Vertical red lines and red text provide the final line identifications, while vertical
   dashed lines show all species initially identified as candidates.
   \label{fig:lineidentification_detail}}
\end{figure*}

With a resolving power of roughly 1000 through the NIRSpec wavelength coverage and 1000-3000 for MIRI, for typical 
conditions in PDR regions, we expect the lines in our extracted spectra to be only marginally resolved,
complicating finding and identifying closely spaced lines in our spectra.
To find lines for identification, we proceed as follows. For NIRSpec, each grating 
was divided into subsections in wavelength, each roughly 100 pixels wide ($\sim 0.05-0.18$~\micron\ depending on the grating 
- see Fig. \ref{fig:lineidentification_detail} for a sample subsection).  In each of these
subsections, non-consecutive maxima - deviating by more the 5-$\sigma$ from the median - were iteratively identified.  
Each subsection was then fit with a function consisting of a linear continuum and $N$ Gaussians, where $N$ is the number of 
maxima identified.  The central wavelengths of the Gaussians were fixed at the position of the maxima for this initial
search.  Lines with Gaussian widths that exceeded 3 pixels were tagged as possible multiple unresolved lines for the
manual follow-up. The assumption of a linear continuum was valid for much of the NIRSpec wavelength coverage, especially 
over the narrower wavelength ranges selected for line identification; however, 
around broad carbonaceous features, all lines were tagged for manual follow-up. 

With the initial automatic findings complete, 
all lines were reviewed manually and refined.  Where the width of the line was tagged in the automatic fitting, there
were generally clear, unresolved multiple lines - in those cases, the `line' was refit with two components.  During the 
manual review, each identified line was removed from the spectrum, and the residuals were examined for potentially weaker lines; 
if identified, these weaker lines were fit and added to the line list.  A similar procedure was followed for MIRI; however,
given that there were many fewer lines present in the MIRI spectra, the initial automatic finding was sufficient, and 
no updates were necessary in the manual follow-up. 
This procedure resulted in finding several hundred lines (see Table \ref{tab:linecounts}) in each region,
each with a central wavelength, width, and area derived from the Gaussian fit.  

\subsection{Identifying the Lines\label{ssec:line_identity}}
With lines "found", it remains to identify each line with a particular material and transition. For identification, 
we utilized the line list assembled by the PDRS4ALL team \citep{2024A&A...685A..74P}\footnote{publis20240406.lst available at 
\url{https://pdrs4all.org/seps/\#line-list}}.  This is a comprehensive list of atomic and molecular transitions between 
$\sim$0.7-29\micron. With MIRI, generally, only a single line in the input list matched our data within a resolution
element, so only a single automated pass through the line list was necessary. However, with the medium-resolution
gratings employed with NIRSpec, the input line list is dense enough that a single line in our spectra may match
several lines in the list within a resolution element. In this case, for each observed line, any theoretical line within
a resolution element of the observed line center was recorded as a possibility with the closest line in terms of 
wavelength designated as the most probable. If multiple \HH\ lines were found, we weighted the selection of the specific
transition by the upper energy level (favoring lower excitation energy) and Einstein A-coefficient of the transition. 
The final identification of the line
with a specific \HH\ transition was made based on whether a given \HH\ transition was part of a clearly identified ro-vibrational
series. If no such series had already been identified, all candidates were left as possibilities 
A similar process was followed for lines
with multiple species identified (generally \HH\ with some other possibility - atomic, CH$^+$, HD, etc), if the \HH\ line
identification was part of a clearly identified ro-vibrational series, we selected the \HH\ line as the most probable 
identification.  Otherwise, all candidates were left as possibilities and reported.  After this initial automated pass
was made, each line was manually inspected and refitted, the central line position refined, and the identification repeated
if the updated line center shifted sufficiently.  In practice, no line identifications were modified in this final step. 
The procedure is outlined graphically in Fig. \ref{fig:lineidentification_detail}, and a detailed example is provided in 
Appendix \ref{ap:linedetail}

\begin{table}[bth]
\begin{threeparttable}[b]
 \caption{Total Line Counts}\label{tab:linecounts}
 \begin{tabular*}{0.45\textwidth}{@{\extracolsep{\fill}}lccccc}
 \toprule
 Region\tnote{a} & H$_2$\tnote{b} & H\,{\sc{ii}} & He\,{\sc{i}} & Other\tnote{c} & Total \\ 
 \midrule
           & \multicolumn{5}{c}{\bf NGC 7023} \\
           \cmidrule{2-6}
    ATM    & 217(25) &  8 & 10 & 32 & 267 \\
    DF1    & 256(31) &  9 & 16 & 48 & 329 \\
    DF2    & 215(30) &  6 & 13 & 41 & 275 \\
    MOL    & 149(27) &  8 & 12 & 35 & 204 \\
    DF3    & 235(31) &  7 & 14 & 37 & 293 \\
           & \multicolumn{5}{c}{\bf Horsehead} \\
           \cmidrule{2-6}
    H\,{\sc{ii}}    &    4(4) & 14 &  1 &  5 &  24 \\
    DF1    & 211(26) & 21 & 16 & 47 & 295 \\
    MO1    & 215(20) & 19 & 14 & 64 & 312 \\
    DF2    & 244(28) & 19 & 22 & 64 & 360 \\
    MO2    & 157(12) & 11 & 13 & 37 & 218 \\
    \hline
 \end{tabular*}

 \begin{tablenotes}
   \item [a] See Fig. \ref{fig:img_regions} for physical location of region.
   \item [b] The number in parentheses indicates the number of pure rotational H$_2$ lines included in the count.
   \item [c] Atomic, ionized and some molecular species (e.g., CH$^+$, CO)
 \end{tablenotes}

\end{threeparttable}
\end{table}

This line identification process resulted in 200-400 lines per region.  The majority of these lines are H$_2$ lines (75-85\%) 
- hence, the much smaller total number of lines (24) detected in the Horsehead H\,{\sc{ii}} region.
Of those H$_2$ lines, 
10-15\% are pure rotational lines, with the remainder being various ro-vibrational transitions - see Table \ref{tab:linecounts} 
for a region-by-region breakdown of the line counts. In addition to H$_2$ lines, we identify hydrogen recombination lines 
from the first few transitions of the Paschen through Humpreys series (see Sect. \ref{ssec:hi}), He\,{\sc{i}} lines, various atomic
lines (e.g., [C\,{\sc{i}}], [Fe\,{\sc{iii}}], [Ne\,{\sc{ii}}], [Cl\,{\sc{ii}}], [S\,{\sc{iii}}]), and, in selected regions, CO lines and CH$^+$ lines
(see Sect. \ref{ssec:molgas}). 

The final line list for each template region is available through CDS (Sect. \ref{sec:dataavail}),
and a sample of the table format is provided in 
Table \ref{tbl:linelist_sample}. More detailed discussions of selected collections of lines and features
follow in Sect. \ref{sec:discussion}.

\subsection{Line Contribution to Imaging Bandpasses\label{ssec:lines_contribution}}

A major source of uncertainty in interpreting narrow and broad-band images of complex sources 
is understanding the contribution of different emission features of interest to the measured flux in the band 
\citep{2025A&A...698A..86C,2025ApJ...986..156L}.
In this program, we have broad-band imaging and spectra in identical regions 
of the sources, allowing a direct measurement of the line contribution to band flux.  As an aid 
in interpreting our broad-band imaging in NGC~7023 and
the Horsehead (see e.g. Sect. \ref{sssec:outflow}), we provide here the measured contribution of line
emission to band flux in the 10 regions defined in Sect. \ref{ssec:extraction}. While the exact values 
of the line contribution are only strictly valid for the overlap regions, they provide a guide to a more
robust estimate in similar regions where only imaging data are available. 

To estimate the line contribution in each region for each of the 21 imaging filters (12 NIRCam, 9 MIRI), 
line-only and continuum-only spectra were constructed for each region. Using the full response curve 
for each filter
each spectrum is integrated across the bandpass response. The line contribution to the band is computed 
as the ratio of the integrated line flux to the total band flux (line $+$ continuum). 
The results are presented as a percentage in Tables \ref{tab:nrclinefrac} and \ref{tab:mirlinefrac} for NIRCam 
and MIRI filters, respectively. Line contributions in each region are also separated into \HH\ and 
all other line fractions.

\begin{table*}
\begin{threeparttable}
 \caption{Percent Line Contribution NIRCam filters\tnote{a}\label{tab:nrclinefrac}}
 {\footnotesize
 \begin{tabular*}{\textwidth}{@{\extracolsep{\fill}}lccccc}
   \toprule
         & \multicolumn{5}{c}{NGC 7023}  \\
         & ATM               & DF1              & DF2              & MOL              & DF3              \\
   \cmidrule{2-6}
   F140M & 10.2  (7.7,2.5)   & 24.4 (20.2,4.2)  & 16.2 (11.6,4.6)  &  6.8  (4.3,2.5)  & 24.3 (19.9,4.5)  \\
   F164N &  0.8  (0.7,0.1)   &  6.5  (6.3,0.2)  &  0.4  (0.2,0.2)  &  0.0             &  8.1  (8.0,0.1)  \\
   F187N & 21.1 (10.0,11.1)  & 34.2 (20.8,13.4) & 35.8 (10.7,25.1) & 28.9  (0.9,28.0) & 41.5 (17.3,24.2) \\
   F210M &  8.7  (8.6,0.1)   & 33.4 (33.2,0.2)  & 20.6 (20.2,0.4)  &  7.9  (7.4,0.5)  & 26.3 (25.2,1.1)  \\
   F212N & 21.6 (21.6,0.0)   & 67.9 (67.9,0.0)  & 50.9 (50.9,0.0)  & 23.3 (23.3,0.0)  & 53.2 (53.1,0.1)  \\
   F250M & 13.2 (12.9,0.3)   & 46.1 (45.1,1.0)  & 35.5 (34.4,1.1)  & 18.1 (16.9,1.2)  & 39.7 (38.0,1.7)  \\
   F300M &  4.8  (4.7,0.1)   & 17.9 (16.5,1.4)  & 15.2 (13.5,1.7)  &  8.5  (7.6,0.9)  & 23.5 (21.1,2.4)  \\
   F335M &  0.7  (0.5,0.2)   &  1.0  (1.0,0.0)  &  1.2  (1.0,0.2)  &  1.1  (1.0,0.1)  &  1.7  (1.7,0.0)  \\
   F323N &  2.1  (2.1,0.0)   &  7.8  (7.8,0.0)  &  7.2  (7.2,0.0)  &  4.3  (4.3,0.0)  &  9.8  (9.8,0.0)  \\
   F405N &  1.1  (1.1,0.0)   &  5.1  (5.1,0.0)  &  6.2  (6.2,0.0)  &  5.6  (5.4,0.2)  &  9.1  (9.2,0.0)  \\
   F430M &  0.6  (0.6,0.0)   &  3.7  (3.7,0.0)  &  3.8  (3.6,0.2)  &  2.4  (2.4,0.0)  &  5.1  (5.1,0.0)  \\
   F470N &  2.3  (2.2,0.1)   & 16.9 (16.9,0.0)  & 15.7 (15.7,0.0)  & 11.8 (11.8,0.0)  & 18.1 (18.1,0.0)  \\
         &                   &                  &                  &                  &                  \\
         & \multicolumn{5}{c}{Horsehead} \\
         & HII\tnote{b}      & DF1              & MO1              & DF2              & MO2              \\
   \cmidrule{2-6}
   F140M &   0.0             & 70.2 (63.9,6.3)  & 63.6 (55.9,7.6)  & 65.0 (57.7,7.3)  & 39.6 (34.2,5.4)  \\
   F164N &   0.0             & 67.2 (65.7,1.5)  & 18.0 (18.0,0.0)  & 23.9 (23.9,0.0)  &  2.8  (2.8,0.0)  \\
   F187N & 100.0 (0.0,100.0) & 97.1 (5.4,91.7)  & 98.3  (4.8,93.5) & 91.6 (17.0,74.6) & 73.6  (3.8,69.8) \\
   F210M & 100.0 (0.0,100.0) & 84.7 (73.7,11.0) & 71.6 (59.6,12.0) & 73.8 (68.8,5.0)  & 38.7 (33.6,5.1)  \\
   F212N &   0.0             & 94.3 (94.2,7.9)  & 82.8 (82.8,0.0)  & 90.1 (90.1,0.0)  & 62.6 (62.6,0.0)  \\
   F250M &   0.0             & 89.4 (79.0,10.4) & 83.7 (71.1,12.5) & 85.4 (84.2,1.2)  & 69.0 (58.0,11.0) \\
   F300M &   0.0             & 63.4 (37.4,26.0) & 64.9 (35.0,29.8) & 70.4 (42.2,28.2) & 81.4 (40.9,40.5) \\
   F335M &   6.4 (0.0,6.4)\tnote{c} & 11.4 (9.3,2.1)   &  8.1  (6.0,2.1)  &  8.6  (7.1,0.5)  &  7.5  (5.7,1.8)  \\
   F323N &  15.8 (0.0,15.8)\tnote{c} & 39.7 (39.7,0.0)  & 24.5 (24.4,0.0)  & 31.4 (31.4,0.0)  & 21.7 (21.7,0.0)  \\
   F405N &  93.5 (0.0,93.5)  & 83.7  (2.8,80.9) & 81.9  (0.0,81.8) & 57.0  (4.6,52.4) & 75.1  (0.0,75.1) \\
   F430M &   0.0             & 22.6 (20.8,1.8)  & 18.0 (13.1,4.9)  & 18.0 (17.1,0.9)  & 18.0 (15.8,2.2)  \\
   F470N &   0.0             & 55.7 (55.7,0.0)  & 26.7  (0.0,26.7) & 34.7  (0.8,33.9) & 17.9  (0.0,17.9) \\
   \midrule
 \end{tabular*}
 }
  \begin{tablenotes}
    \item [a] Line contribution percentages are presented as $f_{total}(f_{H2},f_{other})$ where $f_{H2}$ is the
    percentage of signal contributed by \HH\ in the band pass, $f_{other}$ is all other lines in the band pass and
    $f_{total} = f_{H2}+f_{other}$.
    \item [b] Continuum is consistent with 0 with the exception of note c.
    \item [c] These filters overlap with the aromatic 3.3~\micron\ feature (see Sect. \ref{sssec:outflow}) which is considered continuum for this calculation.
  \end{tablenotes}
\end{threeparttable}
\end{table*}

\begin{table*}
\begin{threeparttable}
 \caption{Percent Line Contribution MIRI filters\tnote{a}\label{tab:mirlinefrac}}
 \begin{tabular*}{\textwidth}{@{\extracolsep{\fill}}lccccc}
   \toprule
          & \multicolumn{5}{c}{NGC 7023}  \\
          & ATM             & DF1             & DF2             & MOL             & DF3            \\
   \cmidrule{2-6}
   F560W  & 0.05(0.05,0.0)  & 0.9(0.9,0.0)    & 0.8(0.8,0.0)    & 0.7(0.7,0.0)    & 0.9(0.9,0.0)   \\
   F770W  & 0.04(0.04,0.0)  & 0.6(0.6,0.0)    & 0.6(0.6,0.0)    & 0.5(0.5,0.0)    & 0.7(0.7,0.0)   \\
   F1000W & 0.4(0.4,0.0)    & 3.5(3.5,0.0)    & 3.3(3.3,0.0)    & 2.8(2.8,0.0)    & 4.8(4.8,0.0)   \\
   F1130W & 0.0             & 0.0             & 0.0             & 0.0             & 0.0            \\
   F1280W & 0.3(0.3,0.0)    & 0.7(0.7,0.0)    & 0.8(0.8,0.0)    & 0.8(0.8,0.0)    & 2.3(2.3,0.0)   \\
   F1500W & 0.11(0.03,0.08) & 0.05(0.04,0.01) & 0.05(0.04,0.01) & 0.12(0.09,0.03) & 0.11(0.1,0.01) \\
   F1800W & 1.0(1.0,0.0)    & 1.7(1.7,0.0)    & 1.7(1.7,0.0)    & 1.8(1.8,0.0)    & 5.0(5.0,0.0)   \\
   F2100W & 0.0             & 0.0             & 0.0             & 0.0             & 0.0            \\
   F2550W & 0.0             & 0.05(0.04,0.01) & 0.05(0.04,0.01) & 0.0             & 0.0            \\
          &                 &                 &                 &                 &                \\
          & \multicolumn{5}{c}{Horsehead} \\
          & HII\tnote{b}    & DF1           & DF2            & MO1            & DF3             \\
   \cmidrule{2-6}
   F560W  &   0.0           &  6.0(6.0,0.0) &  2.3(2.3,0.0)  &  2.6(2.6,0.0)  &  0.0            \\
   F770W  & 100.0(7.4,92.6) &  7.9(4.2,3.7) &  4.8(1.9,2.9)  &  3.1(2.3,0.8)  &  4.1(1.7,2.4)   \\
   F1000W &   8.8(8.8,0.0)  & 19.8(19.8,0.0)& 12.3(12.3,0.0) & 13.5(13.5,0.0) & 12.1(12.1,0.0)  \\
   F1130W &   0.0           &  0.0          &  0.0           &  0.0           &  0.0(0.0,0.0)   \\
   F1280W &  85.4(1.9,83.5) & 27.0(7.2,19.8)& 26.0(8.3,17.7) & 13.4(8.7,4.7)  & 31.2(12.7,18.5) \\
   F1500W &   2.2(0.1,2.1)  &  0.7(0.3,0.4) &  0.6(0.3,0.3)  &  0.5(0.4,0.1)  &  2.6(2.6,0.0)   \\
   F1800W &  41.9(1.9,40.0) & 15.8(8.0,7.9) & 14.7(8.1,6.6)  & 12.9(10.9,2.0) & 69.6(38.3,31.3) \\
   F2100W &  19.3(0.0,19.3) &  4.7(0.0,4.7) &  3.7(0.0,3.7)  &  1.3(0.0,1.3)  & 19.9(0.0,19.9)  \\
   F2550W &   0.0           &  0.0          &  0.0           &  0.0           &  0.0            \\
   \midrule
 \end{tabular*}

  \begin{tablenotes}
    \item [a] Line contribution percentages are presented as $f_{total}(f_{H2},f_{other})$ where $f_{H2}$ is the
    percentage of signal contributed by \HH\ in the band pass, $f_{other}$ is all other lines in the band pass and
    $f_{total} = f_{H2}+f_{other}$.
    \item [b] Continuum is poorly defined short-wards of 15~\micron\, resulting in uncertain calculation.
  \end{tablenotes}
\end{threeparttable}
\end{table*}

Depending on the region and filter, the line `contamination' varies from 0-100\%.  At the upper range, 
line contribution is 100\% only in the Horsehead H\,{\sc{ii}} region where the continuum is negligible, at least 
outside of the 3.3\micron\ aromatic emission band - see Sect. \ref{ssec:carb_ss}.  Line contamination 
in the MIRI filters is much smaller, owing to the decrease in the number of ro-vibrational lines with increasing
wavelength and the broader MIRI filters.  Comparing NGC~7023 and the Horsehead, the line contribution tends to 
be higher in the Horsehead, partly due to a larger number of lines but mostly owing to the relatively weaker
continuum emission.  The regions we've designated "DF" tend to have higher line contributions relative to the 
"MO" regions. This is not surprising as the region definitions did follow brighter filaments in the broad-band 
imaging, some of which will be line emission. Additionally, the number of lines tends to fall in the "MO" regions, 
which are preferentially deeper into the PDR (in projection), and the H$_2$ lines, which are the predominant line 
emission contribution, tend to weaken in the MO regions relative to the DF regions.

\section{Discussion \label{sec:discussion}}

\subsection{H\,{\sc{i}} \label{ssec:hi}}
\begin{figure}[!ht]
\centering
\includegraphics[width=0.45\textwidth]{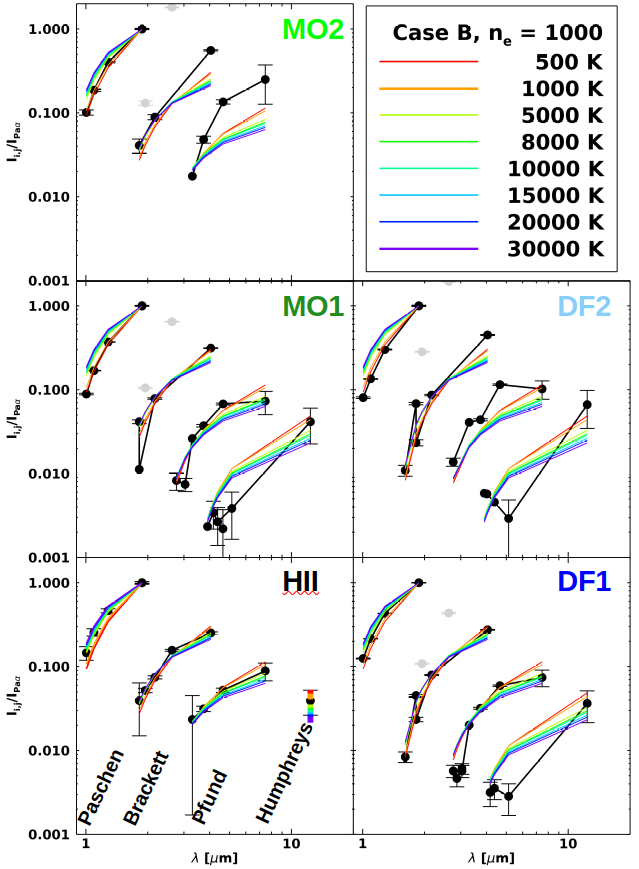}
 \caption{Measured line hydrogen line ratios, normalized to Pa~$\alpha$ for the 5 regions in the Horsehead.  Over plotted in colored lines are the theoretical ratios for a range of temperatures from \citet{2018MNRAS.478.2766P}.  All theoretical ratios are computed for a density of 1000~cm$^{-3}$. Observed series are, left to right, Paschen, Brackett, Pfund, and Humphreys. Except for the H\,{\sc{ii}} region, 4-8 and 4-6 in the Brackett are plotted as light gray symbols as they are confused with nearby H$_2$ lines.
\label{fig:hh_hline_caseB}}
\end{figure}

Given its relatively low temperature, few ionizing photons are produced by HD~200775 at the 
PDR front in the region covered by our IFU data.
As a result, the H\,{\sc{ii}} lines in NGC~7023 are generally quite weak and not well-measured.
While we detect 8-9 H\,{\sc{ii}} lines in each region (roughly the first four lines in the Paschen
and Brackett series), we do not perform any further analysis.
In contrast, with the higher temperature of $\sigma$ Orionis and its well-defined H\,{\sc{ii}} region, 
the H~I lines in the Horsehead are considerably 
stronger and better measured.  We detect 13-20 H~I lines (roughly 3-6 in the Paschen, Brackett, Pfund, 
and Humphreys series) depending on the region (see Fig. \ref{fig:hh_hline_caseB}).  In the Brackett series, 
Brackett 6 and 8 are confused with nearby H$_2$ lines [(1,0)O(2) and (2,1)S(5), respectively] in all regions
except H\,{\sc{ii}} and so are not included in the analysis.  Additionally, lines in the Humphreys series are quite 
weak and noisy and are not considered hereafter. Before analysis, all the line strengths were corrected 
for the foreground extinction towards $\sigma$~Orionis using an $R_V = 3.1$ extinction curve 
\citep[][and references therein]{2023ApJ...950...86G} and $E(B-V) = 0.05$
\citep{1994A&A...289..101B}.

In Fig. \ref{fig:hh_hline_caseB}, we compare our observed line ratios normalized to Paschen-$\alpha$ to 
Case B theoretical line ratios from \citet{2018MNRAS.478.2766P}.  We compare the observed ratios to the 
theoretical ratios for 8 temperatures between 500-30000~K at a single electron density of 
$n_e = 1000~cm^{-3}$, as there is little dependence on the density for the observed transitions. 
There is a clear trend in the best matching theoretical temperature with the region, especially with the 
Paschen series, which includes our best-measured lines. In the H\,{\sc{ii}} region, the observed line ratios are
consistent with high-temperature theoretical ratios, with $T_e \ge 8000~K$.  In DF1, the data are still 
consistent with higher temperatures, but $T_e \sim 5000-8000$~K are preferred.  Deeper into the PDR in 
projection, DF2 and especially MO1 and MO2 are more consistent with 
$T_e \sim 500 - 1000$~K.  While less clear, the lower signal-to-noise Brackett series ratios are also 
consistent with this trend. The complex structure of the Horsehead makes interpretation of the excitation 
temperatures difficult.  While decreasing temperature with depth is expected, it should occur over smaller
spatial scales than we see in projection. The Horsehead nebula is likely illuminated at an oblique angle
from the rear \citep[$\sim6\degr$,][]{2005A&A...437..177H} with a terraced structure \citep[e.g.,][]{2024A&A...687A...4A} and
the illuminated surfaces of the clouds that dominate the hydrogen
emission are more exposed to the radiation field than their projected distance would imply while those
surfaces 'deeper' into the nebula are likely illuminated by a radiation field that has been 
attenuated by material between the H\,{\sc{ii}} region and the cloud surface 
\citep[][and the section below]{2024A&A...687A...4A}.  Disentangling the effects of attenuation and excitation
at the cloud surface seen partially face-on will require detailed radiative transfer modeling. 

\begin{figure*}[h]
\centering
\includegraphics[width=\textwidth]{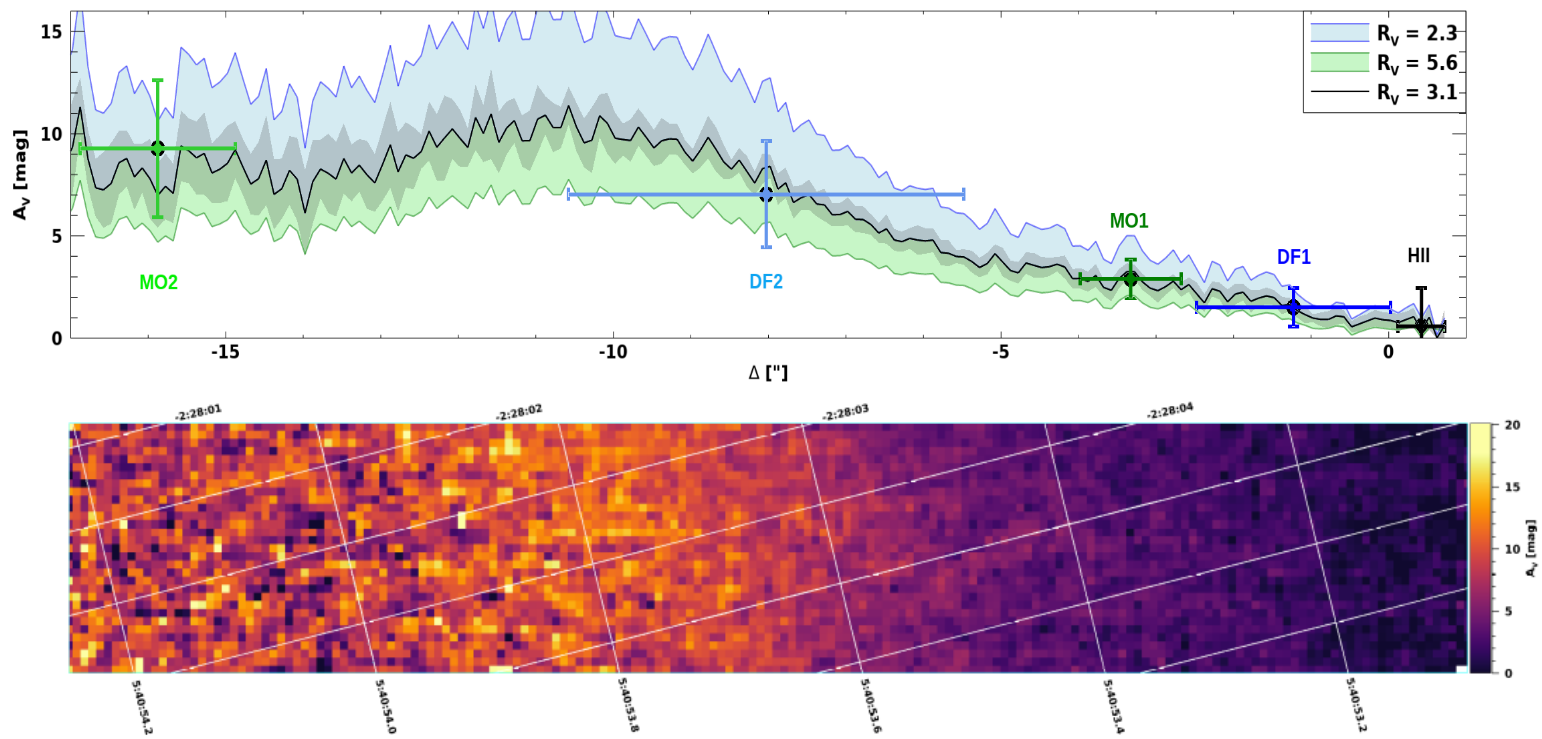}
 \caption{Bottom: spatial distribution of $A_V$ in the Horsehead IFU footprint as estimated using 
 Eq. \ref{eq:av} with $R_V$ = 3.1. Top: Estimates of $A_V$ extracted from each region (points) 
 along with a cut-through 
 the spatial distribution map. Error bars on $A_V$ for the region extractions are dominated by variation 
 over the full region; `error' bars on the distance from the front are defined as the rough width of the 
 region definition - see Fig. \ref{fig:img_regions}.
\label{fig:hh_av_map}}
\end{figure*}

We can use the measured Paschen-$\alpha$ and Brackett-$\alpha$ along with Case B recombination to estimate the 
difference in attenuation between the wavelengths of those transitions, $E(\mathrm{Pa}_\alpha - \mathrm{Br}_\alpha)$.
From the measured $E(\mathrm{Pa}_\alpha - \mathrm{Br}_\alpha)$, with an assumed extinction curve, we can construct
maps of the total visual extinction, $A_V$. To maximize the fidelity of the $A_V$ estimate, well-measured lines
with significant wavelength separation are ideal. With those criteria in mind, we proceed with the 
Pa-$\alpha$ (1.87~\micron) and Br-$\alpha$ (4.05~\micron) lines;  while we have identified hydrogen line pairs
with larger wavelength separations and higher line fluxes in our data, Pa-$\alpha$ and Br-$\alpha$ represent
the best combination of those criteria in these data.  We estimate the total visual extinction as

\begin{equation} 
    A_V = \frac{1.086}{C_{\lambda_2}-C_{\lambda_1}} \left[\ln\left(\frac{I_{\lambda_1}^{obs}}{I_{\lambda_2}^{obs}}\right)-\ln\left(\frac{I_{\lambda_1}^{B}}{I_{\lambda_2}^{B}}\right)\right]
    \label{eq:av}
\end{equation}

\noindent 
where $C_i$ is the extinction at $\lambda_i$, $I_{\lambda_i}^{obs}$ is the observed line strength at $\lambda_i$ and
$I_{\lambda_j}^{B}$ is the Case B line emissivity for $\lambda_j$. For the Case B line ratios, we take values 
from \citet{2018MNRAS.478.2766P} for $n_e = 1000$~cm$^{-3}$ and $T_e = 10000$~K; over the range of reasonable
values of $n_e$ and $T_e$, the uncertainty in $A_V$ is completely dominated by the uncertainty in the 
observed line strengths.  For the extinction, we use the extinction curve defined in \citet{2023ApJ...950...86G}
for an $R_V$ of 3.1. The strength of the Pa-$\alpha$ and Br-$\alpha$ lines 
in our Horsehead data allows us to extract line measurements not only integrated over each
region but also per `spaxel' in the NIRSpec IFU cube.  With those line measurements, we estimate $A_V$ using 
Eq. \ref{eq:av}; the estimates of $A_V$ for each region are given in Table \ref{tbl:av} and plotted in 
Fig. \ref{fig:hh_av_map} along with the spatial map of $A_V$.  The quoted uncertainties in Table \ref{tbl:av} 
and plotted in Fig. \ref{fig:hh_av_map} for the region points include both 
the uncertainties in the observed line as well as the spatial variation in $A_V$ across all the pixels in 
the region. 
For the spatial $A_V$ cut plotted in Fig. \ref{fig:hh_av_map}, in addition to the uncertainty from individual
line measurements per spaxel (gray band), we plot the range of $A_V$ derived for
$2.3 \leq R_V \leq 3.1$ (light blue region)
and $3.1 \leq R_V \leq 5.6$ (light green region).
In the spatial map of $A_V$, especially for $A_V\gtrsim5$, spaxel-to-spaxel variations are dominated by 
the noise in the map rather than reflecting true variation at that spatial scale.

\begin{table}[bth]
\begin{threeparttable}[b]
  \caption{Internal Horsehead Visual Extinction, $A_V$}\label{tab:av_region}
  \begin{tabular*}{0.5\textwidth}{@{\extracolsep{\fill}}cccc}
  \toprule
  Region\tnote{a} & $A_V$ & $\sigma_{line}$ & $\sigma_{spatial}$ \\ 
  \midrule
  H\,{\sc{ii}}             & 0.58  & 0.42 & 1.87  \\ 
  DF1             & 1.51  & 0.10 & 0.93  \\ 
  MO1             & 2.90  & 0.11 & 0.93  \\ 
  DF2             & 7.02  & 0.16 & 2.60  \\ 
  MO2             & 9.28  & 0.19 & 3.35  \\ 
  \hline
  \end{tabular*}

  \label{tbl:av}

 \begin{tablenotes}
   \item [a] See Fig. \ref{fig:img_regions} for physical location of region.
 \end{tablenotes}

\end{threeparttable}
\end{table}

$A_V$ is seen to increase relatively smoothly from near zero in the H\,{\sc{ii}} region into the 
DF2 region and remains roughly 
constant at $\sim10$. 
While the extinction is very small in the H\,{\sc{ii}} region, there is residual extinction observed here - 
while this may reflect the uncertainty in the $A_V$ determination, it may also reflect the 
non-zero amount of dust surviving in the H\,{\sc{ii}} region - see Sect. \ref{sssec:outflow} and \citet{2024A&A...687A...4A}.
While $A_V$ generally follows the line and continuum emission, it is much 
smoother and does not exhibit the sharp transition evident in emission. 
Zannese et al. (in prep) estimated $A_V$ along a similar cut through the 
PDR as shown in Fig. \ref{fig:hh_av_map}, as well as for the same DF1 and DF2 regions defined here,
using an independent technique involving line ratios of \HH\ lines originating from the same
upper energy level. For both the region determinations and the cut across the PDR with R$_V$=3.1,
they find similar results (Table \ref{tbl:av}) with region estimates of 
$A_V \sim 0.3\pm1.3$ and $\sim 6.1\pm1.4$ for DF1 and DF2, respectively, and smoothly rising 
through the PDR cut (Fig. \ref{fig:hh_av_map}), though our estimates are systematically higher which 
may reflect the relative spatial location of the respective emission regions of H\,{\sc{i}} and \HH. 
Based on imaging, \citet{2024A&A...687A...4A} 
proposed a model with a relatively small and constant attenuation layer in front of the illuminated 
molecular gas extending to roughly $5-7\arcsec$ inside of the H\,{\sc{ii}} region/molecular cloud interface (corresponding 
roughly to the DF1 and MO1 regions defined here) with increasing attenuation from $7-12\arcsec$ (see their 
Fig. 13).  We confirm that behavior here with lower and roughly constant $A_V$ from the interface to 
0 to $-5\arcsec$ into the molecular cloud (projected), followed by a steady increase through the DF2 region, 
corresponding to the increasing path length to the illuminated material proposed in \citet{2024A&A...687A...4A}. 
It must be noted that the above 
analysis assumes a `screen-like' geometry, and we do not account for the fact that the dust responsible for 
the implied extinction is mixed with the emitting material - the true column of dust is likely
higher than indicated; more detailed modeling of the dust in the Horsehead is presented in \citet{elyajouri2024}.

\subsection{\HH \label{ssec:h2}}
Of the 200-400 total lines detected in 9 of the 10 regions (excluding the Horsehead HII region), 
75-85\%\ are \HH\ lines.  Of those \HH\ lines, 
$\sim$85\%\ are ro-vibrational lines with the remaining $\sim$15\%\ being pure rotational lines 
(see Table \ref{tab:linecounts}).  Here we provide an inventory of the detected series;
a companion paper will present a detailed analysis of the warm H$_2$ emission at the 
interface between the ionized and molecular gas in the Horsehead (Zannese et al. in prep.).

\begin{figure*}[h]
   \centering
   \includegraphics[width=\textwidth]{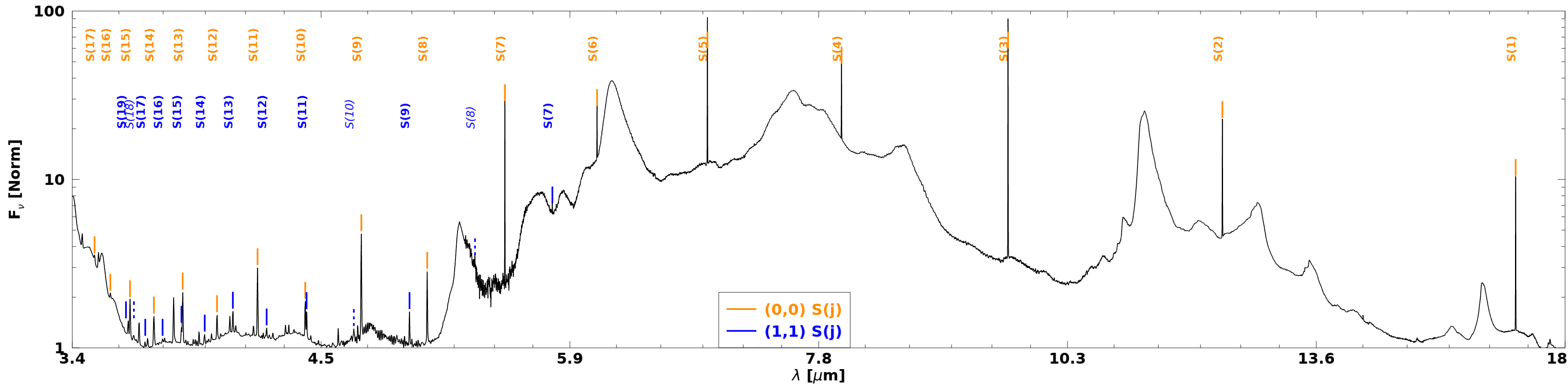}
   \caption{Plot showing the lines in the pure rotational (0,0) and (1,1) series of H$_2$ (orange and blue, respectively) for the DF1 region of NGC~7023. 
   Identifications in italics with a dashed line were not included in final line lists -  the (1,1) S(18,10,8) lines 
   were weak and in complex regions of the spectrum, making extraction uncertain.
   \label{fig:ngc7023df1_h2_purerot_series}}
\end{figure*}

\begin{figure*}[h]
   \centering
   \includegraphics[width=\textwidth]{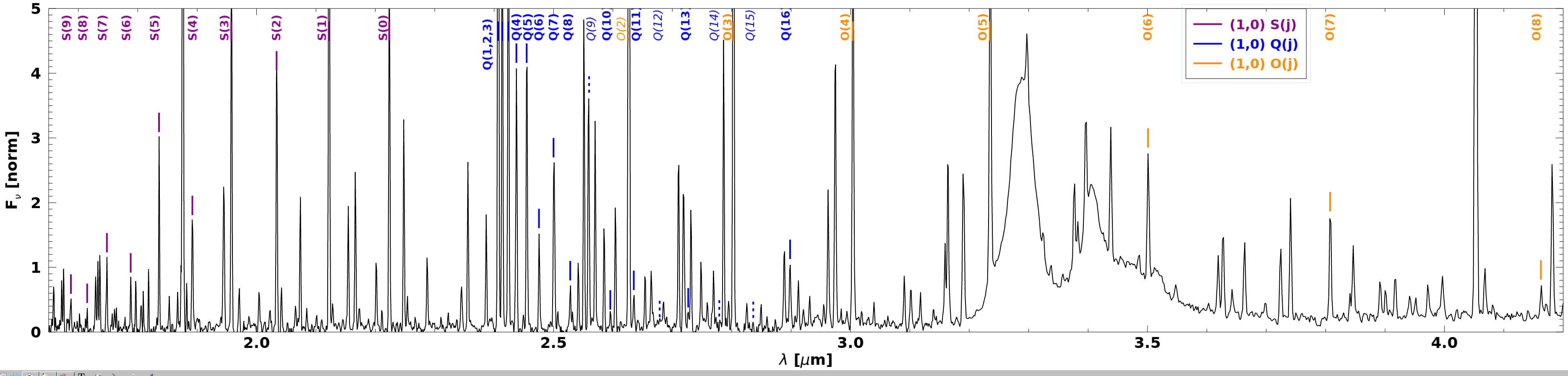}
   \caption{Plot showing the lines in the S, Q, and O branches (purple, blue, and orange, respectively) of the H$_2$
   (1,0) transitions for the DF1 region of the Horsehead. Identifications in italics with a dashed line were not included in final line lists - 
   Q(9) is blended with the H$_2$ (2,1)Q(2) transition and not extracted independently while Q(12), Q(14), and
   Q(15) were very weak and not definitively detected. In addition, the O(2) line is blended with H\,{\sc{i}} Br $\beta$
   and not independently extracted.
   \label{fig:hhdf1_h2_v1-0_sqo_series}}
\end{figure*}

In all regions, except the Horsehead H\,{\sc{ii}} region, we detect the pure rotational 
(0,0) and (1,1) series with tentative identifications of isolated (2,2), (3,3), and (4,4) lines.
In the (0,0) S($J$) series, $J$ values from $1-15$, and in some regions $J=16,17,18$, are detected.
Additionally, $J=9-19$ lines in the (1,1) S($J$) series are identified. Representative 
identifications are shown in Fig. \ref{fig:ngc7023df1_h2_purerot_series} for the NGC~7023 
DF1 region. While a few lines in the (2,2), (3,3), and (4,4) series are formally identified, 
they are generally weak and uncertain.
Of the many ro-vibrational transitions, there are well-defined sequences in the 
vibrational (1,0), (2,1), (2,0), (3,2), and (3,1) series.  These sequences include transitions in 
the S ($\Delta J=-2$), Q ($\Delta J=0$), and O ($\Delta J=2$) branches. Representative sequences 
in the (1,0) vibrational transition are 
shown in Fig. \ref{fig:hhdf1_h2_v1-0_sqo_series} for the Horsehead DF1 region.
Note that while, in general, consecutive lines in the 
specified series are identified, stronger lines from other series or species with stronger 
transitions at similar wavelengths can lead to some lines in the series not being extracted.

We note the formal detection of 4 \HH\ lines in the Horsehead H\,{\sc{ii}} region, specifically the (0,0)~S(5,3,2,1)
lines. Nominally, it is not expected that any \HH\ lines would be present in the harsher exposed environment
of the H\,{\sc{ii}} region. In contrast, we have detected outflow of material from the PDR into the H\,{\sc{ii}} region 
(see \citet{2024A&A...687A...4A} and Sect. \ref{sssec:outflow}) and the outflow may be entraining 
some \HH\ molecules.  However, the detected lines in the low J (0,0) sequence are expected to be largely collisionally
excited rather than radiatively pumped, whereas any \HH\ entrained in the low-density exposed outflow
would more likely be radiatively rather than collisionally excited. In addition, the detected lines in the H\,{\sc{ii}} region
all fall in the MIRI wavelength coverage, where the spatial PSF is considerably larger relative to NIRSpec. Combined, 
these facts would argue that the detection of \HH\ in the H\,{\sc{ii}} region is the result of partial MIRI pixels belonging to 
the PDR emission being included in the defined H\,{\sc{ii}} region. Indeed, there is a significant weakening of any detected 
\HH\ emission in the H\,{\sc{ii}} region if we restrict the extraction region in MIRI to pixels furthest from the interface, 
whereas the aromatic emission (see Sect. \ref{sssec:outflow}) is uniformly distributed throughout the extraction region.

\subsection{Other Molecular Gas Species \label{ssec:molgas}}
In addition to \HH, we identify the presence of two other simple molecules, CH$^+$ and CO. Detection of CH$^+$
was not ubiquitous and is largely confined to the DF1, and DF2 regions in NGC~7023.  CO was detected in  
all regions of NGC~7023 except for ATM.  While there are `formal' automated identifications of 
both molecules in most regions in both objects, except for the above, they are isolated 
(not part of a defined series) `detections' and generally have likely alternative identifications.

\begin{figure}[h]
   \centering
   \includegraphics[width=0.45\textwidth]{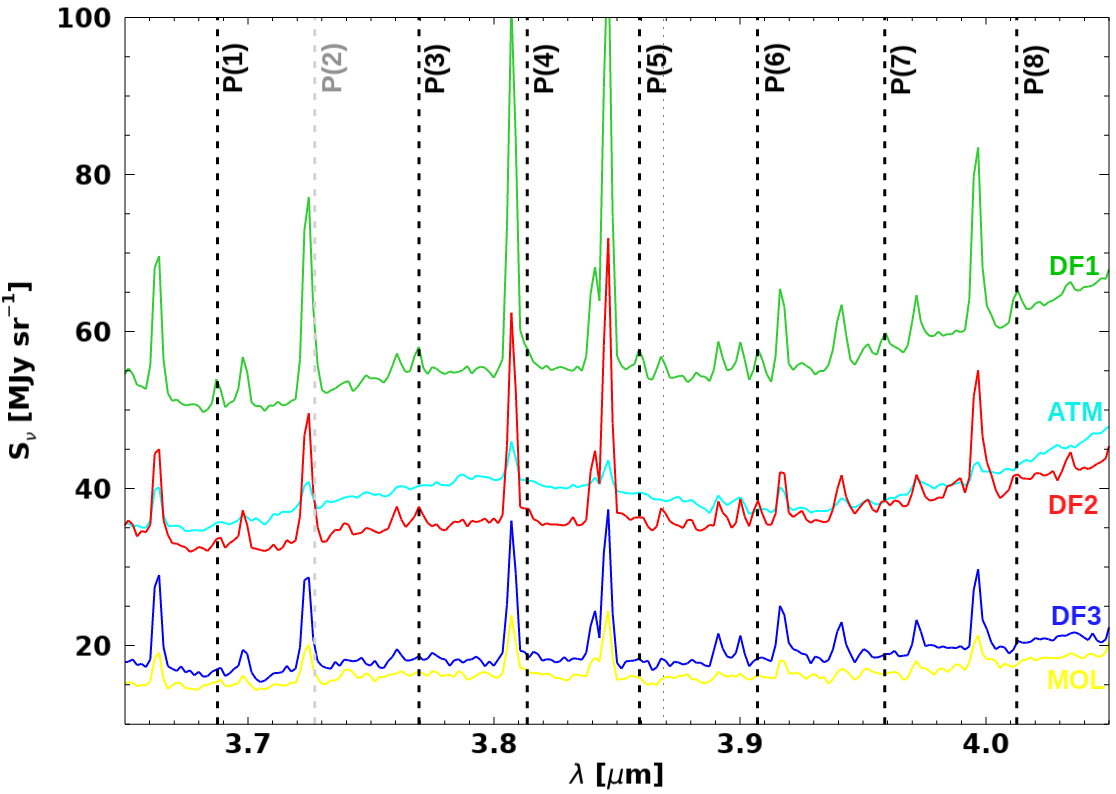}
   \caption{CH$^+$ $\nu=1\rightarrow0$ lines detected in NGC~7023. Detection of CH$^+$ lines restricted to DF1 and 2
   (green and red) regions. All detected lines are P-branch from 1-8.  P(1) is potentially confused with the He\,{\sc{i}} 3D-3Fo
   multiplet and P(2) blended with the much stronger (0,0)S(14) \HH\ line and is not independently detected.}
   \label{fig:chp_7023}
\end{figure}

Transitions in the P branch ($J \rightarrow J+1$) of the $\nu = 1\rightarrow0$ ro-vibrational sequence 
of the hydride CH$^+$ have been detected in NGC~7023. The detections are restricted to the DF1 region 
and, to a lesser degree, the DF2 region.  In the former, P(1-8) lines were detected whereas only P(1,3,4,6) 
were detected in the latter.  We did not detect any transitions in the R branch ($J \rightarrow J-1$).
The detections are shown in Fig. \ref{fig:chp_7023} and integrated line strengths are provided in 
Table \ref{tab:chp7023}.  The limited detection of CH$^+$ indicates that the conditions for significant 
formation of CH$^+$ and excitation of the $\nu = 1$ band are not widespread in NGC~7023 and are absent in 
the Horsehead. \citet{2024A&A...685A..74P} have noted the presence of limited CH$^+$ $\nu = 1\rightarrow0$
P branch and weak R branch emission in the Orion Nebula and a detailed analysis is presented 
in \citet{2025A&A...696A..99Z}.  The absence of significant R branch emission 
is consistent with those data and theoretical considerations \citep{2021ApJ...917...16C}. 
The main formation pathway for $CH^{+}$, $C^{+} + H_{2} \rightarrow CH^{+} + H$, is highly endothermic
requiring high temperatures, and as a collisional process with excited \HH, is strongly density
dependent \citep{2021ApJ...917...15N,2025A&A...696A..99Z}.
Given the strong dependence of the CH$^+$ formation rate on temperature and density, it is not surprising 
that we only weakly detect CH$^+$ in the brightest, densest regions of NGC~7023 \citep{1997ApJ...484..296M,field2010} 
and not at all in the Horsehead,
where gas temperatures and densities are lower \citep{2005A&A...437..177H,2025A&A...696A..99Z}

\begin{table}[bth]
\begin{threeparttable}[b]
 \caption{Detected CH$^+$ lines\label{tab:chp7023}}
 \begin{tabular*}{0.5\textwidth}{@{\extracolsep{\fill}}lccc}
 \toprule
                     &                              & \multicolumn{2}{c}{Region Line Strength\tnote{a}} \\
 Transition\tnote{b} & $\lambda$\tnote{c} [\micron] & DF1                 & DF2                \\ 
 \midrule
  P(1)\tnote{d}      & 3.68758                      & {\it 2.21$\pm$2.61} & {\it 1.22$\pm$107} \\
  P(2)\tnote{e}      & 3.72717                      & $\cdots$             & $\cdots$            \\
  P(3)               & 3.76891                      & 2.19$\pm$0.90       & 2.35$\pm$0.88      \\
  P(4)               & 3.81288                      & 2.62$\pm$1.15       & 1.53$\pm$0.79      \\
  P(5)               & 3.85914                      & 1.98$\pm$0.96       & $\cdots$            \\
  P(6)               & 3.90777                      & 1.61$\pm$0.86       & 1.83$\pm$0.72      \\
  P(7)               & 3.95886                      & 2.06$\pm$1.16       & $\cdots$            \\
  P(8)               & 4.01249                      & 1.66$\pm$1.03       & $\cdots$            \\
 \midrule
 \end{tabular*}
   \begin{tablenotes}
     \item [a] 10$^{-6}$ erg s$^{-1}$ cm$^{-2}$ sr$^{-1}$; See Fig. \ref{fig:img_regions} for physical location of region.
     \item [b] All detected CH$^+$ transitions $\nu=1\rightarrow0$ vibrational modes.
     \item [c] Theoretical wavelength of transition.
     \item [d] Potential blend with He\,{\sc{i}} 3D-3Fo multiplet; uncertain line strength
     \item [e] Blended with H$_2$(0,0)S(14); no extraction
  \end{tablenotes}
\end{threeparttable}
\end{table}

\begin{figure}[h!]
\centering
\includegraphics[width=0.45\textwidth]{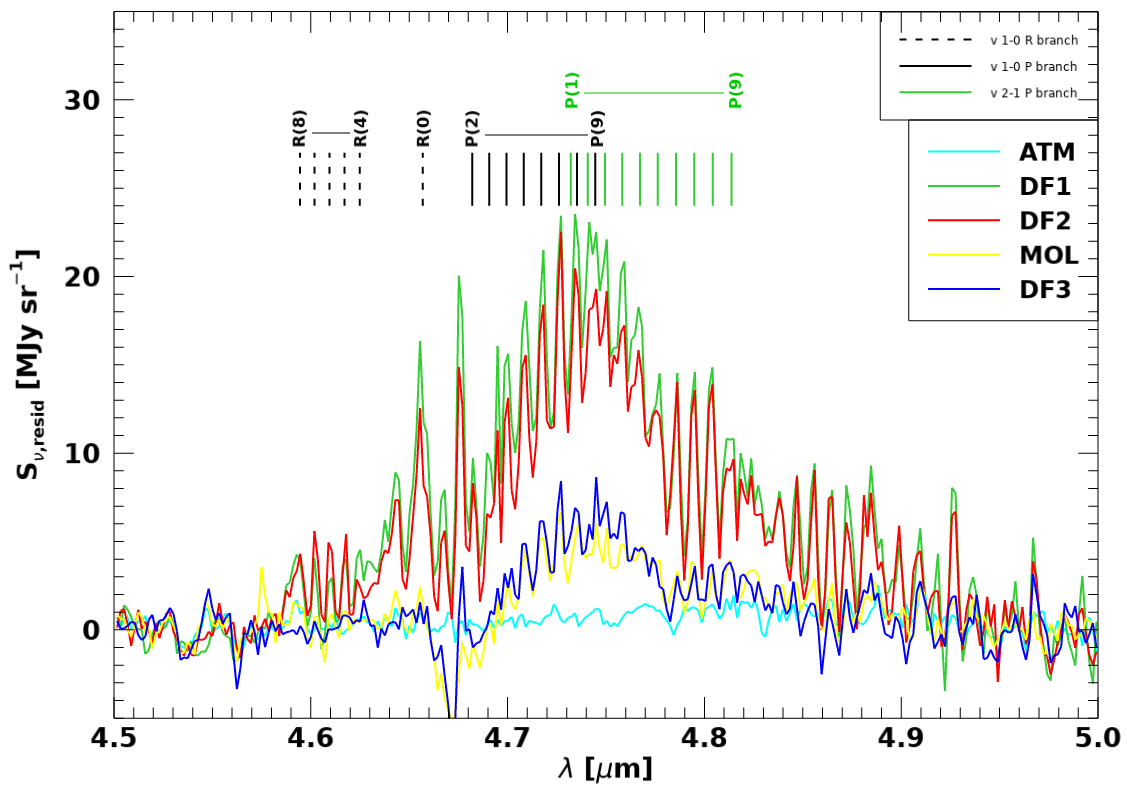}
\caption{Continuum subtracted spectra in the region of prominent CO lines for NGC~7023.  Clearly identified transitions are labeled; beyond $\sim$4.8~\micron, the 1-0 and 2-1 P branch transitions overlap and are not resolved at the NIRSpec medium resolution. 
\label{fig:co_ids}}
\end{figure}

We also find clear CO emission around $\sim$4.7~\micron\ in the $\nu\ 1\rightarrow0$ and $\nu\ 2\rightarrow1$
P and R branches in NGC~7023. \citet{2024A&A...685A..74P} report similar CO detections in Orion.
While these CO lines are fairly well separated in the high-resolution gratings used in \citet{2024A&A...685A..74P}, 
with our medium resolution gratings, the CO $\nu\ 1\rightarrow0$ and $\nu\ 2\rightarrow1$ sequences have 
considerable overlap.  Consequently, our iterative line identification (Sect. \ref{ssec:line_identity}) procedure 
is not well-suited to identifying the CO sequence. However, visual inspection of our template spectra shows a 
clear sequence of line emission superposed on the C-D aliphatic stretch continuum between 4.6-4.8~\micron\ 
(see Fig. \ref{fig:ch_vs_cd_7023} and Sect. \ref{ssec:carb_ss}).  To identify the lines in this spectral region, we 
first fit a pseudo-continuum between 4.5 and 5~\micron\ and subtracted it from the template spectra. The result 
is shown in Fig. \ref{fig:co_ids} for NGC~7023 and a clear sequence of lines is seen in all regions except for 
the ATM template (a similar examination of the Horsehead templates did not yield any unambiguous CO line 
identifications). The spectral resolution is
too low to robustly fit individual lines, so we identify the lines by over-plotting the rest wavelength locations
of known CO lines. There is a good match between the observed line positions and the rest wavelengths of CO lines
for both the $\nu\ 1\rightarrow0$ and $2\rightarrow1$ P-branch transitions with $J\sim 1-9$ (upper energy levels 
between $E_u/k ~ 3000-7000~K$).  Lines on the R-branch of the $\nu\ 1\rightarrow0$ ro-vibrational sequence, while 
identified, are considerably weaker than P-branch lines. For P-branch transitions with $J_l > 9$ 
for both $\nu\ 1\rightarrow0$ and $2\rightarrow1$ ($\lambda \gtrsim 8.3$~\micron), the agreement of the observed 
lines with rest wavelengths is generally poor, and no unique line identifications can be made (Fig. \ref{fig:co_ids}).

While there is a tentative detection of low $J$ lines of HD in the $\nu\ 1\rightarrow0$ R branch in the DF1 region 
of NGC~7023 (and some `scattered' detections of other HD lines in other regions), the lines are weak.
The detections are often in the wings of stronger \HH\ lines or only alternative identifications for other atomic 
or \HH\ lines. Thus, with data at this spectral resolution, we cannot definitively identify HD lines.

\subsection{Ices - NGC~7023 \label{ssec:ices}}

Several ice features of H$_2$O, CO, and CO$_2$ are identified in the more deeply embedded regions 
(MOL, DF3, and, to a lesser degree, DF2) in NGC~7023. Here, we enumerate the detected ices
(proceeding from the shortest wavelengths to the longest).  A similar examination shows no strong 
evidence for ice features
in any region of the Horsehead. 

\begin{figure}[h!]
   \centering
   \includegraphics[width=0.45\textwidth]{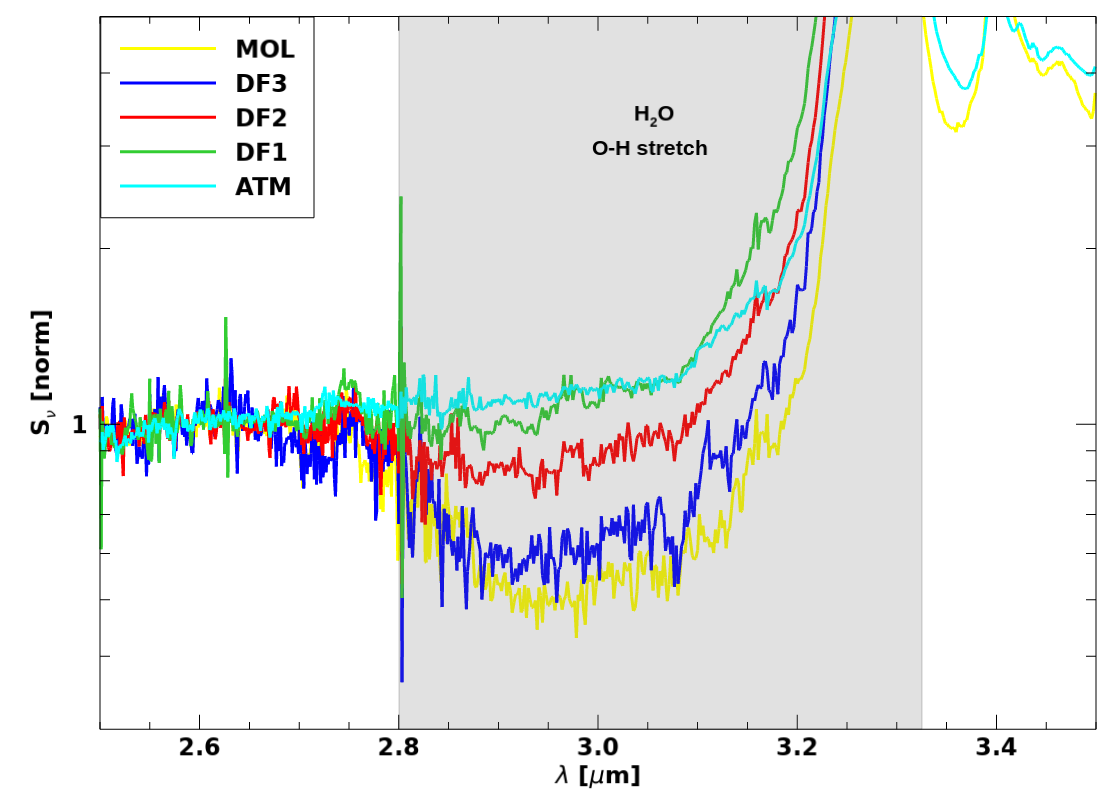}
   \caption{Plot of NGC~7023 spectra in the wavelength regime around 3~\micron.  Significant 
   H$_2$O ice absorption is apparent in the MOL and DF3 regions, with a marginal indication in 
   the DF2 region. The DF1 and ATM regions are consistent with no absorption. Spectra have been 
   normalized at 2.5-2.65~\micron.
   \label{fig:h2oice_3p0}}
\end{figure}

We identify an absorption band at $\sim$3~\micron\ in the MOL, DF3 and potentially DF2, regions 
of NGC~7023  - see Fig. \ref{fig:h2oice_3p0}.
Broad absorption in this spectral region towards dense clouds and disks has been convincingly attributed to 
the O-H stretch in H$_2$O ice \citep{1979A&A....79..256L,1989ApJ...344..413S,1993ApJS...86..713H}, and we 
identify the observed absorption with H$_2$O ice. While the presence of very strong carbonaceous solid-state emission (see Sect. \ref{ssec:carb_ss}) on the red side ($\lambda\gtrsim 3.1$~\micron) of the absorption 
complicates the interpretation, the observed profile is broad, smooth, and roughly centered at $\sim$3~\micron.
These properties are strongly indicative of cold (10-30~K) amorphous H$_2$O ice as warmer crystalline ices 
show more structure in the profile and are shifted towards long wavelengths 
\citep[and references therein]{Boogert15}. We find no evidence for absorption at $\sim$6~\micron\ (H$_2$O
bend) or $\sim$13.6~\micron\ (H$_2$O libration). However, those transitions are expected to be weaker in 
amorphous H$_2$O ice, and the spectra in those wavelength regimes are complicated by ubiquitous 
carbonaceous emission bands. A full multi-component characterization of the spectra (Van De Putte et al., in prep)
will allow a more detailed characterization of the 3~\micron\ H$_2$O ice absorption as well as a more
rigorous search for other, weaker H$_2$O ice absorption features.

\begin{figure*}[h!]
\centering
\begin{tabular}[b]{cc}
 \includegraphics[width=0.45\textwidth]{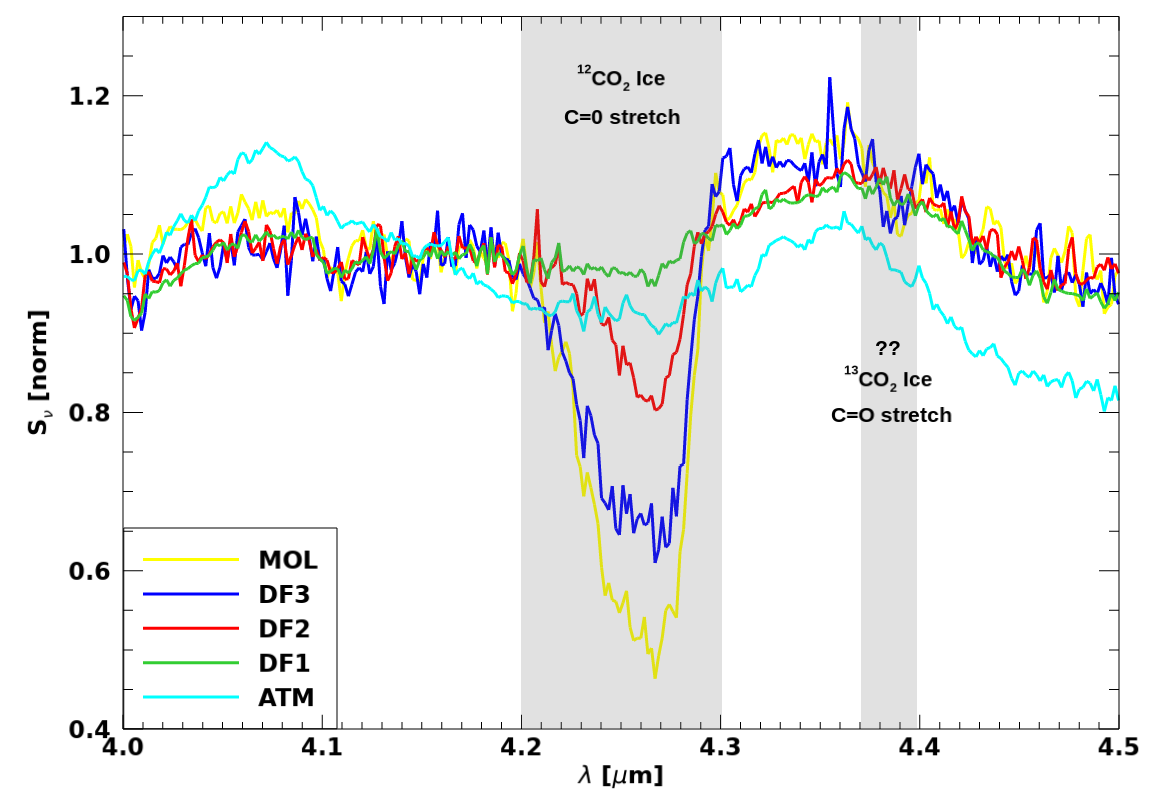}  & \includegraphics[width=0.45\textwidth]{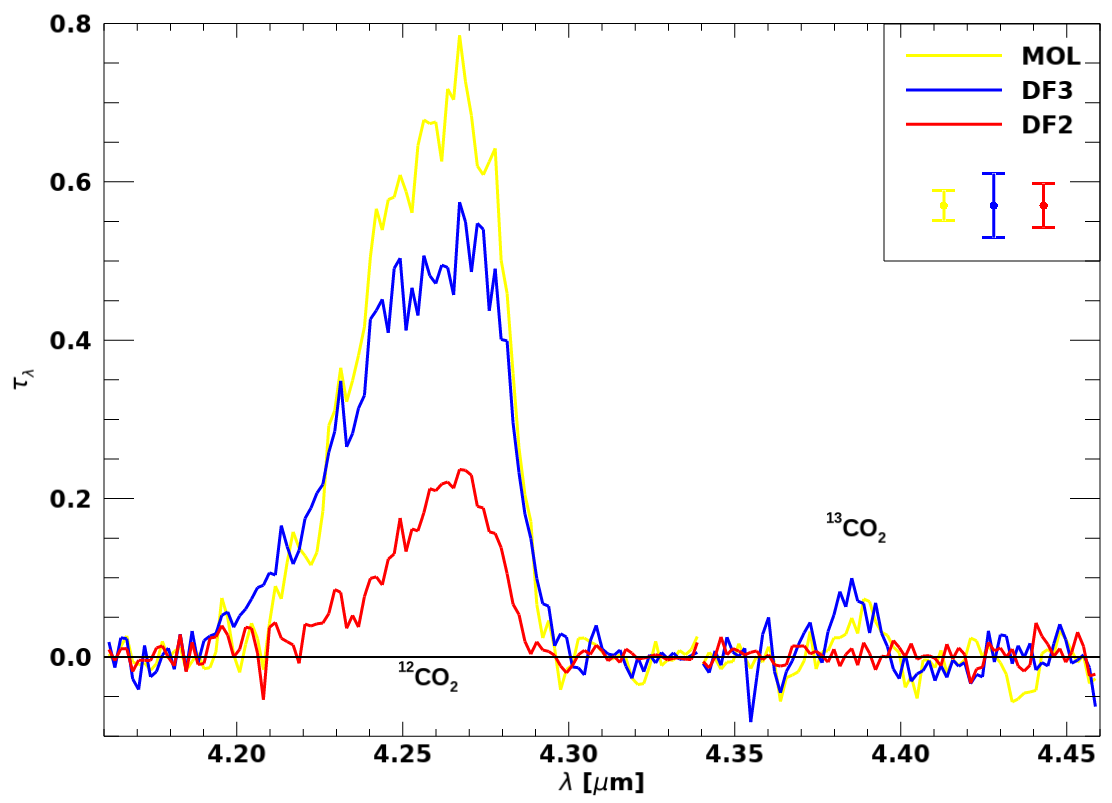} \\ 
\end{tabular}
\caption{Plot of NGC~7023 spectra in the wavelength regime around 4.3~\micron. 
        Left: $^{12}$CO$_2$ ice absorption at $\sim4.26$~\micron\ is detected in the 
        MOL, DF3, and DF2 regions.  Additionally, there is a weak detection of 
        $^{13}$CO$_2$ ice absorption at $\sim4.39$~\micron\ in the MOL and DF3
        regions.  Right: Estimate of the optical depth in the ice absorption bands.
        \label{fig:co2ice_4p25}}
\end{figure*}

\begin{figure*}[h!]
\centering
\begin{tabular}[b]{cc}
 \includegraphics[width=0.45\textwidth]{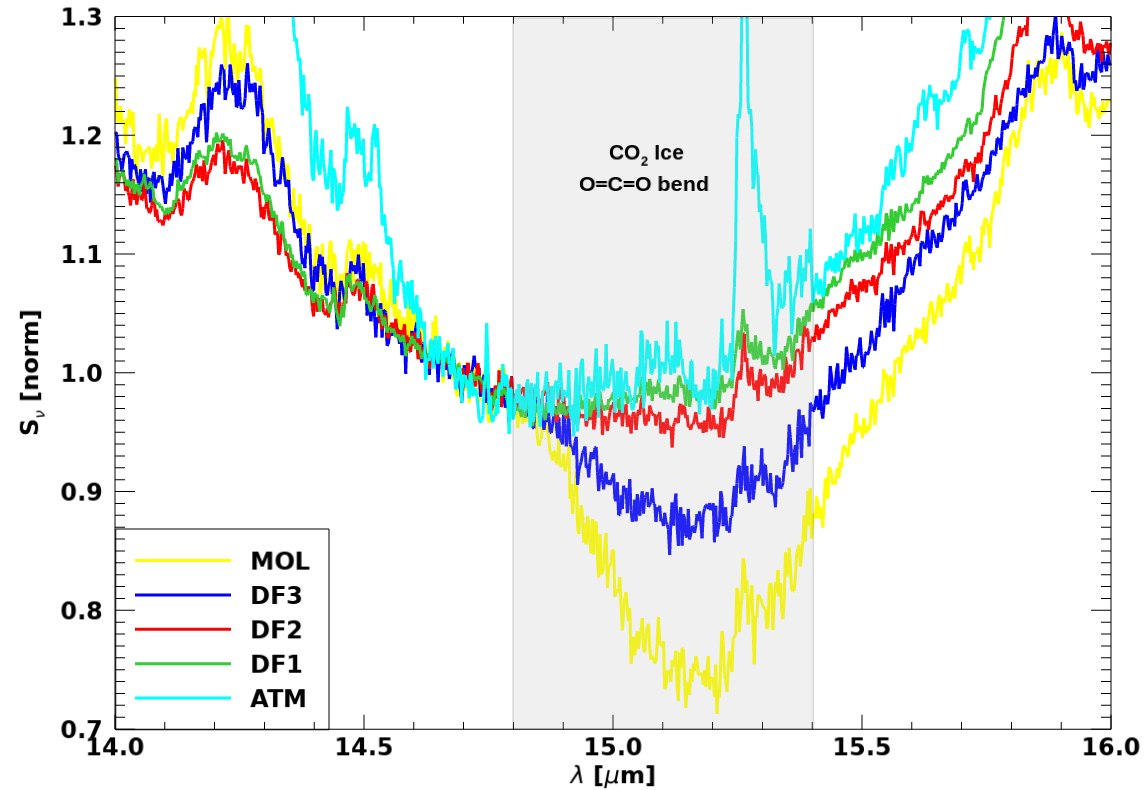}  & \includegraphics[width=0.45\textwidth]{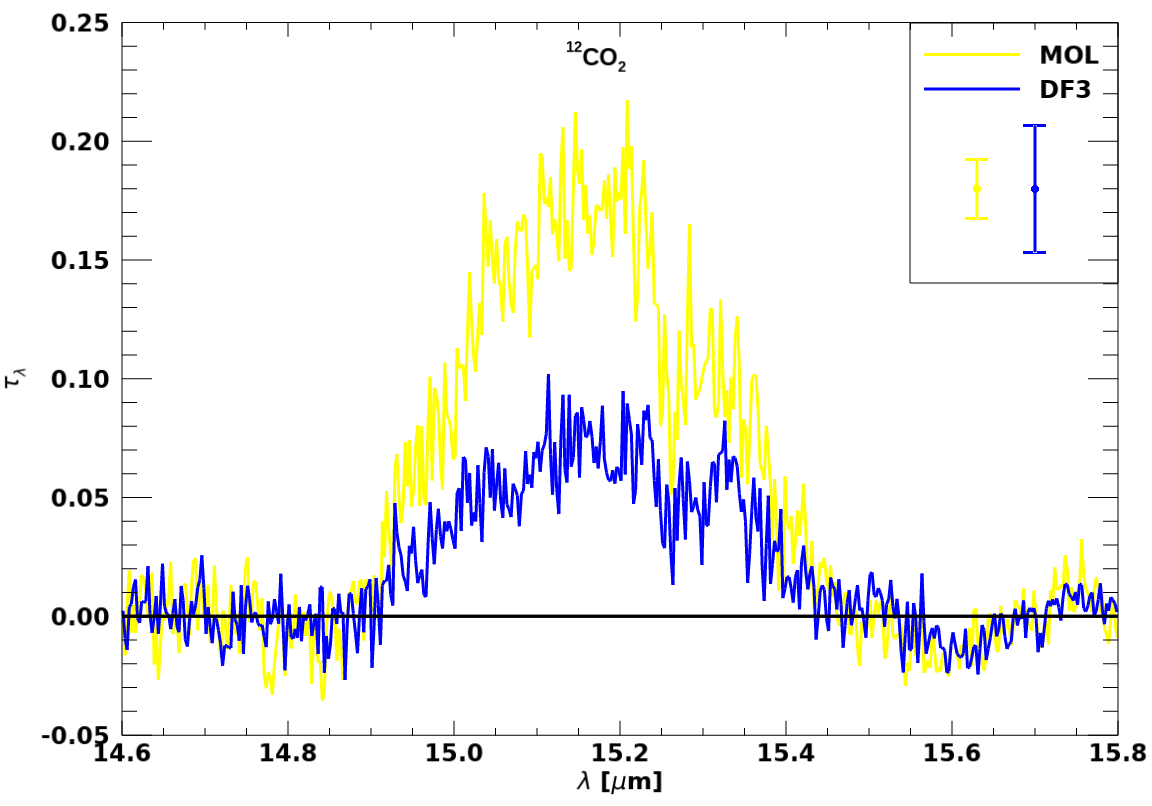} \\ 
\end{tabular}
\caption{Plot of NGC~7023 spectra in the wavelength regime around 15.2~\micron. 
         Left: $^{12}$CO$_2$ ice absorption at $\sim15.2$~\micron\ is detected in the 
         MOL and DF3 regions. The spectra are normalized between 14.6 and 14.8~\micron.
         Right: Estimate of the optical depth in the ice absorption band.
         \label{fig:co2ice_15p2}}
\end{figure*}

We identify three absorption features in our spectra with the presence of CO$_2$ ice - $^{12}$CO$_2$ at 
$\sim$4.26~\micron\ and $\sim$15.2~\micron, as well as $^{13}$CO$_2$ at $\sim$4.39~\micron\ \citep{Boogert15}; 
see Figs. \ref{fig:co2ice_4p25}, \ref{fig:co2ice_15p2}, and respectively. All three
features appear to be broad single peaks.  While the $\sim$4.26~\micron\ feature is observed in the MOL, DF3,
and DF2, the other features are only detected in the MOL and DF3 regions. For the $\sim$4.39~\micron\ $^{13}$CO$_2$
feature in DF2, if we assume a similar scaling to the strength of $\sim$4.26~\micron\ $^{12}$CO$_2$ feature as 
observed in MOL and DF3, it would be detected at $<$1~$\sigma$, so its absence in DF2 is almost certainly an 
issue of sensitivity and does not necessarily reflect a physical difference. In the case of the $\sim$15.2~\micron\ 
feature, the presence of strong carbonaceous emission bands on either side of a potentially weak,
ice feature complicates 
extraction of the ice profile, and we cannot rule out that the ice absorption is absent in DF2 despite the clear
detection at $\sim$4.26~\micron. With respect to the extraction of the $\sim$15.2~\micron\ ice profile, we also 
note that there is a strong, presumably carbonaceous, emission feature superposed on the ice absorption profile
($\sim$15.25~\micron), creating spurious structure in the ice band profile (Fig. \ref{fig:co2ice_15p2}).


The lack of structure in both the $\sim$4.26~\micron\ and $\sim$15.2~\micron\ features is consistent with 
cold CO$_2$ ice in a polar (H$_2$O dominated) phase. In addition to structure, the 4.26~\micron\ feature will 
tend to shift to longer wavelengths with both annealing to higher temperatures and/or the addition of significant
non-polar phases \citep[see e.g.][and references therein]{1999ApJ...522..357G,Boogert15}. Very similar behavior 
is seen with the $\sim$15.2~\micron\ CO$_2$ emission \citep{1999A&A...350..240E}.  While polar CO$_2$ at
$\sim$15.2~\micron\ tends to have a long tail to longer wavelengths that is not apparent in 
Fig. \ref{fig:co2ice_15p2} \citep{Boogert15}, the complex continuum will require a full decomposition to
disentangle details of the ice absorption shape from the underlying continuum. In support of cold H$_2$O 
dominated CO$_2$ ices, the $^{13}$CO$_2$ profile at $\sim$4.39~\micron\ narrows and shifts to shorter wavelengths
($\sim$4.38~\micron) at higher temperatures and for pure CO$_2$ ice \citep{2000A&A...353..349B}. While, given 
the signal-to-noise of our $^{13}$CO$_2$ detection here, we cannot comment on any structure in the profile, 
the profile centroids are clearly longward of $\sim$4.38~\micron\ (Fig. \ref{fig:co2ice_4p25}).

\begin{figure*}[h!]
\centering
\begin{tabular}[b]{cc}
 \includegraphics[width=0.45\textwidth]{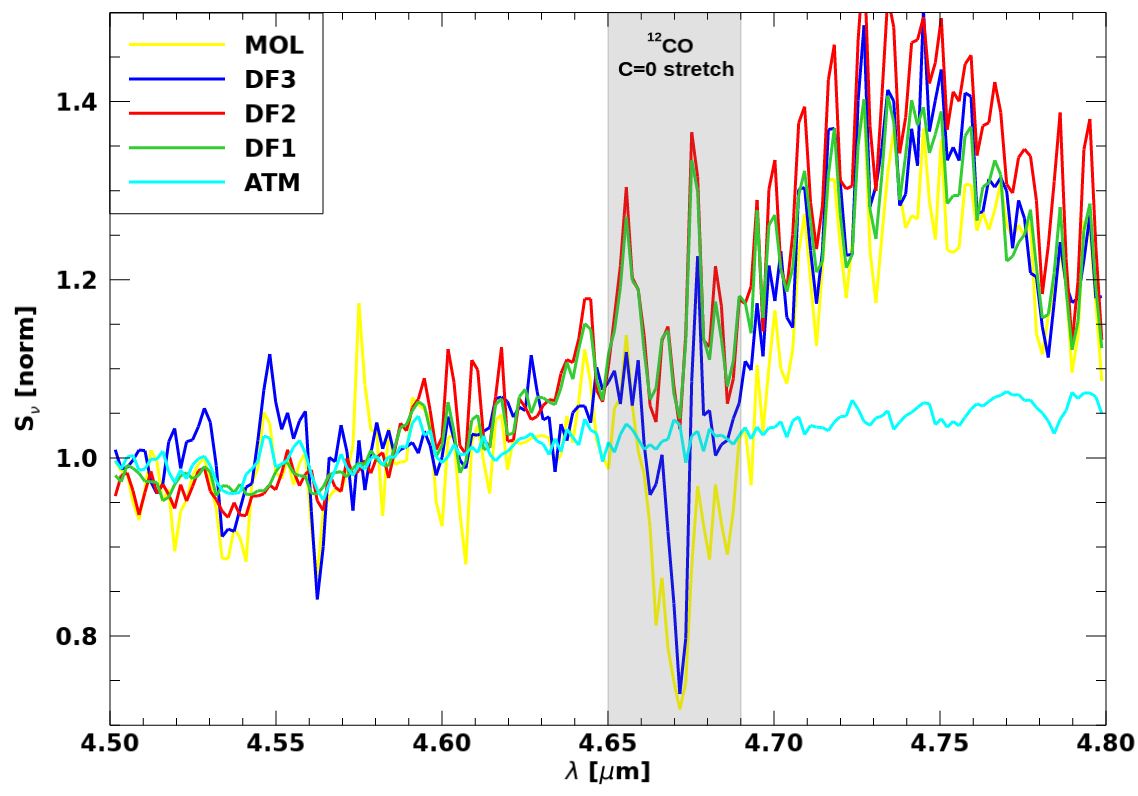}  & \includegraphics[width=0.45\textwidth]{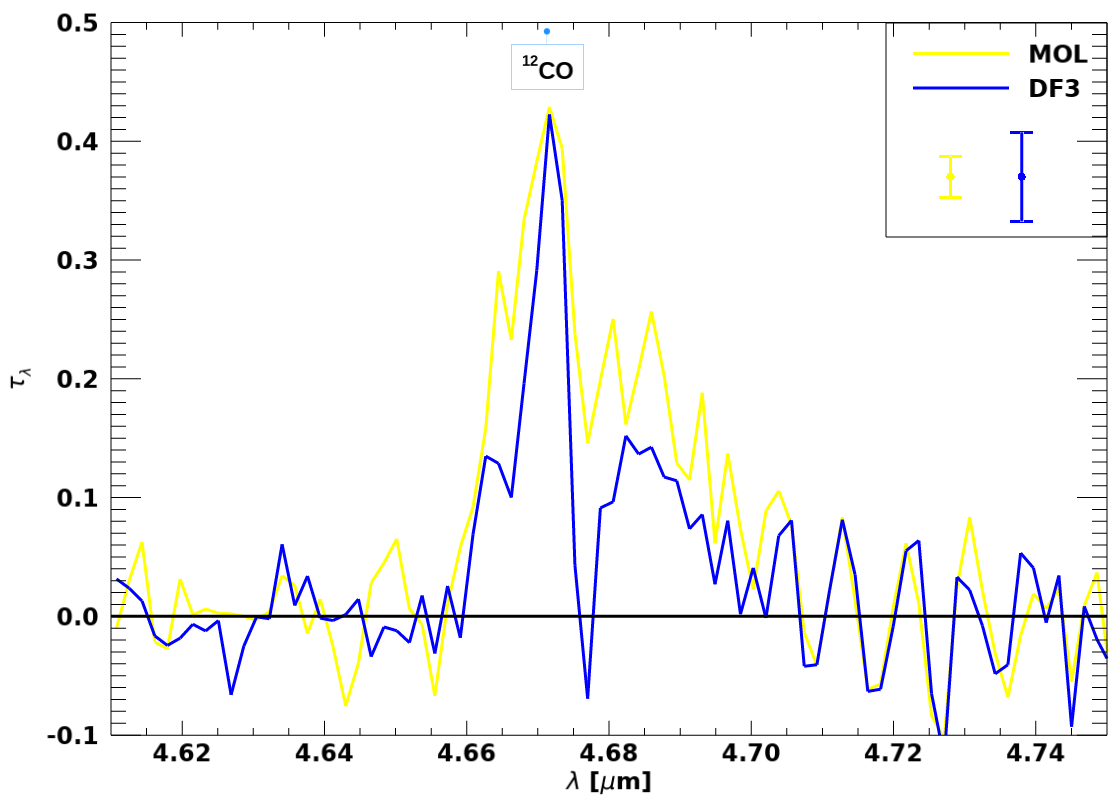} \\ 
\end{tabular}
\caption{Plot of NGC~7023 spectra in the wavelength regime around 4.67~\micron. 
         Left: Weak $^{12}$CO ice absorption at $\sim$4.67~\micron\ is detected in the 
         MOL and DF3 regions, with a potential shoulder out to 4.68~\micron. 
         Right: Estimate of the optical depth in the ice absorption band.
         \label{fig:coice_4p67}}
\end{figure*}

Finally, we identify $^{12}$CO ice absorption at $\sim$4.67~\micron\ in the MOL and DF3 regions. Qualitatively, 
our observed profile has a relatively sharp peak at $\sim$4.67~\micron\ with a potentially broad long wavelength
wing out to $\sim$4.7~\micron; see Fig. \ref{fig:coice_4p67}.  There is no apparent shoulder on the short 
wavelength side of the feature. In the literature, observed profiles of $^{12}$CO at $\sim$4.67~\micron\ have been decomposed
into a narrow feature at $\sim$4.67~\micron, a broad wing at $\sim$4.68~\micron, and a broad feature at
$\sim$4.62~\micron. These components have been identified with pure CO ice, CO ice mixtures (H$_2$O/
CH$_3$OH), and CO plus NH$_3$ 
\citep{1984ApJ...276..533L,1988ApJ...329..498S,2000A&A...353..349B,2003A&A...408..981P}. 
We see no evidence of any 
CO:NH$_3$ component and conclude that the observed profile is consistent with cold CO:H$_2$O ice mixtures. 
Concerning the gas phase detection of CO (see Sect. \ref{ssec:molgas}), the relative weakness of the CO 
(R/P branches) in MOL/DF3, coupled with strength of said lines in DF1/DF2 and commensurate absence of any 
ice signature may be indicative of CO gas freezing out onto grains in the transition from the DF2 region
to DF3/MOL regions.

\subsection{Carbonaceous Solid State Features \label{ssec:carb_ss}}

\subsubsection{Outflow - Horsehead Nebula\label{sssec:outflow}}
\citet{2024A&A...687A...4A} report on the discovery of the entrainment of dust in the photo-evaporative flow 
from the Horsehead PDR based on residual emission extending into the H\,{\sc{ii}} region in both the NIRCam F335M and 
MIRI F770W imaging filters. Here, we confirm the presence of material extending into the H\,{\sc{ii}} region and further 
elucidate its composition. 

\begin{figure*}[hbt!]
\centering
\begin{tabular}[b]{cc}
 \includegraphics[width=0.45\textwidth]{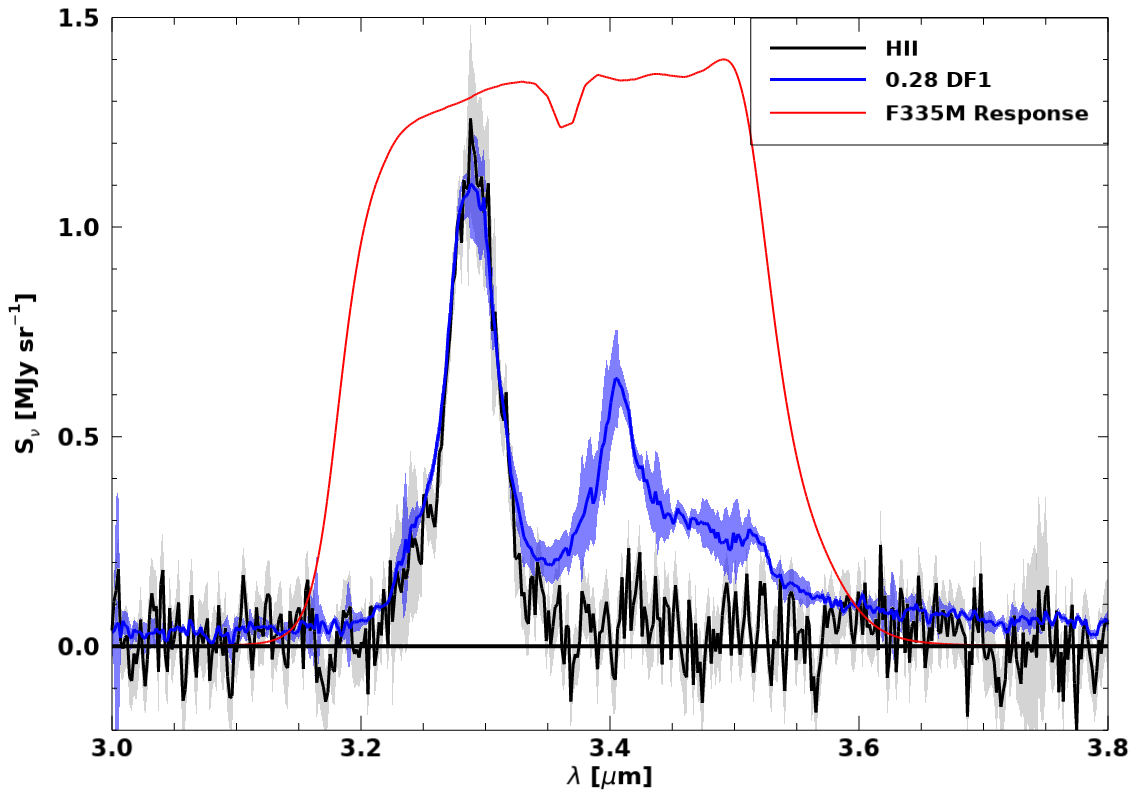}  & \includegraphics[width=0.45\textwidth]{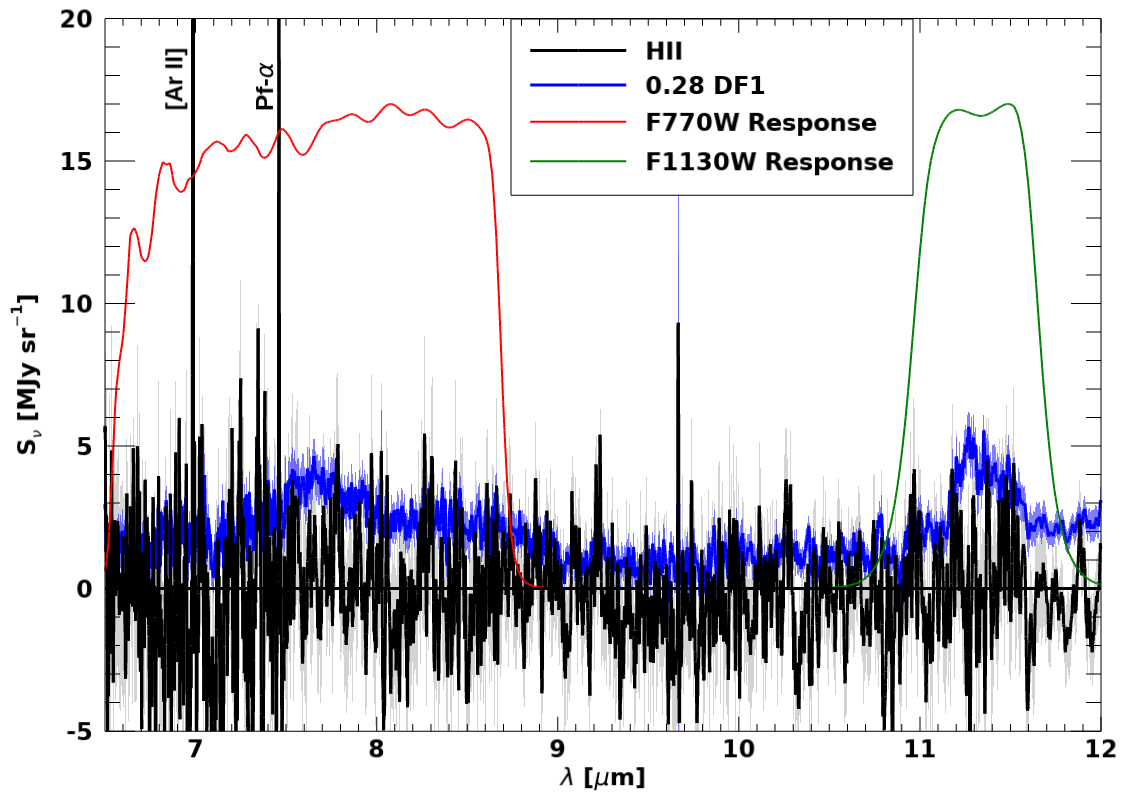} \\ 
\end{tabular}
\caption{Left: H\,{\sc{ii}} and DF1 extractions around the 3.3-3.4~\micron\ features. The DF1 spectrum has been scaled down by 
72\%\ so that the 3.3~\micron\ bands match between the two regions. The F335M filter trace is shown in red. Atomic 
and molecular line emission contributes 6\% and 12\% to the H\,{\sc{ii}} and DF1 region emission, respectively, and the lines
have been removed (Sect. \ref{ssec:lines_contribution}).
Right: H\,{\sc{ii}} and DF1 extractions around the 7.6 and 11.3~\micron\ 
carbonaceous bands. All spectra have been smoothed with a 5-pixel boxcar to emphasize any broad features. 
The DF1 spectrum has been scaled as on the left. MIRI F770W and F1130W filter traces are shown 
in red and green, respectively. In both plots, error bars (3-$\sigma$) are plotted as grey and blue-shaded 
regions for H\,{\sc{ii}} and DF1, respectively. 
\label{fig:hh_outflow}}
\end{figure*}

In Fig. \ref{fig:hh_outflow}, we plot spectra from the wavelength regime covered by the NIRCam F335M band 
for both the H\,{\sc{ii}} region and the DF1 region.  Spectra for both 
regions exhibit broad emission features, commonly identified with carbonaceous material \citep[][PAH]{2008ARA&A..46..289T}.
In Fig. \ref{fig:hh_outflow}, the DF1 region spectrum 
has been scaled by a factor of 0.28 so that the strength of the emission feature centered at 3.3~\micron\ 
matches between regions. There is no continuum emission (e.g., from very small grains) in the F335M bandpass in either region
so all of the emission observed in F335M in the H\,{\sc{ii}} region originates from the 3.3~\micron\ complex and 
two atomic lines, $\lambda$3.2294 [Fe\,{\sc{iii}}] and H\,{\sc{i}} Pfund-$\delta$. The atomic lines were fit and removed
(Sect. \ref{sec:lines}) before further analysis of the outflow.
After line removal, the integrated strength of the 3.3~\micron\ feature in the H\,{\sc{ii}} region
implies a specific intensity of 0.18~MJy sr$^{-1}$ in the F335M filter; this contrasts with a median 
specific intensity of 0.08~MJy sr$^{-1}$ reported by \citet{2024A&A...687A...4A} in the F335M outflow.  However, 
the outflow specific intensity from \citet{2024A&A...687A...4A} is the average over a much larger region while the 
NIRSpec IFU coverage in the H\,{\sc{ii}} region is restricted to a region very near the PDR front with higher intensity. 
The median specific intensity in the F335M imaging in the same region covered by the IFU mosaic is $\sim$0.15~MJy sr$^{-1}$.
With a line contamination in the F335M imaging filter in the H\,{\sc{ii}} region 
of order 6\% (see Sect. \ref{ssec:lines_contribution}), the "pure" 3.3~\micron\ aromatic emission in the F335M filter
is $\sim$0.14~MJy sr$^{-1}$, in good agreement with the IFU measurement.
The confirmation of emission from a population of small carbonaceous material from 
the spectroscopic observations supports the mass estimate for the outflow made in \citet{2024A&A...687A...4A}. The survival
of small grains at the PDR edge to supply the outflow is consistent with the conclusion in \citet{elyajouri2024}
that small grain destruction  in the Horsehead may be less efficient 
than in higher UV field PDRs. 

\begin{figure}[hbt!]
\centering
\includegraphics[width=0.45\textwidth]{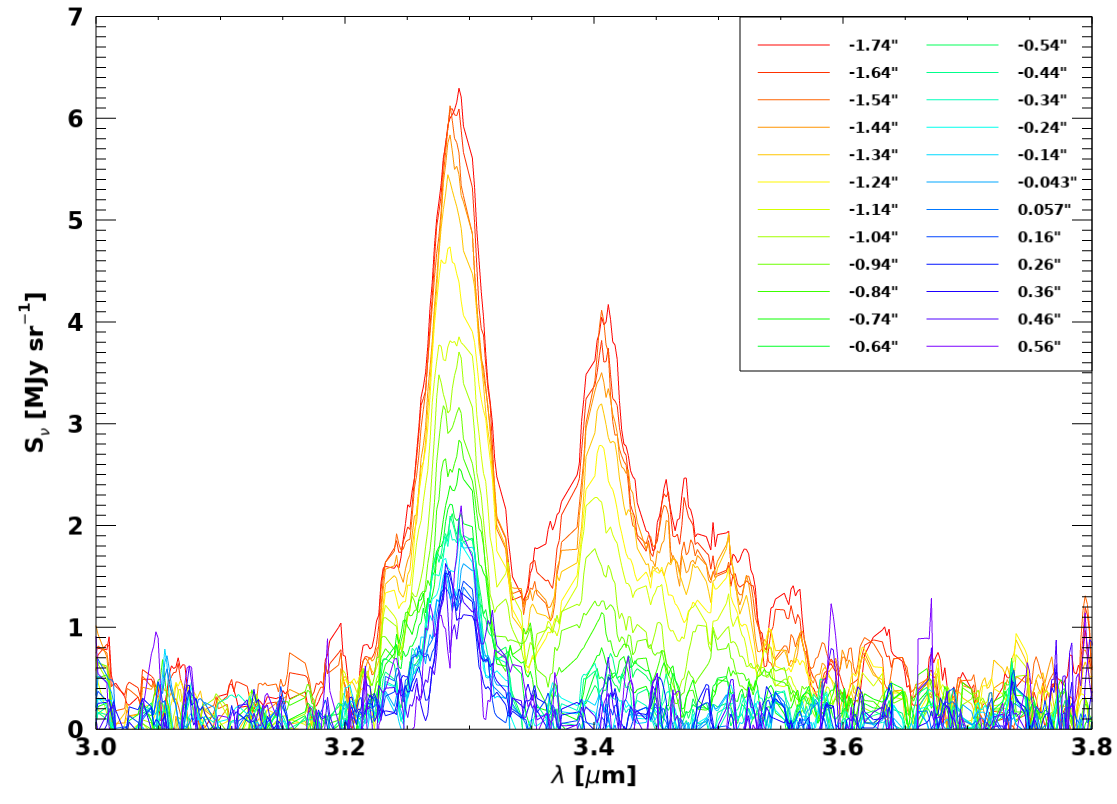}
\caption{Spectra from individual `spaxels' across the PDR front. Colors correspond to distance along the long
axis of the IFU mosaic relative to the dissociation front defined in \citet{2024A&A...687A...4A}, with red being
deepest into the cloud and purple at the edge of the IFU footprint nearest the exciting star. 
\label{fig:hh_outflow_spaxel}}
\end{figure}

\begin{figure*}[hbt!]
\centering
\begin{tabular}[b]{cc}
 \includegraphics[width=0.45\textwidth]{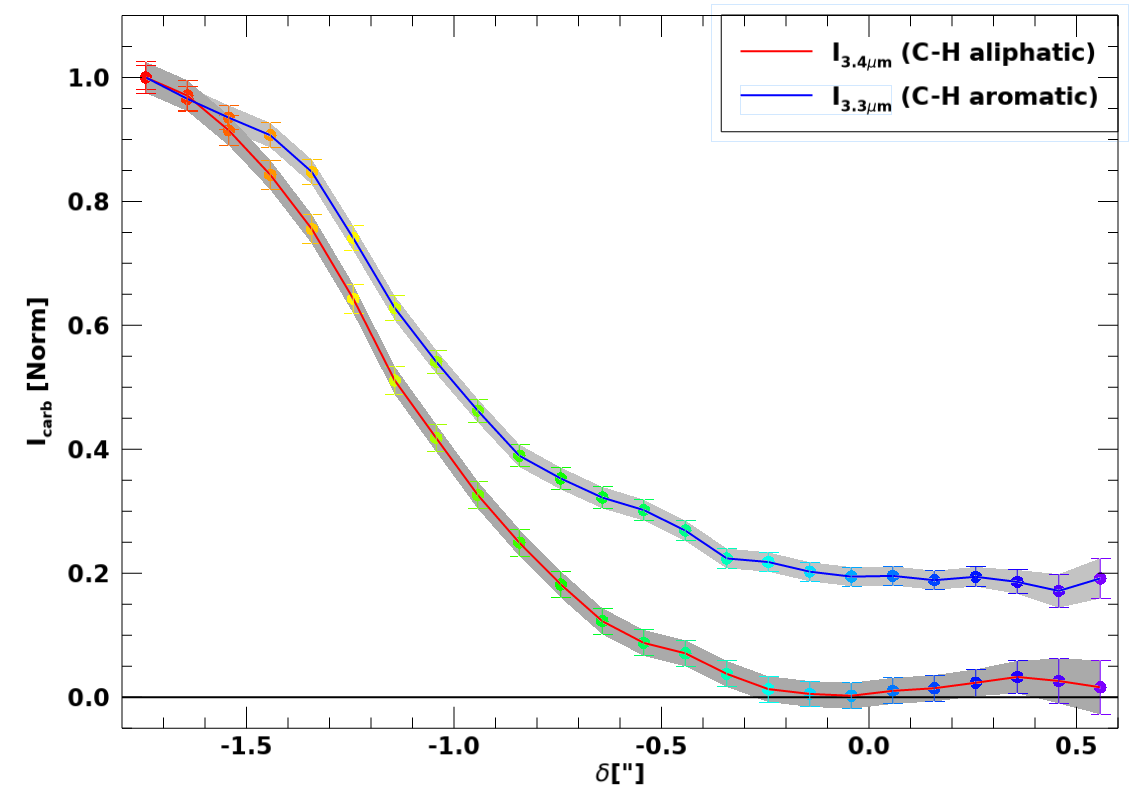}  & \includegraphics[width=0.45\textwidth]{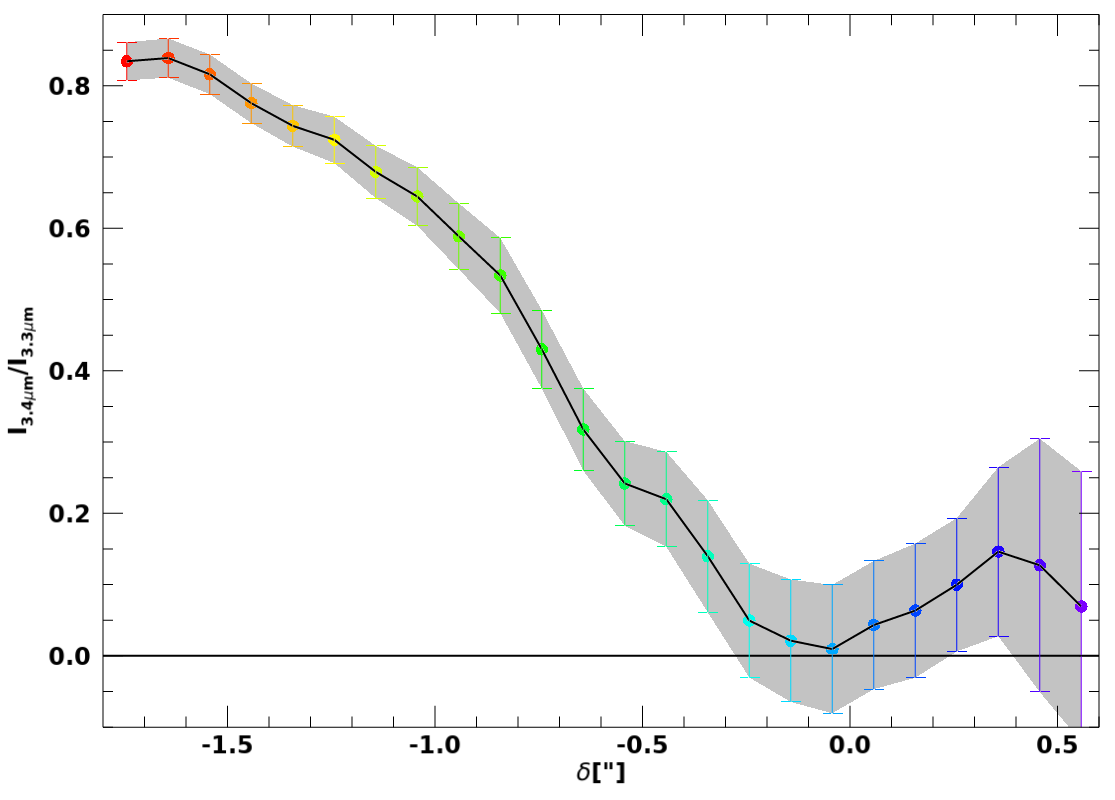} \\ 
\end{tabular}
\caption{Left: Feature strength in the 3.4~\micron\ aliphatic feature (red) and 3.3~\micron\ aromatic feature (blue)
extracted from per spaxel spectra shown in Fig. \ref{fig:hh_outflow_spaxel}. Right: ratio of I$_{3.4\tmicron}$ 
to I$_{3.3\tmicron}$. In both panels, the color coding of the individual points follows the colors defined in 
Fig. \ref{fig:hh_outflow_spaxel}. \label{fig:hh_ali_aro_spaxel_comparison}
}
\end{figure*}

In the case of the MIRI F770W imaging filter, the situation is not as clear.  The F770W filter bandpass spans the MIRI channel~1
and channel~2 IFU.  Unfortunately, the H\,{\sc{ii}} region IFU extraction is of low signal to noise and any definitive spectroscopic 
identification of the emission source in the F770W band is challenging - see the right panel of Fig. \ref{fig:hh_outflow}. 
A strong {[Ar\,{\sc{ii}}]} line at 6.98527~\micron\ and Pfund-$\alpha$ at 7.46040~\micron\ 
contribute 100\%\ of the emission in the MIRI F770W bandpass in the region covered by the H\,{\sc{ii}} IFU extraction
(Sect. \ref{ssec:lines_contribution}) with no detected emission from the 7.7~\micron\ C-C aromatic stretch. In addition, 
we detect no emission from the 11.3~\micron\ C-C out-of-plane bending modes of aromatic carbonaceous materials.
In the same fashion as for the 3.3-3.4~\micron\ complex, we can estimate the detectability of emission in either 
feature in the H\,{\sc{ii}} region by scaling the emission from the DF1 region. As we have no reference emission feature 
detected in the H\,{\sc{ii}} region from carbonaceous material in this wavelength regime, we scale the DF1 region by the
same factor as used for the 3.3~\micron\ feature above, 0.28. The results (right-hand panel of 
Fig. \ref{fig:hh_outflow}) indicate that, if the scaling of the DF1 spectrum is reasonable, any emission present 
may have been marginally detected. The absence of such a detection in our data may reflect a size
segregation, with only the smallest carbonaceous carriers being entrained in the flow. However, given the low SN in the 
H\,{\sc{ii}} extraction region, we draw no definitive conclusions regarding the presence or absence of these modes in the outflow
material. That said, the non-detection of any carbonaceous signatures in the MIRI F770W bandpass in the 
H\,{\sc{ii}} region indicate that the emission detected in \citet{2024A&A...687A...4A} is dominated by
[Ar\,{\sc{ii}}] and Pfund-$\alpha$ atomic line emission and does not represent the detection of the flow at longer 
wavelengths.

\subsubsection{Aliphatic Stability\label{sssec:alistability}}
It is noteworthy that, in the evaporative flow, only 3.3~\micron\ emission is observed, whereas the DF1 region
shows another component at 3.4~\micron\ with a shoulder extending to $\sim$3.6~\micron\ (Fig. \ref{fig:hh_outflow}).
As the 3.3~\micron\ and 3.4~\micron\ features are attributed to C-H stretch modes in aromatic (PAH) and aliphatic 
bonds, respectively \citep{2008ARA&A..46..289T}, the absence of 3.4~\micron\ emission in the H\,{\sc{ii}} region implies
that the outflow is purely aromatic. In Fig. \ref{fig:hh_outflow}, it is clear that, were the carriers
of the aliphatic 3.4~\micron\ feature and plateau present in the outflow in a similar proportion to 
the 3.3~\micron\ aromatic carriers as in the DF1 region, they would be easily observed and that therefore their 
absence reflects a true physical difference at the boundary between the molecular cloud and the outflow.

While the comparison between the DF1 and H\,{\sc{ii}} full region extractions demonstrates the depletion of the aliphatic bonds 
at the PDR edge, there is sufficient SN in the 3.3 and 3.4~\micron\ bands to extract per spaxel spectra across the front
at the resolution of a single NIRSpec spatial pixel ($\sim0\farcs1\simeq50$~AU).  Per spaxel spectra are shown in 
Fig. \ref{fig:hh_outflow_spaxel} and are color coded from deepest in the PDR ($\sim-1\farcs7$) to the edge of the NIRSpec 
IFU footprint in the H\,{\sc{ii}} region ($\sim+0\farcs5$). Here, we have defined positive towards the exciting star, and 0 is taken 
to be the position of the front from \citet{2024A&A...687A...4A}.  The integrated strengths of the 3.3~\micron\ aromatic 
and 3.4~\micron\ aliphatic emission and their ratio are shown in Fig. \ref{fig:hh_ali_aro_spaxel_comparison}. 
From $-1\farcs7$ to $\sim -1\farcs6$, both aromatic and aliphatic emission 
strengths decrease but maintain an approximately constant ratio.  From $-1\farcs6$ to $-0\farcs5$, both continue to decrease,
but the 3.4~\micron\ aliphatic feature weakens more quickly, especially starting at $\sim-0\farcs75$, where the ratio 
steepens (right panel of Fig. \ref{fig:hh_ali_aro_spaxel_comparison}).
 The 3.3~\micron\ aromatic feature stops decreasing in strength
at $\sim -0\farcs3$ as we transition out of the PDR and into the H\,{\sc{ii}} region, and the 
3.4~\micron\ aliphatic feature reaches 0 strength around $\delta \sim -0\farcs25$, reflecting a small spatial offset 
in the emission. 

As the edge of the PDR defined by the DF1 region should be the source of the material in the outflow, the absence of the 
aliphatic carrier in the outflow demonstrates unambiguously the rapid destruction of those carriers in the transition
from the shielded PDR environment to the UV exposed, unshielded H\,{\sc{ii}} environment. Indeed, the fact that the aliphatic
emission begins decreasing more quickly than the aromatic emission and reaches 0 inside the front (see Figs.
\ref{fig:hh_ali_aro_spaxel_comparison}, \ref{fig:horsehead_ali_to_aro_H}) indicates exposure to even moderate levels
of UV radiation in the outer, less shielded, regions of the PDR can impact the stability of the aliphatic bonds in the
carbonaceous species. This behavior is consistent with theoretical expectations and lab results
\citep{2013ApJS..205....8S,2021A&A...652A..42M} that demonstrate that aromatic structures are 
expected to be stable relative to aliphatic structures in the face of UV radiation. While this expectation has been noted
observationally as well \citep[and references therein]{2023ApJS..268...50Y,2023ApJS..268...12Y,2025ApJ...986..156L}, here we
resolve the transition in situ across a narrow PDR front at a resolution of $\sim$100~au.

\begin{figure}[!ht]
\centering
\includegraphics[width=0.45\textwidth]{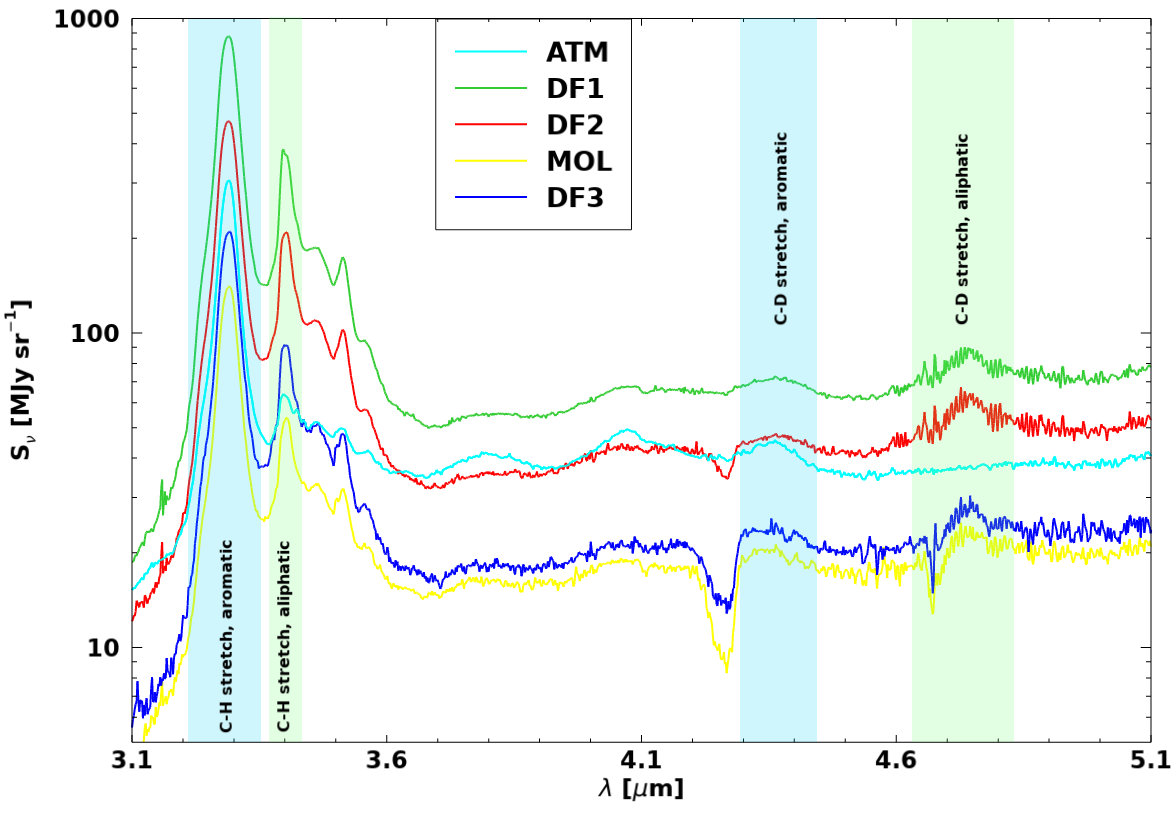}
\caption{Spectra for all region extractions in NGC7023 from 3.1-5.1~\micron. Colors are as for the region 
definitions in Fig. \ref{fig:img_regions}. Locations of C-H/C-D stretches in aromatic and aliphatic modes are
indicated by light blue and green boxes, respectively. All spectra have strong emission lines subtracted.
\label{fig:ch_vs_cd_7023}}
\end{figure}

\begin{figure}[!ht]
\centering
\includegraphics[width=0.45\textwidth]{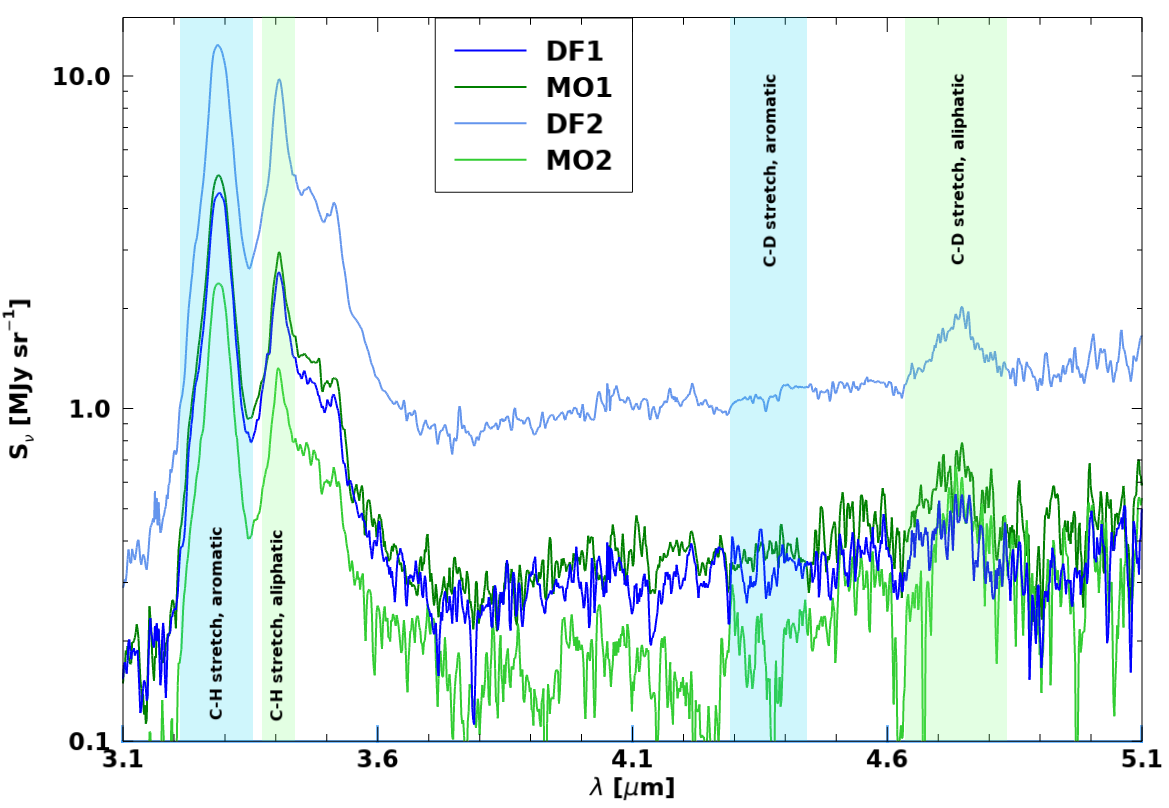}
\caption{As in Fig. \ref{fig:ch_vs_cd_7023} for the Horsehead. \label{fig:ch_vs_cd_hh}}
\end{figure}

While the reduction of the aliphatic bonds relative to the aromatic bonds as the UV field increases is
unambiguously demonstrated at the HII/DF1 interface in the Horsehead, similar behavior is less directly
observed in all region extractions for both NGC 7023 and the Horsehead. In Figs. \ref{fig:ch_vs_cd_7023}
and \ref{fig:ch_vs_cd_hh}, we plot spectra between 3.1 and 5.1~\micron\ for all extraction
regions in NGC~7023 and the Horsehead, respectively.  This spectral region contains the aforementioned C$-$H stretching
modes at 3.3~\micron\ and 3.4~\micron\, along with a plateau 
on the long wavelength shoulder of the 3.4~\micron\ band and features at 4.4~\micron\ and 4.7~\micron. 
The latter features have been attributed to the `deuterated' counterparts to the C$-$H stretch modes in aromatic and 
aliphatic structures, respectively, where the H atoms in the carrier have been substituted with deuterium 
and will be discussed in more detail in Sect. \ref{sssec:dsubstitution}.

In NGC~7023, the 3.4~\micron\ aliphatic band is significantly weakened with respect to the 3.3~\micron\ 
aromatic band in the transition from DF1 to ATM.
With reference to Fig. \ref{fig:img_regions}, the ATM region in NGC~7023 is closest to HD~200775 and in the 
lower-density bubble created by energetic photons. The reduced strength of the 3.4~\micron\ band is consistent with the
preferential destruction of aliphatic bonds in less shielded environments.  
More quantitatively, we can estimate the strengths of both the 3.3 and 3.4~\micron\ features by decomposing the region 
spectra between 3 and 5~\micron\ into a polynomial continuum and eight Drude profiles with fixed central wavelengths corresponding
to the aromatic and aliphatic features of interest. 
We can estimate the fraction of aliphatic relative to aromatic C$-$H bonds in the carriers using the observed
intensity ratios and intrinsic band strengths from \citet{2023ApJS..268...12Y} where they ascribe the later feature to 
aliphatic chains on PAH molecules (hydrogenation).
Both the feature strengths for the aromatic and aliphatic components, along with the estimated aliphatic fraction, are
reported in Table \ref{tbl:regfeaturestr}. In general, the aliphatic fraction is roughly constant at $\sim$0.15-0.25 
in both objects.  The exceptions are the least shielded regions in both - ATM in NGC~7023 ($\sim$0.06) and, 
as illustrated above,
HII in the Horsehead - see Figs. \ref{fig:hh_outflow}, \ref{fig:hh_outflow_spaxel}, \ref{fig:hh_ali_aro_spaxel_comparison}).

\begin{figure}[h]
\centering
\includegraphics[width=0.5\textwidth]{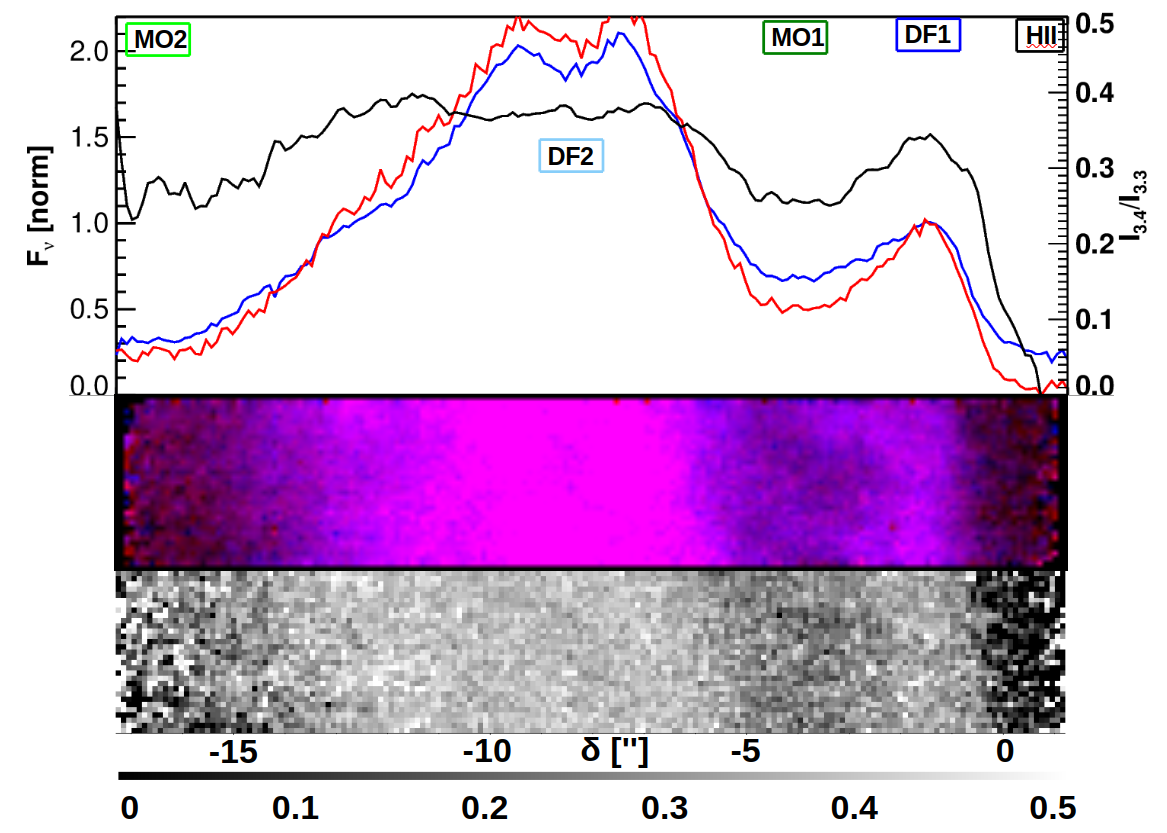}
\caption{Horsehead per pixel maps of the aliphatic (3.4~\micron) and aromatic (3.3~\micron) integrated band strength. Middle panel: 
aromatic in blue, aliphatic in red.  Bottom panel: ratio of aliphatic to aromatic band strengths. Top panel: Cuts along the
long mosaic axis. Aromatic and aliphatic cuts are in blue and red, respectively (as for the middle panel), and the band ratio 
is plotted in black - see the y-axis on the right.
\label{fig:horsehead_ali_to_aro_H}}
\end{figure}

\begin{figure}[h]
\centering
\includegraphics[width=0.5\textwidth]{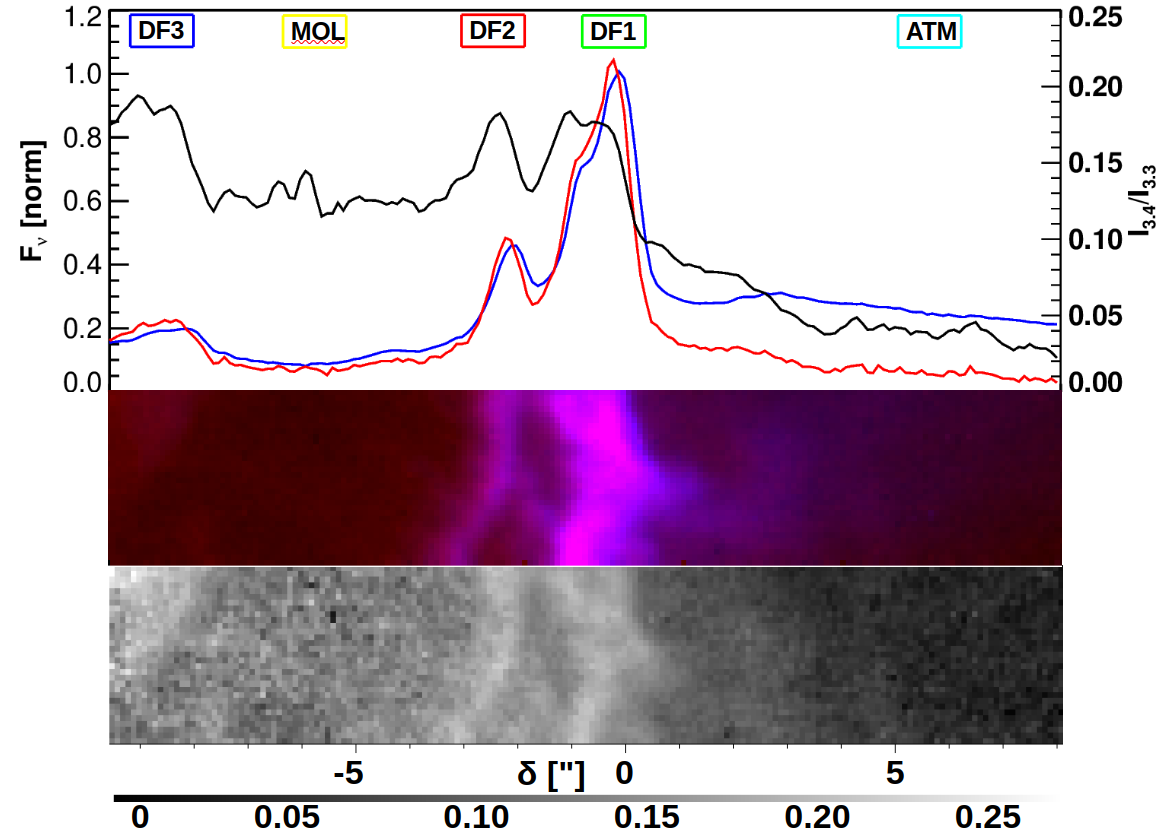}
\caption{Per pixel maps of the aliphatic (3.4~\micron) and aromatic (3.3~\micron) integrated band strength. Middle panel: 
aromatic in blue, aliphatic in red.  Bottom panel: ratio of aliphatic to aromatic band strengths. Top panel: Cuts along the
long mosaic axis. Aromatic and aliphatic cuts are in blue and red, respectively (as for the middle panel), and the band ratio 
is plotted in black - see y-axis on the right.  
\label{fig:ngc7023_ali_to_aro_H}}
\end{figure}

Given the strength of the 3.3 and 3.4~\micron\ features, in addition to restricting ourselves to the defined regions, 
we can generate per-pixel maps of the feature strengths to map the distribution more carefully.
We generate maps following a similar procedure as described above for the regions but for each spaxel in the three-dimensional
data cube. However, owing to the lower signal-to-noise compared to the region extractions, we do not perform a full decomposition
for each spaxel but rather fit a local continuum around each feature and integrate across the feature profile.
A more detailed decomposition of the per spaxel spectra will be presented in upcoming
papers (Van De Putte et al. in prep).

The per-pixel maps are shown in Figs.~\ref{fig:horsehead_ali_to_aro_H} and \ref{fig:ngc7023_ali_to_aro_H} for 
the Horsehead and NGC~7023, respectively; we leave the spaxel comparison in terms of observed intensities rather than 
converting to aliphatic fraction, given the higher uncertainty in the per spaxel ratios relative to the region 
extractions. In those figures, we show 2-color 
images of the aromatic and aliphatic feature strength, along with their ratio, as well as cuts along the long axis 
of the mosaic. The per-pixel maps largely 
confirm the results from the integrated regions. In the less shielded regions (ATM and HII for NGC~7023 and 
the Horsehead, respectively), there is significantly less aliphatic emission - in the case of the Horsehead, 
no aliphatic emission (Fig. \ref{fig:hh_outflow}) and at the PDR front, both aliphatic and aromatic 
emission rise rapidly to reach a relatively constant ratio through the rest of the cloud - 0.3-0.4 for the Horsehead
and 0.15-0.2 for NGC~7023, despite significant differences in the total emission.

While the I$_{3.4}$/I$_{3.3}$ ratio is relatively constant behind the main fronts in both objects, there is a
trend of lower ratios in the presumably more shielded regions - MO1 and MO2 in the Horsehead and MOL in NGC~7023 - 
compared to the brighter DF regions. This is seemingly in contradiction to the observation that the aliphatic bonds
are more easily destroyed in the less shielded regions as discussed above (NGC~7023-ATM and Horsehead-HII).  If the 
aliphatic bonds are more fragile, one might expect a higher ratio in shielded regions. However, the emission in the 
features depends not only on the relative stability of the bonds but also the excitation of the mode and the relative 
number of bonds. Aromatic compounds tend to have absorption bands extending to longer UV wavelengths relative to the 
UV absorption bands in aliphatic materials, potentially leading to less excitation of aliphatic modes in the UV shielded 
regions.  However, this would require discrete aliphatic and aromatic carriers, as in any mixed material, the absorbed 
energy will be distributed roughly in proportion to the number of modes without respect to the aromaticity of the mode.
In contrast, if the composition of the material changes in the denser environments through, e.g. 'grain' growth, in such
a way so as to increase the relative proportion of aromatic to aliphatic bonds, the same trend would be observed.

In the Horsehead, while there is no appreciable aliphatic component in the H\,{\sc{ii}} region, at the ionization front,
both aliphatic
and aromatic emissions rise rapidly with, as noted above, a small spatial offset between the components (Figs.
\ref{fig:hh_outflow_spaxel}, \ref{fig:horsehead_ali_to_aro_H}).
In NGC~7023, at the ATM-DF1 boundary, the aliphatic emission is significantly offset spatially from 
the aromatic emission, peaking $\sim$0.15" inside the aromatic emission ($\sim$ 2.8$\times$10$^{-4}$~pc).  
The aliphatic emission peaks are also narrower, not extending as deeply into the PDR as the aromatic emission 
peaks. For the Horsehead, if the lack of aliphatic emission in the outflow is indeed indicative of preferential
destruction of the C-H bonds in the aliphatic members in a harsh UV environment, their rapid appearance at 
the front is consistent with a very narrow ionization front and rapid density increase
\citep{2024A&A...687A...4A,2023A&A...677A.152H} providing shielding over a small spatial scale.
For NGC~7023, with its more intense UV field at the PDR front, the larger spatial offset in the aromatic and
aliphatic components might suggest efficient destruction of the C-H bonds in the aliphatic members, deeper  
into the transition region.  

We also note the behavior of the "plateau" on the red side of the 3.4~\micron\ aliphatic feature. The origin of the 
plateau has eluded a definitive identification and has been ascribed to a variety of materials including
amorphous hydrogenated carbon, vibrational modes from aliphatic side members, and/or anharmonicity of PAH
\citep[and references therein]{2022A&A...657A.128J,2023ApJ...959...74B,2024A&A...685A..75C}. Comparing the strength
of the plateau relative to the 3.3~\micron\ aromatic and 3.4~\micron\ aliphatic features in NGC~7023, \citet{2015A&A...577A..16P}
found that the plateau was strongly correlated with the strength of the former and only very weakly with the latter,
suggesting that the plateau emission was dominated by an aromatic material.  In contrast, here we find 
that the plateau is more strongly correlated with the 3.4~\micron\ aliphatic emission.  In the Horsehead, in the transition 
from DF1 to the HII outflow, the plateau disappears completely along with the 3.4~\micron\ aliphatic emission
(Figs. \ref{fig:hh_outflow}, \ref{fig:hh_outflow_spaxel}), suggesting a common origin for the carriers. 
In NGC~7023, while our spectral mosaic is offset slightly from the P1-4 region definitions in
\citet{2015A&A...577A..16P}, our ATM regions sample a similar environment as P1.  In our ATM region, both the plateau 
and the 3.4~\micron\ aliphatic feature are weaker relative to the 3.3~\micron\ aromatic feature (Fig. \ref{fig:ch_vs_cd_7023})
compared to the other regions. \citet{2015A&A...577A..16P} perform a per spaxel decomposition of their data to derive
the correlation, whereas the behavior of the plateau relative to the aromatic and aliphatic bands presented from our data is 
qualitative and confined to our region extractions. Per spaxel and feature decompositions of the plateau and features 
in NGC~7023 will be analyzed in a forthcoming paper to further explore the nature of the plateau.

\subsubsection{Deuteration of Aliphatic and Aromatic Bonds\label{sssec:dsubstitution}}
In Figs. \ref{fig:ch_vs_cd_7023} and \ref{fig:ch_vs_cd_hh}, in addition to the 
aforementioned aromatic and aliphatic
C$-$H stretch features at 3.3 and 3.4~\micron, there are weaker features detected at 4.4 and 4.7~\micron.  These
are generally attributed to the `deuterated' counterparts to the C$-$H stretch modes at 3.3 and 3.4~\micron, respectively,
where deuterium has been substituted 
\citep[deuterium fractionation,][]{2005A&G....46b..29M}
for one or more hydrogen atoms in the bonds
\citep{2004ApJ...614..770H,2023ApJS..268...12Y}. Note that while we refer to the deuterated aromatic band as 
the "4.4~\micron" feature, in our data the band is centered closer to 4.38~\micron\ with a width of $\sim$0.1~\micron.
While both deuterated aliphatic and aromatic features 
have been detected observationally \citep[e.g.,][]{2014ApJ...780..114O,2016A&A...586A..65D,2024A&A...685A..74P}, 
they are generally tentative and of low signal-to-noise, especially in the 4.4~\micron\ aromatic band.
Both deuterated aliphatic and aromatic features are seen in most of our region spectra. While per spaxel extractions
will be done in a future paper, here, we extract feature strengths based on our defined regions.
As described above, we decompose the region spectra and extract integrated strengths of the 
4.4 and 4.7~\micron\ features where possible.  In the case of the H\,{\sc{ii}} region in the Horsehead, no features beyond the 
3.3~\micron\ feature are seen.  In addition, there is no measurable feature in the deuterated aromatic feature at
4.4~\micron\ in any region of the Horsehead and the deuterated aliphatic feature at 4.7~\micron, while well 
detected ($\geq3\sigma$) in the DF2, MO1, and MO2 regions, is quite weak in DF1 ($\sim2\sigma$).
In NGC~7023, both deuterated aliphatic and aromatic features are well detected ($\geq 3\sigma$) in all regions 
except for the aliphatic feature at 4.7~\micron\ in the ATM region, which is absent. 
All measured feature intensities are given in Table \ref{tbl:regfeaturestr}. 
In addition, following the prescription of 
\citet{2020ApJS..251...12Y} and \citet{2023ApJS..268...12Y}, who generate intrinsic band strengths assuming an
underlying PAH structure with associated alaphatic chains/side members, we compute the ratio of the number
of deuterated to hydrogen carbon bonds and report those ratios in the table. Note that we are computing the 
ratio of deuterated bonds on the corresponding `parent' - e.g., the fraction of aliphatic deuteration relative
to the number of aliphatic C-H bonds on the parent material, and similarly for the
deuterated aromatic bonds. 

\begin{table*}[bth]
\begin{threeparttable}[b]
 \caption{Aromatic and Aliphatic Band Strengths}\label{tab:data}
 \begin{tabular*}{\textwidth}{@{\extracolsep{\fill}}lccccccc}
 \toprule
                 & \multicolumn{3}{c}{\bf Aromatic}                       & \multicolumn{3}{c}{\bf Aliphatic} \\
                 \cmidrule{2-4}                                             \cmidrule{5-7}
 Region\tnote{a} & I(3.3~\micron)\tnote{b}  & I(4.4~\micron)\tnote{b} & N$_{D}$/N$_{H}$\tnote{c} & I(3.4~\micron)\tnote{b} & I(4.7~\micron)\tnote{b} & N$_{D}$/N$_{H}$\tnote{c} & N$_{H,aliph}$/N$_{H,arom}$ \\ 
 \midrule
           & \multicolumn{6}{c}{\bf NGC 7023} \\
           \cmidrule{2-8}
    ATM    &  9.72$\pm$0.11  & 0.36$\pm$0.04 & 0.047$\pm$0.025 & 1.14$\pm$0.11   & $\cdots$        & $\cdots$        & 0.062$\pm$0.012 \\
    DF1    & 27.66$\pm$0.64  & 0.48$\pm$0.13 & 0.022$\pm$0.013 & 8.65$\pm$0.57   & 1.01$\pm$0.11   & 0.169$\pm$0.087 & 0.166$\pm$0.030 \\
    DF2    & 15.79$\pm$0.31  & 0.19$\pm$0.06 & 0.015$\pm$0.009 & 4.97$\pm$0.37   & 0.92$\pm$0.06   & 0.269$\pm$0.137 & 0.167$\pm$0.031 \\
    MOL    &  5.02$\pm$0.05  & 0.11$\pm$0.02 & 0.028$\pm$0.015 & 1.18$\pm$0.07   & 0.38$\pm$0.02   & 0.305$\pm$0.154 & 0.125$\pm$0.024 \\
    DF3    &  7.45$\pm$0.08  & 0.16$\pm$0.03 & 0.027$\pm$0.015 & 3.26$\pm$0.13   & 0.38$\pm$0.03   & 0.169$\pm$0.086 & 0.233$\pm$0.041 \\
           & \multicolumn{6}{c}{\bf Horsehead} \\
           \cmidrule{2-8}
    H\,{\sc{ii}}    & 0.045$\pm$0.002 & $\cdots$       & $\cdots$     & $\cdots$        & $\cdots$              & $\cdots$              & $\cdots$     \\
    DF1    & 0.169$\pm$0.004 & $\cdots$       & $\cdots$     & 0.063$\pm$0.005 & {\it 0.002$\pm$0.001} & {\it 0.046$\pm$0.033} & 0.198$\pm$0.037 \\
    MO1    & 0.191$\pm$0.002 & $\cdots$       & $\cdots$     & 0.071$\pm$0.004 & 0.006$\pm$0.001       & 0.123$\pm$0.065       & 0.197$\pm$0.035 \\
    DF2    & 0.470$\pm$0.008 & $\cdots$       & $\cdots$     & 0.243$\pm$0.012 & 0.018$\pm$0.002       & 0.108$\pm$0.055       & 0.274$\pm$0.049 \\
    MO2    & 0.095$\pm$0.002 & $\cdots$       & $\cdots$     & 0.028$\pm$0.004 & 0.007$\pm$0.001       & 0.363$\pm$0.196       & 0.156$\pm$0.035 \\
    \hline
 \end{tabular*}

 \label{tbl:regfeaturestr}
 \begin{tablenotes}
   \item [a] See Fig. \ref{fig:img_regions} for physical location of region.
   \item [b] Observed energy flux in 10$^{-14}$ erg s$^{-1}$ cm$^{-2}$ arcsec$^{-2}$.
 \end{tablenotes}

\end{threeparttable}
\end{table*}

The interpretation of the deuterated features is perhaps less straightforward relative to the aromatic and aliphatic
ratios for 'pure' C$-$H bonds as they depend not only on whether aromatic and/or aliphatic bonds exist in the carriers
and the excitation mechanism for the bond,
but also on the ease with which deuterium substitution occurs in a given environment.  Nevertheless, a somewhat consistent
picture arises here. Where aliphatic bonds exist, deuterium substitution is quite efficient across a variety of
environments. With reference to Table \ref{tbl:regfeaturestr}, where a 3.4~\micron\ aliphatic feature is observed, a 
corresponding deuterated feature is observed with an estimated deuterium fraction of $\sim$0.1 to nearly 0.4.
There are two exceptions, the Horsehead DF1 region and the NGC~7023 ATM region.  For the Horsehead DF1, the detection and measurement
of the 4.7~\micron\ deuterated member is poor and has low signal-to-noise.  In the case of the NGC~7023 ATM region,
there is no detection; however, we note that the aliphatic C$-$H signature is quite weak in ATM relative to
the aromatic emission, resulting in an estimated ratio of only $\sim$0.06 aliphatic bonds relative to aromatic. In
contrast, in the other regions in both objects, the ratio of aliphatic to aromatic bonds is 
$\sim$0.2 (column 8 of Table \ref{tbl:regfeaturestr}). 
Above, we have attributed the relative weakness of the aliphatic emission in the NGC~7023 ATM and Horsehead
HII regions as being due to exposure to harsher radiation 
fields in less shielded environments preferentially destroying (or inhibiting the formation of) aliphatic bonds.
In that context, the lack of deuterium substitution in the ATM region implies either a similar inhibition or a 
threshold of available aliphatic C$-$H bonds, below which deuterium substitution is inefficient.
In Table~\ref{tbl:regfeaturestr}, column 7, we can note that the deuterated fraction in the aliphatic bonds is correlated 
with the inferred region shielding, being highest in the MOL regions of both NGC~7023 and the Horsehead, 
implying that the former explanation - deuterium substitution is inhibited in higher UV fields - may be 
the primary driver of the strength of the deuterated aliphatic feature at $\sim$4.7~\micron. In contrast, there
is no such trend with the aromatic deuteration seen in NGC~7023; the highest deuterium fraction 
in the aromatic bonds occurs in the less shielded ATM region. 

Relative to the aliphatic component, deuterium substitution in the aromatic bonds appears to be less efficient.  This is seen 
in Table \ref{tbl:regfeaturestr}, column 4, where the aromatic deuterated fraction is only $\sim$0.03 compared to 
$\sim$0.2 for the aliphatic bonds. Indeed, in the Horsehead, we detect no deuterium substitution for the aromatic 
bonds at all - no 4.4~\micron\ feature is detected in any Horsehead regions. This may be partially due to an 
observational bias; if we assume an aromatic deuteration fraction of $\sim$0.03 as observed in NGC~7023, the expected
strength for the 4.4~\micron\ feature in the Horsehead regions would be near the threshold of detectability. 
In addition, the intrinsic band strength of the aliphatic C-D stretch is $\sim$2 times stronger than the 
aromatic C-D stretch in PAHs \citep[e.g.][]{2023ApJS..268...12Y,2020ApJS..251...12Y}.
However, given the observed strength of the 3.3~\micron\ feature in the Horsehead DF2 region, we would expect 
to detect the deuterated 4.4~\micron\ feature for a similar $\sim$0.03 deuterium fraction. 
Therefore, there may be a reduced aromatic deuteration efficiency in the Horsehead relative to NGC~7023. 
\citet{2025ApJ...983..136Y} have recently computed deuteration fractions
for the template regions defined by \citet{2024A&A...685A..74P} in Orion. When cast in terms of the metric 
defined there, our results agree well with the Orion results with $N_{{\text{D}},4.7}/N_{{\text{H}},3.3}$ 
of $\sim$0.03.

While there is a clear detection of deuterium uptake in the carbonaceous material, we find no evidence of 
deuterated \HH\ in the gas phase.  There are a handful of lines formally identified as HD lines
in our templates (see Sect. \ref{ssec:line_identity}) based largely on central wavelength matching.
In all cases, a manual check yields H$_2$ lines well within a spectral
resolution element of the measured line center, often part of a well-identified H$_2$ series in our spectra
(see Fig. \ref{fig:lineidentification_detail}). It is widely considered that the dominant H$_2$ formation pathway 
in clouds is formation catalyzed by grain surfaces \citep[e.g.][and references therein]{Wakelam17}. 
\citet{2006ASPC..348...58D} has shown that, for 
carbonaceous species with C-H bonds, it is energetically favorable for deuterium to replace H atoms 
in the C-H bonds, leading to high D to H ratios in such materials when interacting with deuterium
from the gas phase. Additionally, in such grains/molecules, for an H atom on the 
surface, it is energetically favorable to interact with another H atom, leaving the C-D bond intact, 
thus catalyzing the formation of \HH\ rather than HD. The absence of observed
HD gas lines coupled with the large deuteration fraction of the carbonaceous material bonds seen here is
consistent with the energetics of deuteration described in \citet{2006ASPC..348...58D}. 

\section{Summary \label{sec:summary}}
We have presented spectral imaging of the prototypical PDRs in the Horsehead nebula and the 
NW filament of NGC~7023 obtained with the NIRSpec and MIRI IFUs on JWST. Our data cover
a wavelength regime from $\sim$0.97-28~\micron\ with a spectral resolving power ranging
 $\sim$1000 ($0.97-5.3$~\micron, NIRSpec IFU medium gratings) and $3000-1000$ ($4.9-28$~\micron,
MIRI IFU). IFU data were obtained as a mosaic stripe with a size of $\sim3\arcsec\times18\arcsec$ across
the PDR fronts\footnote{The exact size of the mosaic depends on the specific instrument and grating; 
in this paper, we have restricted the analysis to spatial regions with complete coverage 
across the full wavelength regime} at an angular resolution ranging from $\sim0\farcs1-0\farcs35$
which corresponds to a physical resolution of $\sim$$2-7\times10^{-4}$~pc at the distance
of both systems ($\sim$400~pc). 

Spectra in five aperture-matched regions were extracted in both PDRs, corresponding roughly 
to the dissociation fronts where dense material is exposed to the UV radiation of the exciting
star, more deeply embedded molecular-like regions, as well as atomic and H\,{\sc{ii}} regions. 
These region extractions reveal a plethora of spectral lines both atomic and molecular,
numbering 200-300 in each region, 
along with solid-state features of ice and carbonaceous materials. The molecular lines include 
a nearly complete series of pure rotational \HH\ in the lower energy levels of the 0, 1, and 2 
vibrational states
along with several ro-vibrational \HH\ series, as well as CO and CH$^+$.  While we have provided
integrated strengths for these lines from the aperture-matched extraction regions as an illustration of the 
power of these data in understanding the structure of the PDRs, many are strong enough to 
allow extraction at a much higher spatial resolution. Mapping the behavior of the excitation of the \HH\ 
lines across the front at the highest resolutions provides powerful constraints on the conditions
and structure at the PDR front and is the subject of ongoing work with these data (Zannese et al. in prep.).
Line lists and spectra for all regions are provided electronically through the CDS (Sect. \ref{sec:dataavail}).

We identified several features of H$_2$O, CO, and CO$_2$ ice in the more deeply embedded
regions of NGC~7023 (MOL, DF3, and DF2). All ice detections are consistent with cold ($\lesssim$50~K) amorphous materials.

The ubiquitous broad emission bands between 3 and 20~\micron\ are prominent in our 
extracted spectra and spectral cubes. A full decomposition of the bands and a study of their 
substructure will be presented in forthcoming papers. For this overview paper, we focused on 
the spectral region between 3-4~\micron\ as an illustration of the unique diagnostic power of 
this dataset. Using integrated region spectra and per spaxel spectra as a function of distance
into the Horsehead PDR front, we confirmed the entrainment of carbonaceous dust/molecular clusters
in the photo-evaporative flow from the PDR surface.  Additionally, spectral information allowed us
to determine that the composition of the carbonaceous component of the photo-evaporative flow is 
entirely aromatic, consisting of emission only in the 3.3~\micron\ band. Combined with per spaxel 
and region extraction of the 3.3~\micron\ and 3.4~\micron\ band strengths, we confirm that the 
aromatic component of the carbonaceous material is more robust to exposure to strong UV fields 
and can survive deeper into unshielded environments relative to the aliphatic carrier. In the more
shielded regions inside of the PDR front, the aliphatic-to-aromatic ratio is 
$N_{\text{H,aliph}}/N_{\text{H,arom}} \sim 0.2$.

Deuterium substitution for H in one or more of the C-H bonds responsible for the 3.3 and 3.4~\micron\ 
bands has also been detected.  The substitution of heavier deuterium for the hydrogen atom in the bond
results in the band emission shifting to longer wavelengths. The aromatic band at 3.3~\micron\ shifts 
to $\sim$4.4~\micron\ while the aliphatic band at 3.4~\micron\ shifts to $\sim$4.7~\micron. Extracting 
the deuterated features in the regions where they are detected, we find that aliphatic deuterium
substitution is either more efficient or possibly more robust to destruction by UV radiation.  
Where detected, find we $N_{\text{D,aliph}}/N_{\text{H,aliph}} \sim 0.1-0.3$ compared to 
$N_{\text{D,arom}}/N_{\text{H,arom}} \sim 0.03$.

\section{Data Availability\label{sec:dataavail}}
Tables \ref{tbl:spectra_example} and \ref{tbl:linelist_sample} are only available in electronic form at the CDS via anonymous ftp to cdsarc.u-strasbg.fr 
or via \url{http://cdsweb.u-strasbg.fr/cgi-bin/qcat?J/A+A/}.

\begin{acknowledgements}
This work is based on observations made with the NASA/ESA/CSA James Webb Space Telescope. 
The data were obtained from the Mikulski Archive for Space Telescopes at the Space Telescope 
Science Institute, which is operated by the Association of Universities for Research in 
Astronomy, Inc., under NASA contract NAS 5-03127 for JWST. These observations are associated 
with GTO program \#01192.
K.M. is supported by JWST–NIRCam contract no. NAS5-02015 to the University of Arizona.
K.D.G, D. VDP, and A.N-C are partially supported by NASA grant 80NSSC21K1294.
MB acknowledges funding from the Belgian Science Policy Office (BELSPO) through the PRODEX 
project “JWST/MIRI Science exploitation” (C4000142239).
Part of this work was performed at the French MIRI center with the support of 
CNES and the ANR-labcom INCLASS between IAS and ACRI-ST, and also supported 
by the Programme National ``Physique et Chimie du Milieu Interstellaire'' (PCMI) of 
CNRS/INSU with INC/INP co-funded by CEA and CNES.
Data analyses were performed using IDL$^{\circledR}$. IDL$^{\circledR}$ is a registered trademark
of NV5 Global, Inc.\footnote{\url{https://www.nv5geospatialsoftware.com/Products/IDL}}
This work made use of Astropy:\footnote{\url{http://www.astropy.org}} a community-developed core
Python package and an
ecosystem of tools and resources for astronomy \citep{astropy:2013, astropy:2018, astropy:2022}.

\end{acknowledgements}

\bibliography{aa54851-25}{}
\bibliographystyle{aasjournal}

\begin{appendix} 
\section{Spectra\label{ap:spectralplots}}  

In addition to scaling the NIRSpec IFU flux to the imaging, see Sect. \ref{ssec:wcsup}, 
small shifts were made to individual MIRI IFU spectra where necessary to match 
across the full wavelength range.  No intra-channel shifts where necessary 
e.g., the short, medium, and long bands were in good agreement for each channel 
for both objects.
For NGC~7023, a multiplicative factor 1.02 was applied to the channel 1 spectrum 
to match the NIRSpec G395M grating on the short wavelength end and the 
MIRI channel 2 on the long wavelength end. For the Horsehead, small additive
inter-channel shifts were applied to the channel 3 and 4 spectra so they matched
in the overlap region as well as with MIRI channel 2.

Finally, all orders are merged by averaging in the overlap between
spectra with the exception of G395M to MIRI CH1 where the MIRI
CH1 was simply truncated. The resolution in the overlap region
was taken to be that of the longest wavelength grating, e.g. data
from the short wavelength overlap grating were interpolated onto
the long wavelength grating grid.

Plots of the full spectra for each object and region are provided here in 
three wavelength blocks for each object. Each figure panel shows a plot for 
each region of an object for a given wavelength regime. 
The figures were generated using the provided spectral extraction tables.
Electronically provided tables follow the naming convention OBJECT\_REGION\_spectrum.tbl, 
and a sample of the table format is given in Table \ref{tbl:spectra_example}.
The continuum columns, $C_{\nu}$, were generated from 

\begin{equation}
    C_{\nu} = F_{\nu} - \sum_{i=1}^{nline} G(\lambda_{i},\sigma_{i},A_{i})
\end{equation}

\noindent
where $F_{\nu}$ is the total extracted flux, including lines, for each region,
$G$ is the Gaussian function, and $\lambda_i,\sigma_i$ and $A_i$ are the fit parameters
reported in columns 1, 3, and 5 of Table \ref{tbl:linelist_sample}, respectively, and
the sum is over all lines detected in the region (see Table \ref{tab:linecounts}). 

\begin{figure*}
   \centering
   \begin{tabular}[b]{cc}
     \includegraphics[width=\columnwidth]{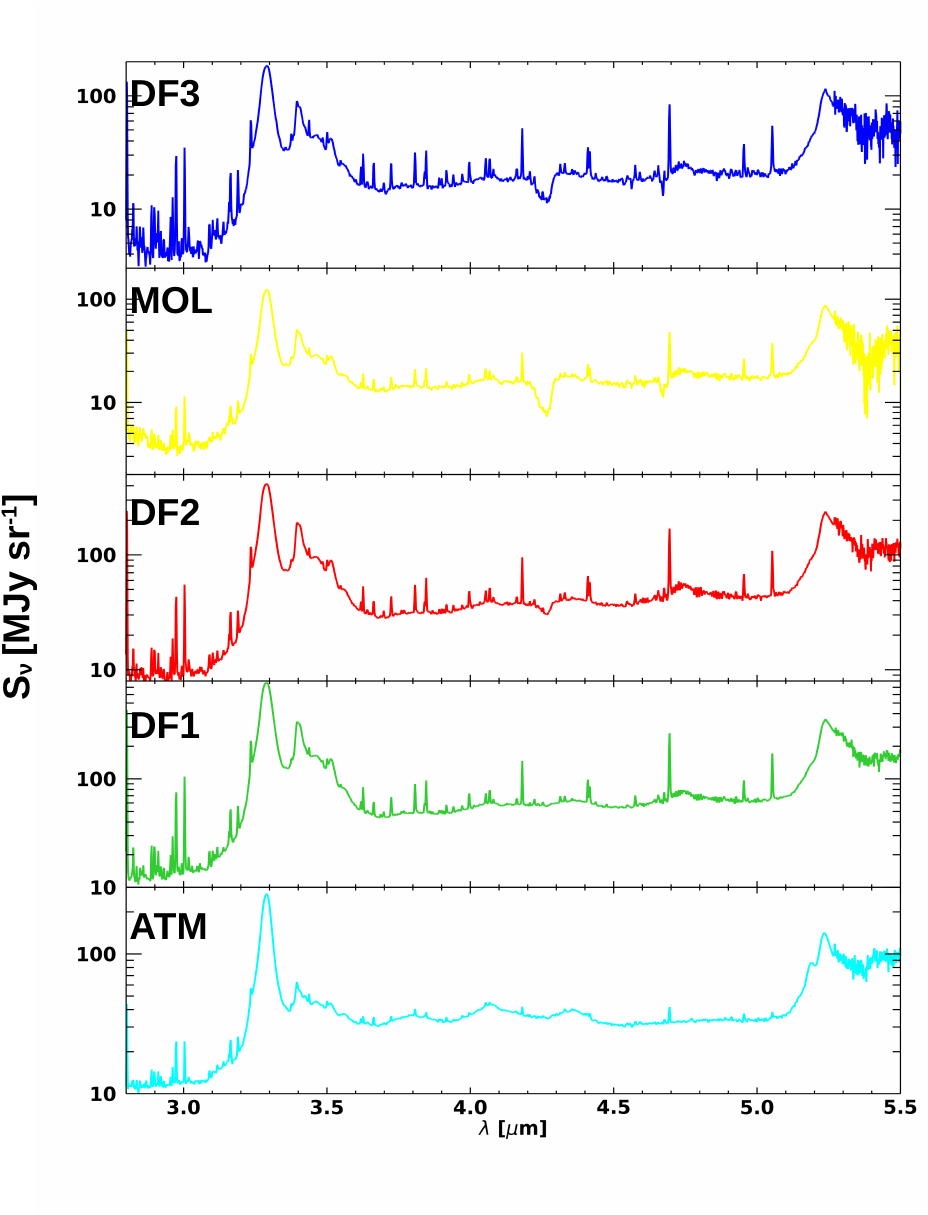} & \includegraphics[width=\columnwidth]{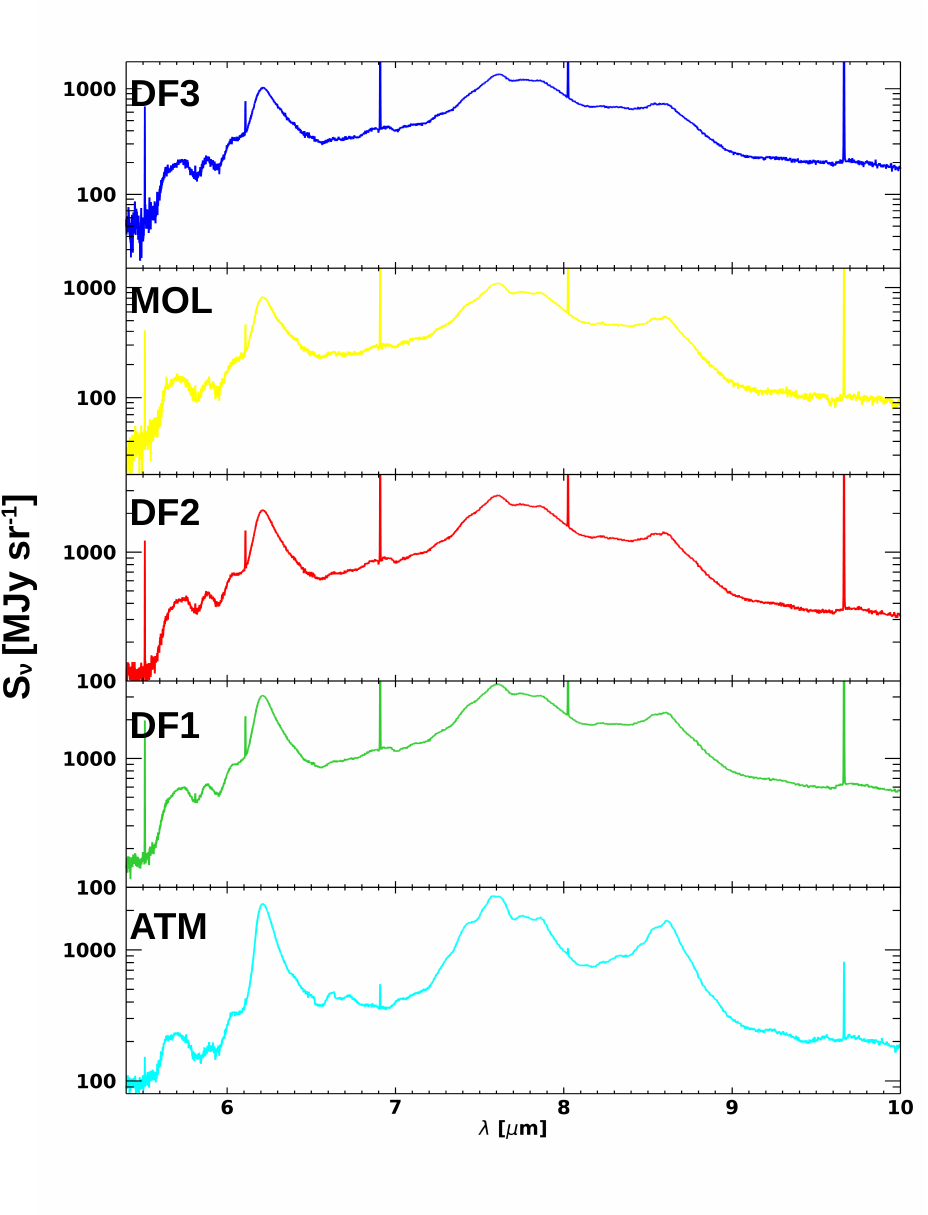} \\
   \end{tabular}
   \caption{As in Fig. \ref{fig:ngc7023_1p0_3p0_spec} for wavelengths between 
   $2.8 < \lambda < 5.5$~\micron\ (left) and between $5.4 < \lambda < 10.0$~\micron\ (right) 
   for regions in NGC~7023. 
   \label{fig:ngc7023_2p8_5p5_spec}}
\end{figure*}

\begin{figure*}
   \centering
   \includegraphics[width=\columnwidth]{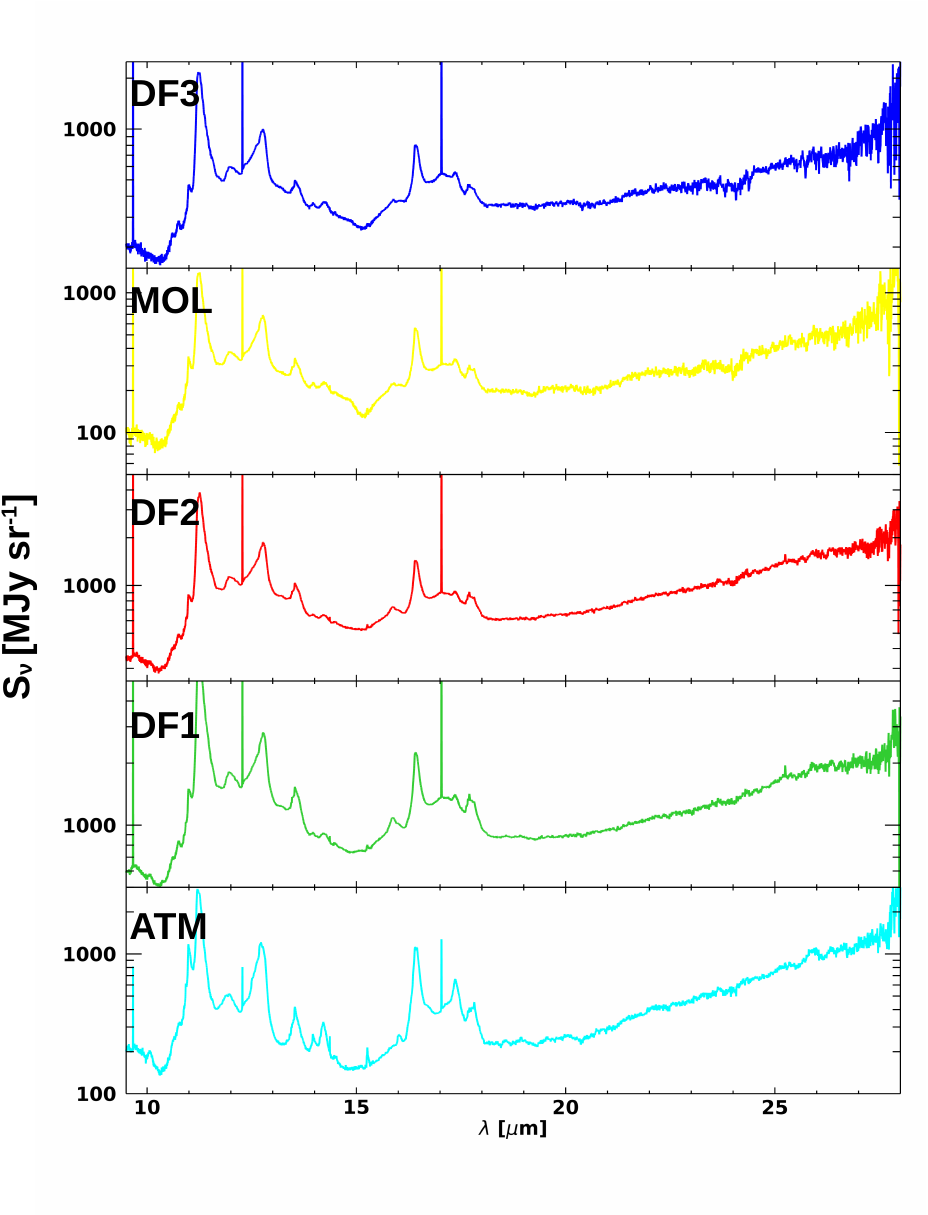}
   \caption{As in Fig. \ref{fig:ngc7023_1p0_3p0_spec} for wavelengths between 
   $9.5 < \lambda < 28.0$~\micron\ for regions in NGC~7023. 
   \label{fig:ngc7023_9p5_28p0_spec}}
\end{figure*}

\begin{figure*}
   \centering
   \begin{tabular}[b]{cc}
     \includegraphics[width=\columnwidth]{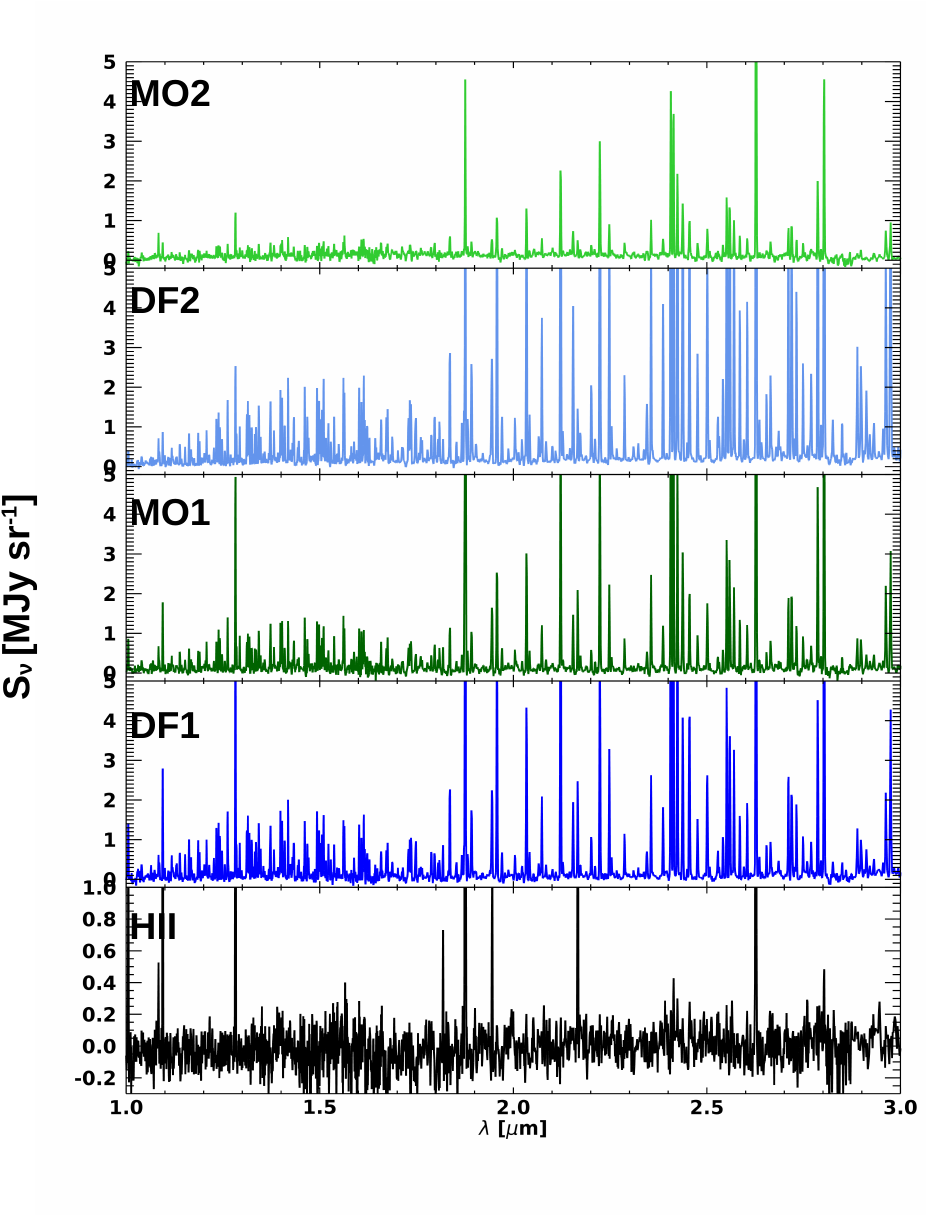} & \includegraphics[width=\columnwidth]{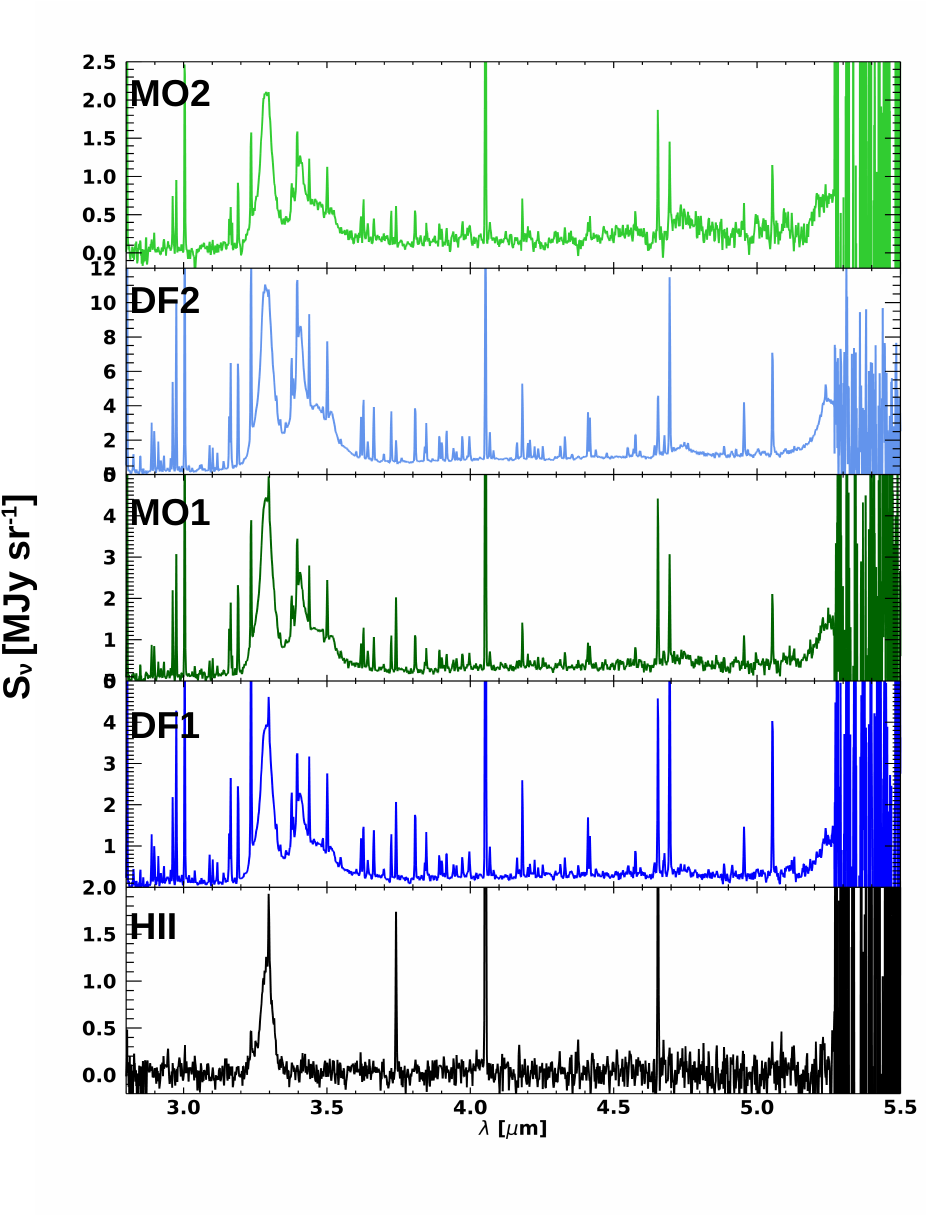} \\
   \end{tabular}
   \caption{As in Fig. \ref{fig:ngc7023_1p0_3p0_spec} for the five regions defined for the Horsehead over the  
   wavelength range $1.0 < \lambda < 3.0$~\micron (left) and $2.8 < \lambda < 5.5$~\micron\ (right). Regions are color code as in 
   Fig. \ref{fig:img_regions} with MO2, DF2, MO1, DF1, and H\,{\sc{ii}} displayed top to bottom.
   \label{fig:horsehead_1p0_3p0_spec}}
\end{figure*}

\begin{figure*}
   \centering
   \begin{tabular}[b]{cc}
     \includegraphics[width=\columnwidth]{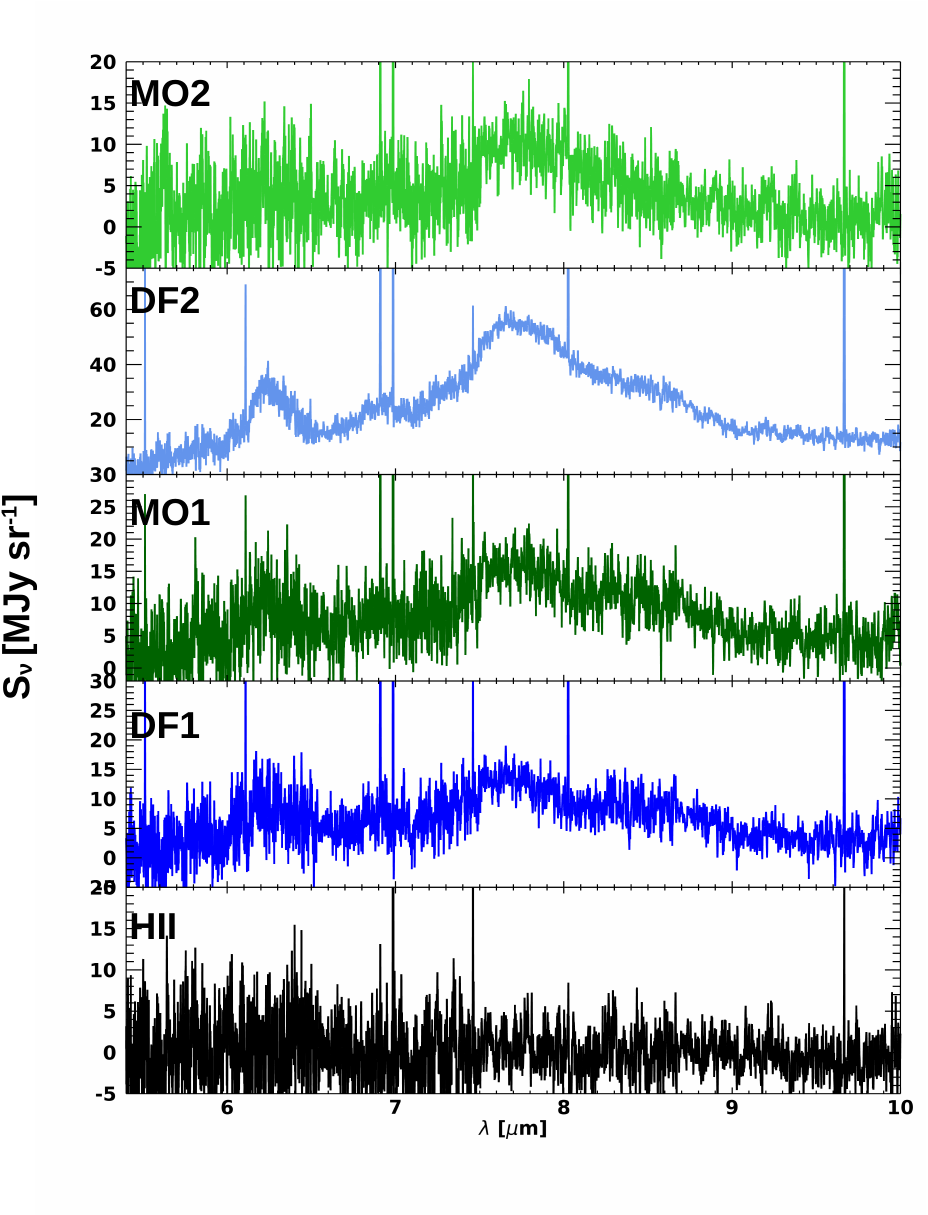} & \includegraphics[width=\columnwidth]{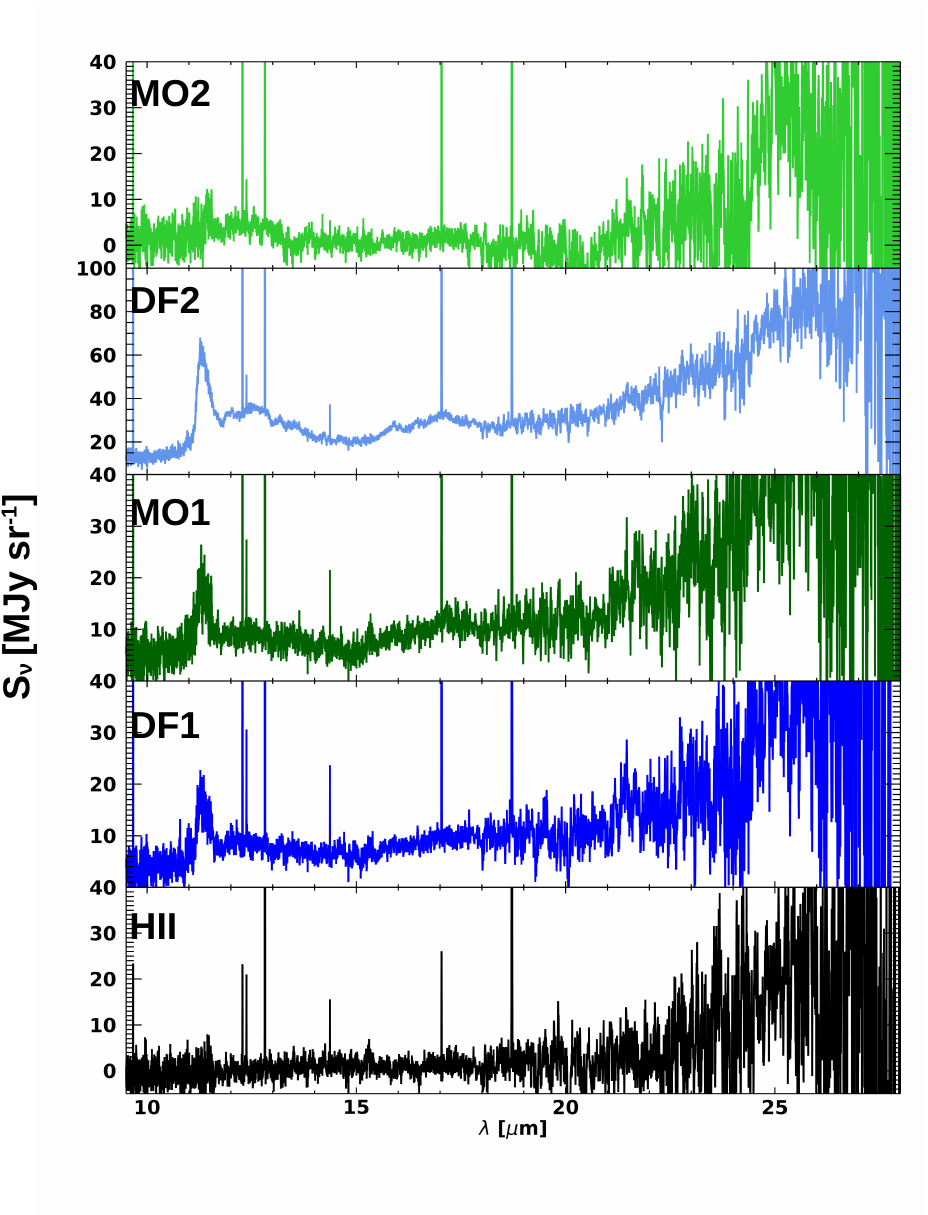} \\
   \end{tabular}
   \caption{As in Fig. \ref{fig:horsehead_1p0_3p0_spec} for wavelengths between 
   $5.4 < \lambda < 10.0$~\micron\ (left) and $9.5 < \lambda < 28.0$~\micron\ (right) for regions in the Horsehead.
   \label{fig:horsehead_9p5_28p0_spec}}
\end{figure*}

\begin{table}[bth]
\begin{threeparttable}[b]
 \caption{Spectral Extraction Sample}\label{tbl:spectra_example}
 \begin{tabular*}{0.45\textwidth}{@{\extracolsep{\fill}}ccccc}
 \toprule
 $\lambda$\tnote{a} & F$_{\nu}$\tnote{b} & $\sigma_{F\nu}$\tnote{b} & C$_{\nu}$\tnote{c} & $\sigma_{C\nu}$\tnote{c} \\
 \midrule
 $\cdots$  &           &                 &           &          \\
 4.40134   &  61.823   & 0.48544         &  61.823   &  1.1719  \\
 4.40313   &  61.177   & 0.47591         &  61.176   &  1.1490  \\
 4.40492   &  61.028   & 0.46505         &  60.881   &  1.1227  \\
 4.40671   &  65.194   & 0.51344         &  60.697   &  1.9657  \\
 4.40850   &  89.511   & 0.77764         &  60.905   &  2.9772  \\
 4.41030   &  98.202   & 0.83242         &  60.402   &  3.1869  \\
 4.41209   &  71.347   & 0.55234         &  60.607   &  2.1146  \\
 4.41388   &  66.718   & 0.52846         &  59.966   &  2.0232  \\
 4.41566   &  85.206   & 0.65192         &  60.018   &  2.4958  \\
 4.41745   &  84.771   & 0.64103         &  59.695   &  2.4541  \\
 4.41924   &  65.547   & 0.48533         &  59.463   &  1.8580  \\
 4.42103   &  60.776   & 0.46027         &  60.417   &  1.1112  \\
 4.42282   &  59.730   & 0.45990         &  59.725   &  1.7607  \\
 4.42461   &  59.044   & 0.46504         &  59.044   &  1.7804  \\
 4.42640   &  58.216   & 0.45652         &  58.216   &  1.7477  \\
 4.42819   &  58.756   & 0.46093         &  58.756   &  1.1128  \\
 4.42999   &  58.778   & 0.47315         &  58.778   &  1.1423  \\
 $\cdots$  &           &                 &           &          \\
 \hline
 \end{tabular*}

 \begin{tablenotes}
   \item [a] Wavelength, \micron.
   \item [b] Flux and uncertainty in MJy sr$^{-1}$ 
   \item [c] Continuum and uncertainty in MJy sr$^{-1}$ 
 \end{tablenotes}

\end{threeparttable}
\end{table}


\section{Line Extraction\label{ap:linedetail}}

Here we describe each step in the process of the line identification procedure outlined in Sect. \ref{ssec:line_finding} and 
Sect. \ref{ssec:line_identity} for a single subsection of the G140M data from the DF1 region in NGC~7023. Refer to 
Fig. \ref{fig:lineidentification_detail} for a visual guide.

\begin{enumerate}
  \item The initial 5-$\sigma$ cut "found" the (4,2)S(1), (3,1)Q(1), (3,1)Q(3), (2,0)O(3), and (3,1)Q(5) lines; note that
    no actual identification was done at this point, just finding lines. \label{lines:item1}
  \item After removal of the lines found in \ref{lines:item1}, the next 5-$\sigma$ pass identified the (3,1)Q(2), (5,3)S(4), 
    (3,1)Q(4) and (5,3)S(3) lines, with the (3,1)Q(2) being particularly broad. \label{lines:item2}
  \item A manual examination/refinement led to identifying the (3,1)Q(2) line as two separate lines, resulting in 10 line 
    candidates for that slice. \label{lines:item3}
\end{enumerate}

With the 10 candidates in this slice of the wavelength coverage, the template line list was parsed and all candidates were
associated with a list of candidate lines. At the identification step, only (3,1)Q(1) line was an unambiguous
identification as the line match was very close in wavelength, relatively low upper energy level ($\sim$17000~K), and there 
were no other lines within several 
resolution elements. For all other lines, refinement based on the wavelength difference, upper energy level, A coefficients, and 
cross-reference to other lines was performed.

\begin{itemize}
  \item (4,2)S(1): for this line, the 2 nearby \HH\ lines had significantly higher excitation energies, 
  smaller A coefficients, and relatively marginal wavelength matches.
    While the nearby He\,{\sc{i}} 3S-Po line is somewhat far from the observed line center, the line fit is somewhat broad.  So, while we identify 
    this line as (4,3)S(1), it may be a blend with the He\,{\sc{i}} line. 
  \item O\,{\sc{i}} 3P-3So multiplet: Given the relative weakness of this line and its blend with two stronger nearby lines,
    the identification is not secure. Alternative identifications with nearby \HH\ lines are tentatively excluded due 
    to low A coefficients and large excitation energies; however, the final identification remains very uncertain.
  \item (3,1)Q(2): The observed line is a good match; the (2,0)O(9) has a lower A coefficient, higher excitation, and is 
    a relatively poor wavelength match. The He\,{\sc{i}} candidate is at the edge of the window of the allowed wavelength match, so is 
    not favored.
    Additionally, the unambiguous identification of the (3,1)Q(1) line supports the identification
    as part of a 'known' series. 
  \item (3,1)Q(3): A well-identified line and part of the (3,1)Q(J) series.  The only nearby line is weaker and somewhat 
    offset from the observed line center.
  \item (5,3)S(4): Well matched in wavelength and strongest \HH\ line near the observed wavelength.
  \item (3,1)Q(4): Well matched in wavelength and, while it does not have the highest A-coefficient of the other \HH\ lines
    within a resolution element of the observed line, it is part of the (3,1)Q(J) series and has lower excitation energy than
    other candidates.
  \item (2,0)O(3): Two very good wavelength matches in \HH. However, the (8,4)O(15) line has a factor of 10 smaller A coefficient
     and much higher excitation.
    There are also nearby HD lines that may contribute to the line profile. 
  \item (3,1)Q(5): Best wavelength match and a part of the (3,1)Q(J) series. Other nearby \HH\ lines are weaker.  There may be 
    some 'contamination' from O\,{\sc{i}} and/or He\,{\sc{i}} lines. 
  \item (5,3)S(3): Best wavelength match; there may be some contribution from N\,{\sc{i}}. 
\end{itemize}

For the NIRSpec identifications, while we have identified the best candidates for all observed lines in a manner 
parallel to the above outline, at the NIRSpec resolution, we cannot unambiguously associate a small subset 
of lines with a unique transition. There are likely blends in some of our identifications and where the 
identification is less secure, we try to include those alternatives in the line lists. In the future,
detailed modeling and correctly differentiating these lines may be possible. 

\begin{figure*}[h!]
\centering
\begin{tabular}[b]{cc}
 \includegraphics[width=0.45\textwidth]{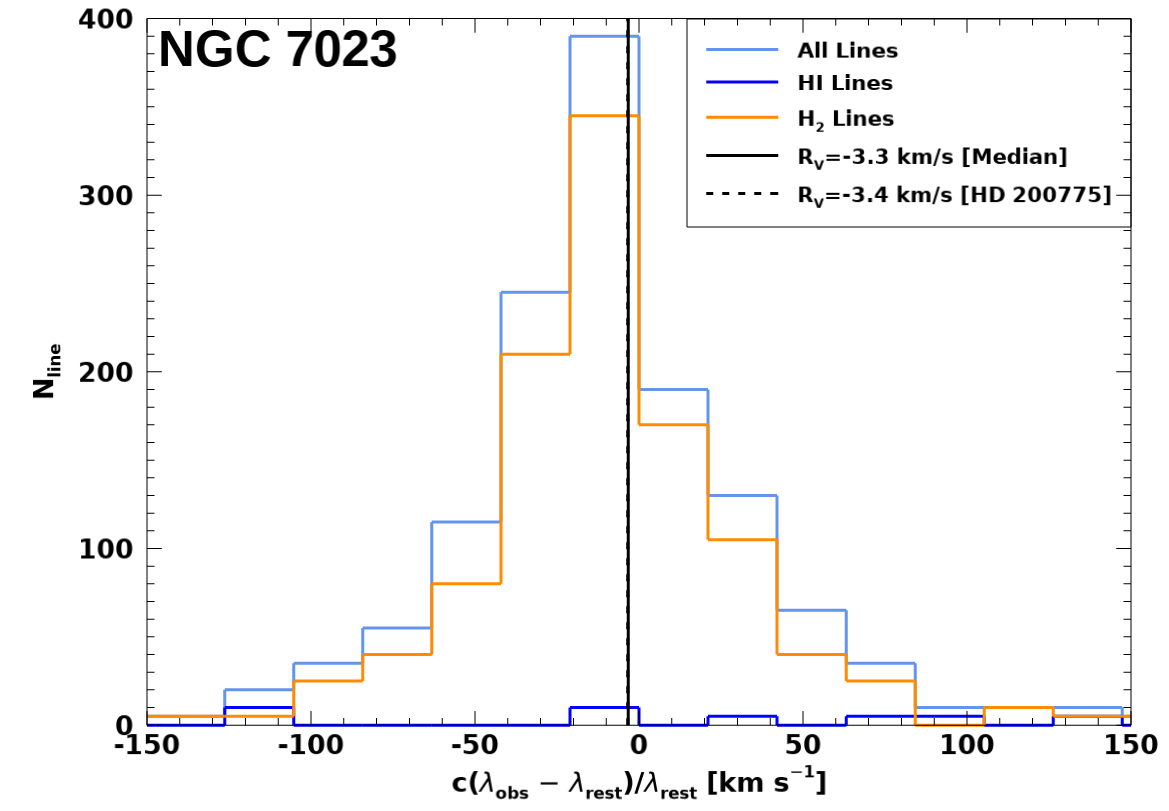}  & \includegraphics[width=0.45\textwidth]{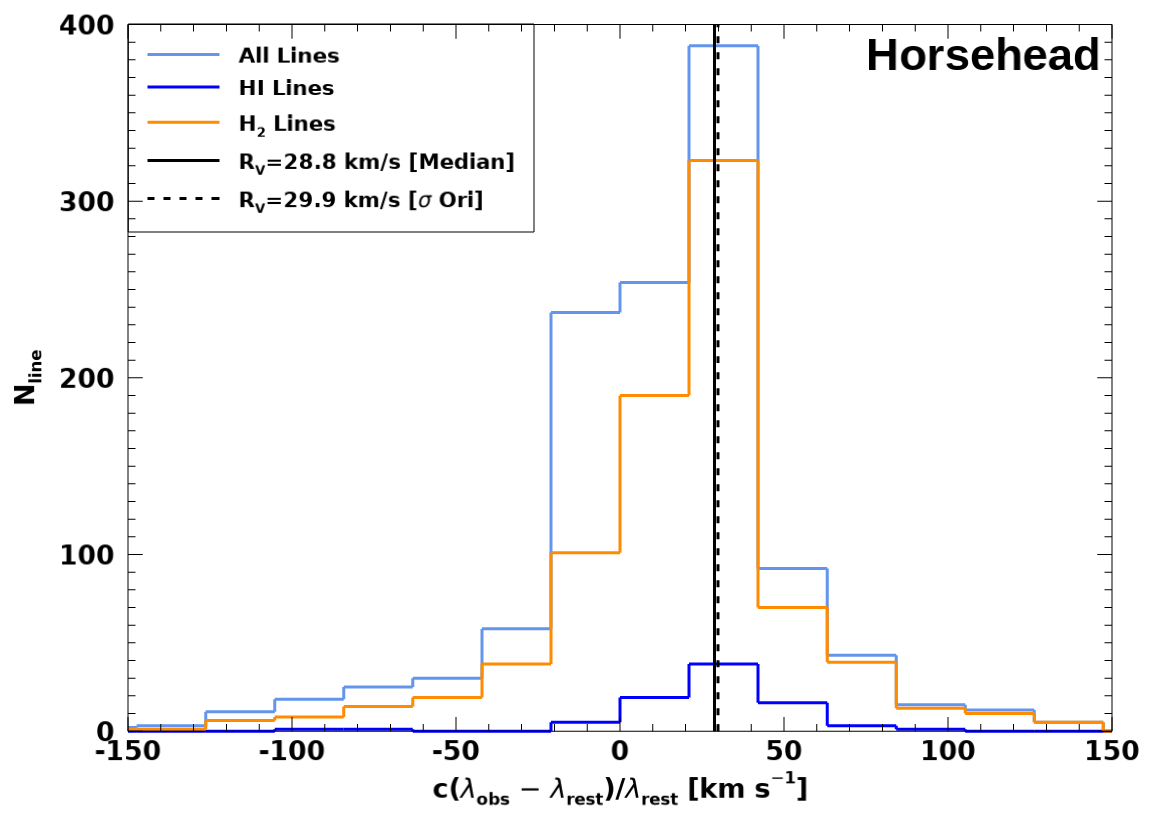} \\
\end{tabular}
\caption{Velocity offset of measured line center from rest wavelength of identified line. NGC~7023 on the left, and the
  Horsehead on the right. Median observed deviation plotted as solid black line and systemtic velocity as the dashed vertical
  black line.
        \label{fig:fracwave}}
\end{figure*}

We plot the velocity offset of our measured line centers with respect to the rest wavelength of the identified
transition in Fig. \ref{fig:fracwave}. For the bulk of our measured lines (in the NIRSpec wavelength regime), the
velocity resolution is $\sim$300 km~s$^{-1}$. In contrast, the derived velocity shifts for are generally much
smaller than a single resolution element and in very good agreement with the systemic velocity of the exciting
stars: $+29.9$ and $-3.4$~km~s$^{-1}$ for $\sigma$~Ori and HD~200775, respectively
\citep{2007AN....328..889K,2006AstL...32..759G}. This close agreement indicates the the instrumental line
profile of both NIRSpec and MIRI is well resolved at the detector.

All line identifications are available as electronic tables for each object and region (10 total). Online
line identification tables follow the naming convention  OBJECT\_REGION\_linelist.tbl
As sample of the table format is given in Table \ref{tbl:linelist_sample}.

\begin{table*}[bth]
\begin{threeparttable}[b]
 \caption{Line List Sample\tnote{a}}\label{tbl:linelist_sample}
 {\footnotesize
 \begin{tabular*}{\textwidth}{ccccccccccc}
 \toprule
 $\lambda_{obs}$\tnote{b} & E$_{\lambda,obs}$\tnote{b} & $\sigma$\tnote{b} & E$_\sigma$\tnote{b} & A\tnote{c} & E$_A$\tnote{c} & F$_{line}$\tnote{d} & E$_{F,line}$\tnote{d} & $\lambda_{0}$\tnote{b,e} & Species & Transition \\ 
 \midrule
 0.98273 & 1.4E-5 & 5.1E-4 & 1.5E-5 & 9.2E-3 & 2.2E-4 & 2.9E-5 & 6.8E-7 & 0.98268 & [CI] & 3P-1D1-2          \\ 
 0.98538 & 5.7E-6 & 5.1E-4 & 5.2E-6 & 3.2E-2 & 3.4E-4 & 1.0E-4 & 1.0E-6 & 0.98530 & [CI] & 3P-1D2-2          \\ 
 1.00254 & 1.2E-4 & 4.5E-4 & 1.4E-4 & 7.5E-4 & 1.8E-4 & 2.2E-6 & 5.4E-7 & 1.00254 & H2   & v=6,J=2$\rightarrow$v=3,J=0  \\ 
 1.00574 & 4.6E-5 & 8.2E-4 & 4.7E-5 & 4.8E-3 & 2.3E-4 & 1.4E-5 & 6.9E-7 & 1.00521 & HI   & 3-7               \\ 
 1.04023 & 7.5E-5 & 1.1E-3 & 7.4E-5 & 3.8E-3 & 2.4E-4 & 1.1E-5 & 6.8E-7 & 1.03945 & H2   & v=5,J=2$\rightarrow$v=2,J=4  \\ 
 1.05335 & 2.1E-5 & 3.3E-4 & 6.6E-5 & 2.5E-3 & 1.1E-4 & 6.6E-6 & 3.2E-7 & 1.05357 & H2   & v=2,J=11$\rightarrow$v=0,J=9 \\ 
 1.06405 & 2.0E-5 & 6.6E-4 & 2.0E-5 & 7.0E-3 & 1.8E-4 & 1.8E-5 & 5.1E-7 & 1.06414 & H2   & v=2,J=9$\rightarrow$v=0,J=7  \\ 
 1.06910 & 2.3E-5 & 7.4E-4 & 2.2E-5 & 7.4E-3 & 2.0E-4 & 1.9E-5 & 5.5E-7 & 1.06921 & NI   & 2Po-2S3/2-1/2     \\ 
 $\cdots$&        &        &        &        &        &        &        &         &      &                   \\ 

 \midrule
 \end{tabular*}
 }

 \begin{tablenotes}
   \item [a] Values in the sample table have been truncated and alternate identifications removed for space reasons. Full tables
   without these spacing shenanigans are provided via CDS.
   \item [b] \micron
   \item [c] Integrated line in MJy~sr$^{-1}$~$\text{\micron}$
   \item [d] Line flux in erg~s$^{-1}$~cm$^{-2}$~sr$^{-1}$
   \item [e] Central wavelength of identified line
 \end{tablenotes}

\end{threeparttable}
\end{table*}

\end{appendix}

\end{document}